\documentclass[11pt, a4paper]{article}
\usepackage{jheppub_kim}
\usepackage{pdflscape}
\usepackage{amsmath}
\usepackage{amssymb}
\usepackage{dcolumn}
\usepackage{bm}
\usepackage{epsfig}
\usepackage{amsfonts}
\usepackage{subfigure}
\usepackage{graphicx}
\usepackage{threeparttable}

\newcommand{\be}{\begin{equation}}
\newcommand{\ee}{\end{equation}}
\newcommand{\bea}{\begin{eqnarray}}
\newcommand{\eea}{\end{eqnarray}}

\def\be{\begin{equation}}
\def\ee{\end{equation}}
\def\bea{\begin{eqnarray}}
\def\eea{\end{eqnarray}}

\begin{document}

\title{Qualitative analysis of Kantowski-Sachs metric in a generic class of
$f(R)$ models}

\author[a]{Genly Leon}
\author[b]{and Armando A. Roque}

\affiliation[a]{Instituto de F\'{\i}sica, Pontificia Universidad  Cat\'olica
de Valpara\'{\i}so, Casilla 4950, Valpara\'{\i}so, Chile}
\affiliation[b]{Grupo de Estudios Avanzados, Universidad de Cienfuegos,
Carretera a Rodas, Cuatro Caminos, s/n. Cienfuegos, Cuba}

\emailAdd{genly.leon@ucv.cl}
\emailAdd{arestrada@ucf.edu.cu}

%\pacs{04.50.Kd, 98.80.-k, 95.36.+x}

\abstract{In this paper we investigate, from the dynamical systems
perspective, the evolution of a Kantowski-Sachs metric in a generic class of
$f(R)$ models. We present conditions (i. e., differentiability conditions,
existence of minima, monotony intervals, etc.) for a free input function
related to the $f(R)$, that guarantee the asymptotic stability of
well-motivated physical solutions, specially, self-accelerated solutions,
allowing to describe both inflationary- and late-time acceleration stages of
the cosmic evolution. We discuss which $f(R)$ theories allows for a cosmic
evolution with an acceptable matter era, in correspondence to the modern
cosmological paradigm. We find a very rich behavior, and amongst others the
universe can result in isotropized solutions with observables in agreement
with observations, such as de Sitter, quintessence-like, or phantom
solutions. Additionally, we find that a cosmological bounce and turnaround
are realized in a part of the parameter-space as a consequence of the 
metric choice.}
\keywords{modified gravity, dark energy, Kantowski-Sachs metric, dynamical
analysis}
%\pacs{98.80.-k, 95.36.+x, 04.50.Kd}

\maketitle

\date{\today}
\section{Introduction}

Several astrophysical and cosmological measurements, including the recent
WMAP nine year release, and the Planck measurements, suggest that the
observable universe is homogeneous and isotropic at the large scale and that
it is currently experiencing an accelerated expansion  phase
\cite{1,2,00,05,06,07}. The explanation of the isotropy and homogeneity of
the universe and the flatness problem lead to the construction of the
inflationary paradigm 
\cite{Starobinsky:1980te,Guth1981,Linde1982,Albrecht1982} \footnote{Reference
\cite{Starobinsky:1980te} is the pioneer model with a de Sitter
(inflationary) stage belonging to the class of modified gravity theories.
These models contain as special case the $R+R^2$ inflationary model which
appears to produce the best fit to the recent WMAP9 and Planck data on the
value of $n_s$ \cite{06,07,Marchini:2013lpp,Marchini:2013oya,Ade:2013zuv}.}.
Usually in the inflationary scenarios the authors start with a homogeneous
and isotropic Friedmann-Robertson-Walker metric (FRW) 
and then are examined the evolution of the cosmological perturbations.
However, the more strong way to proceed is to put from the beginning an
arbitrary metric and then examine if the metric tends asymptotically to the
flat FRW geometry \footnote{This discussion acquires interest since it may be
relevant for the explanation of the anisotropic ``anomalies'' reported in the
recently announced Planck Probe results \cite{Ade:2013zuv}.}. This is a very
difficult program even using a numerical approach, see for example the
references \cite{Goldwirth1989,Goldwirth1990,Deruelle1995}. Thus, several
authors investigated the special case of homogeneous but anisotropic  Bianchi
\cite{Ellis:1968vb,Misner1973,Peebles1993} (see \cite{Tsagas07} and
references therein) and the Kantowski-Sachs metrics
\cite{chernov64,KS66,Kofman:1985dw,Burd:1988ss,Yearsley:1996yg,
Collins:1977fg,Weber:1984xh,Gron:1986ua,Demianski:1988js,Szydlowski:1988jy,
Bombelli:1990ze,
Campbell:1990uu,Chakraborty:1990jp,Mendes:1990eb,VargasMoniz:1992za,
Cavaglia:1994ku,Nojiri:1999iv,Simeone:2000sz,Sanyal:2001pe,Simeone:2002fp,
Li:2003bq,Modesto:2004wm,Kao:2006yk,Chiou:2008eg,Camci:2000bu,Byland:1998gx,
Leon:2010pu,Fadragas:2013ina}. The simplest well-studied but still very
interesting Bianchi geometries are the Bianchi I
\cite{Gurovich:1979xg,Henneaux:1980ft,Muller:1985mr,Starobinsky:1987zz,
Burd:1988ss,Yearsley:1996yg,Berkin:1992ub,Aguirregabiria:1993pm,
Saha:2003xv,Clifton2005,Clifton:2006kc,Barrow:2005dn,Leach2006,
Bandyopadhyay2007,Sharif:2009xa,
MartinBenito:2009qu,Aref'eva:2009vf,Fadragas:2013ina} and  Bianchi III
\cite{Louko:1987gg,Burd:1988ss,Tikekar:1992ia,Christodoulakis:2006vi,
Fadragas:2013ina}, since other
Bianchi models (for
instance the Bianchi IX one), although more realistic, they are much more
complicated. These geometries have been examined analytically, exploring
their rich behavior,
for different matter content of the universe and for different cosmological
scenarios (e.g.,  
 \cite{Misner1973,Peebles1993,Henneaux:1980ft,Muller:1985mr,Burd:1988ss,
Yearsley:1996yg,Berkin:1992ub,Aguirregabiria:1993pm,Barrow:1994nt,
Saha:2003xv,Clifton2005,Clifton:2006kc,Barrow:2005dn,Leach2006,
Bandyopadhyay2007,Sharif:2009xa,
MartinBenito:2009qu,Aref'eva:2009vf,Louko:1987gg,Tikekar:1992ia,
Christodoulakis:2006vi,Collins:1977fg,Weber:1984xh,Sahni1986,
Demianski:1988js,Campbell:1990uu,Chakraborty:1990jp,Mendes:1990eb,
Nojiri:1999iv,Li:2003bq,
Chiou:2008eg,Leon:2010pu,Byland:1998gx,Ibanez:1995zs,Aguirregabiria:1996uh,
vandenHoogen:1996vc,Chimento:1998cf,
vandenHoogen:1998cc,Solomons:2001ef,Middleton:2010bv,Czuchry:2012ad,
Chimento:2012is,
Keresztes:2013tua,Fadragas:2013ina}). 

On the other hand, to explain the acceleration of the expansion one choice is
to introduce the concept of
dark energy
(see \cite{Sahni:1999gb,Peebles:2002gy,Copeland:2006wr} and references
therein), which could be
the simple cosmological constant, a quintessence scalar field 
\cite{Ratra:1987rm,Wetterich:1987fm,Liddle:1998xm,Dutta:2009yb}, a phantom
field \cite{Boisseau:2000pr,Caldwell:1999ew,Nojiri:2003vn,Onemli:2004mb,
Saridakis:2008fy}, or the quintom scenario
\cite{Feng:2004ad,Guo:2004fq,Zhao:2006mp,Lazkoz:2006pa,Lazkoz:2007mx,
Setare:2008pz,Cai:2009zp,Leon:2012vt}. The second one is to consider Extended
Gravity models, specially the $f(R)$- models (see
\cite{Carroll2005,Nojiri:2006ri,Capozziello:2007ec,SotiriouFaraoni2010,
Antonio2010,Nojiri:2010wj,Motohashi:2010zz,Capozziello:2011et} and references
therein), as alternatives to Dark Energy. Other modified (extended)
gravitational scenarios that have gained much interests due to their
cosmological features are the extended nonlinear massive gravity scenario
\cite{Cai:2013lqa,Gannouji:2013rwa,Leon:2013qh,Cai:2012ag}, and the
Teleparalell Dark Energy model \cite{Xu:2012jf,Geng:2011ka,Geng:2011aj}.
However, we follow the mainstream and investigate $f(R)$ models. 

$f(R)$-models have been investigated widely in the literature.  In
particular, Bianchi I models in the context of quadratic and $f(R)$
cosmology were first investigated in \cite{Gurovich:1979xg}, where the
authors showed that anisotropic part of the spacetime metric can be
integrated explicitly. In the reference \cite{Barrow:2006xb} was done a
phase-space analysis of a gravitational theory involving all allowed
quadratic curvature invariants to appear in the Lagrangian assuming for the
geometry the Bianchi type I and type II models, which incorporate both shear
and 3-curvature anisotropies. The inclusion of quadratic terms  provides a
new mechanism for constraining the initial singularity to be isotropic.
Additionally, there was given the conditions under which the de Sitter
solution is stable, and for certain values of the parameters there is a
possible late-time phantom-like behavior. Furthermore, there exist vacuum
solutions with positive cosmological constant which do not approach de Sitter
at late 
times, instead, they inflate anisotropically \cite{Barrow:2006xb}. In the
references \cite{Bamba:2010iy,Motohashi:2011wy,Elizalde:2011ds} were
investigated the oscillations of the $f(R)$ dark energy around the phantom
divide line, 
$w_{DE} = -1$. The analytical condition for the existence of this effect was
derived in \cite{Motohashi:2011wy}. In \cite{Elizalde:2011ds} was
investigated the phantom divide crossing for modified gravity both during the
matter era and also in the de Sitter epoch. The unification of the inflation
and of the cosmic acceleration in the context of modified gravity theories
was investigated for example in \cite{Nojiri:2003ft}.  However, this
model is not viable due the violation of the stability conditions. The first
viable cosmological model of this type was first constructed in
\cite{Appleby:2009uf}, requiring a more complicated $f(R)$ function.  In
\cite{Nojiri:2006gh} was developed a general scheme for modified $f(R)$
gravity reconstruction from any realistic FRW cosmology; another
reconstruction method using cosmic parameters instead of the time law for the
scale factor, was presented and discussed in \cite{Carloni:2010ph}. Finally,
in the reference \cite{Lima:2013gua} is described the cosmological 
evolution predicted by three distinct $f(R)$ theories, with emphasis on the
evolution of linear perturbations.  Regarding to linear perturbations in
$f(R)$ viable cosmological models, the more important effect, which is the
anomalous growth of density perturbations was, prior to reference
\cite{Lima:2013gua}, considered in \cite{Hu:2007nk,Starobinsky:2007hu}.

In this paper we investigate from the dynamical systems perspective the
viability of cosmological models based on Kantowski-Sachs metrics for a
generic class of $f(R)$, allowing for a cosmic evolution with an acceptable
matter era, in correspondence to the modern cosmological paradigm. We present
sufficient conditions (i.e., differentiability conditions, existence of
minima, monotony intervals, an other mathematical properties for a free input
function) for the asymptotic stability of well-motivated physical solutions,
specially, self-accelerated solutions, allowing to describe both inflationary
and late-time acceleration. The procedure used for the examination of
arbitrary $f(R)$ theories was first introduced  in the reference
\cite{Amendola:2006we} and the purpose of the present investigation is to
improve it  and extend it to the anisotropic $f(R)$ scenario. Particularly,
in \cite{Amendola:2006we} the authors demonstrated that the cosmological
behavior of the flat Friedmann- Robertson-Walker $f(R)$ 
models can be understood, from a geometric perspective, by analyzing the
properties of a curve $m(r)$ in the plane $(r,m)$, where $$m=\frac{R
f''(R)}{f'(R)}=\frac{d \ln{f'(R)}}{d \ln R}$$ and $$r=-\frac{R
f'(R)}{f(R)}=-\frac{d \ln{f(R)}}{d \ln R}.$$  However, as discussed in
\cite{Carloni:2007br}, the approach in \cite{Amendola:2006we} is incomplete,
in the sense that the authors consider only the condition $1+r+m(r)=0$ to
define the singular values of $r$ and they omit some important solutions
satisfying $r=0$ and/or $\frac{\dot R}{H R}=0.$ This leads to changes
concerning the dynamics, and to some inconsistent results when comparing with
\cite{Carloni:2007br}. It is worthy to mention that using our approach it is
possible to overcome the previously commented difficulties -remarked in Ref.
\cite{Carloni:2007br}- about the approach of
Ref. \cite{Amendola:2006we} (see details at the end of section \ref{NorM}). 
However, since the analysis in references \cite{Amendola:2006we} and
\cite{Carloni:2007br}, is qualitative, an accurate numerical analysis is
required. This numerical elaboration was done in \cite{Jaime:2012yi} for the
case of $R^n$-gravity, where the authors considered the whole mixture of
matter components of the Universe, including radiation, and without any
identification with a scalar tensor theory in any frame. There was shown
numerically that for the homogeneous and isotropic $R^n$ model, including the
case $n = 2$, an adequate matter
dominated era followed by a satisfactory accelerated expansion is very
unlikely or impossible to happen. Thus, this model seems to be in
disagreement with what is required to be the features of the current Universe
as noticed in \cite{Amendola:2006we}. Regarding the investigations of
the $f(R)\propto R^n,$ it is worthy to note that, in the isotropic vacuum
case, this model can be integrated analytically, see \cite{Muller:1989rp}.

In the reference \cite{Shabani:2013djy} are investigated, from the dynamical
systems viewpoint and by means of the method developed in
\cite{Amendola:2006we}, the so-called $f(R,T)$ theory, where $T$ is the trace
of the energy-momentum tensor. This theory was first proposed in
\cite{Harko:2011kv}. 
In \cite{Shabani:2013djy} was investigated the flat Friedmann- Robertson-
Walker (FRW) background metric. Specially, for the class $f(R,T)=g(R)+h(T),
h\neq 0,$ the only form that respects the conservation equations is
$f(R,T)=g(R)+c_1\sqrt{-T}+c_2$ where $c_i, i=1,2$ do not depend on T, but
possibly depends on $R.$
We recover the results in \cite{Shabani:2013djy} for this particular $f(R,T)$
choice in the isotropic regime. However, our results are more general since
we consider also anisotropy.

In the reference \cite{Abdelwahab:2011dk} are investigated FRW metric on the
framework of $f(R)=R+\alpha R^n$ gravity using the same approach as in
\cite{Carloni:2007br}. Bianchi I universes in $R^n$ cosmologies with torsion
have been investigated in \cite{Carloni:2013hna}. In the reference
\cite{Leon:2010ai} the authors examine the asymptotic properties of a
universe based in a Scalar-tensor theory (and then related through conformal
transformation to $f(R)$-theories). They consider an FRW metric and a scalar
field coupled to matter, also it is included radiation. The authors prove
that critical points associated to the de Sitter solution are asymptotically
stable, and also generalize the results in \cite{Leon:2008de}. The analytical
results in \cite{Leon:2010ai} are illustrated for the important modified
gravity models  $f(R) = R + \alpha R^{2}$ (quadratic gravity) and $f(R) =
R^{n}$. For quadratic gravity, it is proved, using the explicit calculation
of the center manifold of the critical point associated 
to the de Sitter solution (with unbounded scalar field) is locally
asymptotically unstable (saddle point). In this paper we extent these results
to the Kantowski-Sachs metrics. Finally, in the reference \cite{Leon:2010pu}
were investigated $R^{n}$-gravity models for anisotropic Kantowski-Sachs
metric. There were presented conditions for obtaining  late-time
acceleration, additionally, in the range $2 < n < 3$, it is obtained phantom
behavior. Besides, isotropization is achieved irrespectively the initial
degree of anisotropy. Additionally, it is possible to obtain late-time
contracting and cyclic solutions with high probability.  In this paper we
extent the results in \cite{Amendola:2006we,Leon:2010ai,Leon:2010pu} to the
Kantowski-Sachs metric. Particularly, we formalize and extent the geometric
procedure discussed in \cite{Amendola:2006we} in such way that the problems
cited in \cite{Carloni:2007br} do not arise,  and apply the procedure to
``generic'' $f(R)$ models 
for the case of a Kantowski-Sachs metric. By ``generic'' we refer to starting
with a unspecified $f(R)$, and then deduce mathematical properties
(differentiability, existence of minima, monotony intervals, etc) for the
free input functions in order to obtain cosmological solutions compatible
with the modern cosmological paradigm. We extent the results obtained in the
reference \cite{Leon:2010ai} related to the stability analysis for the de
Sitter solution (with unbounded scalar field) for the homogeneous but
anisotropic Kantowski-Sachs metric and we extent to generic  $f(R)$ models of
the results in \cite{Leon:2010pu} that were obtained for $R^n$-cosmologies.
Our results are also in agreement with the related ones in
\cite{Shabani:2013djy}.

The paper is organized as follows. In section \ref{Sect:2} we construct the
cosmological scenario of anisotropic $f(R)$-gravity, presenting the kinematic
and dynamical variables specifying the equations for the Kantowski-Sachs
metric.  Having extracted the cosmological equations in section \ref{Sect:3}
we perform a systematic phase-space and stability analysis of the system. It
is presented a general method for the qualitative analysis of $f(R)$-gravity
without considering an explicit form for the function $f(R)$. Instead we
leave the function $f(R)$ as a free function and use a parametrization that
allows for the treatment of arbitrary (unspecified) $f(R)$ anzatzes.  In
section \ref{Formalism} we present a formalism for the physical description
of the solutions and it is discussed the connection with the cosmological
observables. In the section \ref{Sect:5} we analyze the physical implications
of the obtained results, and  we discuss the cosmological behaviors of a
generic $f(R)$ in a universe with a 
Kantowski-Sachs geometry. In section \ref{Sect:6} are illustrated our
analytical results for a number of $f(R)$-theories. Our main purpose is to
illustrate the possibility to realize the matter era followed by a late-time
acceleration phase. Additionally it is discussed the possibility of a bounce
or a turnaround. Finally, our results are summarized in section \ref{Sect:7}.
 
\section{The cosmological model}\label{Sect:2}
In this section we consider an $f(R)$-gravity theory given in the metric
approach with action \cite{SotiriouFaraoni2010,Antonio2010,Motohashi:2010zz}
\begin{equation} \label{qwert}
 S_{met}=\int_{V}d^{4}x\sqrt{-g}\left[f(R) - 2\Lambda + L_{m} \right],
\end{equation}
where $L_{m}$ is the matter Lagrangian. Additionally, we use the metric
signature ($-1,1,1,1$). Greek indexes run from $0$ to $3$, and we impose the
standard units in which $\kappa^2\equiv 8\pi G = c = 1$. Also, in the following, and
without loss of generality, we set the usual cosmological constant
$\Lambda=0.$  Furthermore, $f(R)$ is a function of the Ricci scalar $R$, that
satisfies the following very general conditions \cite{Appleby:2009uf}:
\begin{enumerate}

\item Existence of a stable Newtonian limit for all values of $R$ where the
Newtonian gravity accurately describes the observed inhomogeneities and
compact objects in the Universe, i.e., for $R\gg R_0\equiv R(t_0)$, where
$t_0$ is the present moment and $R_0$ is the present FRW background value,
and up to curvatures in the center of neutron stars: 
\begin{equation}\label{GR_limit}
|f(R)-R|\ll R,\;|f'(R)-1|\ll 1,\; R f''(R)\ll 1, 
\end{equation}
for $R\gg R_0$. The last of the conditions \eqref{GR_limit} implies that its
Compton
wavelength is much less than the radius of curvature of the
background space-time. Additionally, the conditions
\eqref{GR_limit} guarantee that non-GR corrections to a
space-time metric remain small \cite{Appleby:2009uf}.

\item Classical and quantum stability: 
\begin{equation}\label{stabil}
f'(R)>0, f''(R)>0.  
\end{equation} 
The first condition implies that gravity is attractive and the graviton is
not ghost. Its violation in an FRW background would imply  the
formation of a strong space-like anisotropic curvature singularity with
power-law behavior of the metric coefficients
\cite{Nariai:1973eg,Gurovich:1979xg}. This singularity prevents a transition
to the region where the effective gravitational constant in negative in a
generic, non-degenerate solution. Additionally, note that at the Newtonian
regime, the effective scalaron\footnote{The scalaron is defined by the scalar
field $\Phi\equiv -\sqrt{\frac{3}{2}} \ln f'(R) $ with an effective potential
$V(\Phi(R))\equiv \frac{1}{2 f'(R)^2}\left(R f'(R)-f(R)\right).$}
mass squared is $M_s^2(R)=1/(3F''(R))$. Thus, the second condition in
\eqref{stabil} implies that the scalaron is not a tachyon, i.e., the scalaron
has a finite rest-mass $M_s=(3 f''(R))^{-\frac{1}{2}}.$ If $f''(R)$ becomes
zero for a finite $R=R_c$, then a weak (sudden) curvature singularity forms
generically \cite{Appleby:2009uf}.  

\item  In the absence of matter, exact de Sitter solutions are associated to
positive real roots of the functional equation 
\begin{equation}\label{st_1}
R f'(R)-2 f(R)=0.
\end{equation} 
It is well-known that these kind of solutions (and nearby
solutions) are very important for the description of early inflationary epoch
and the late-time acceleration phase. For the asymptotic future stability of
these solutions near de Sitter ones it is required that
$$f'(R_1)/f''(R_1)>R_1,$$ where $R_1$ satisfies \eqref{st_1}
\cite{Muller:1987hp}.
Specific functional forms satisfying all this conditions have been proposed,
e.g., in the references \cite{Hu:2007nk,Appleby:2007vb,Starobinsky:2007hu}.

\item Finally, we have additionally considered  the condition $R
f'(R)-f(R)\geq 0$ to get a non-negative scalaron potential. 

\end{enumerate}

The fourth-order equations obtained by varying action \eqref{qwert} with
respect to the metric are:
\begin{equation}\label{EinstNew}
 f'(R) R_{\alpha\beta}-\frac{f(R)}{2}g_{\alpha
\beta}-\nabla_{\alpha}\nabla_{\beta}f'(R)+g_{\alpha\beta}\Box
f'(R)=T_{\alpha\beta},
\end{equation}
where the prime denotes differentiation with respect to $R$. In this
expression $T_{\alpha\beta}$ denotes the matter energy-momentum tensor, which
is assumed to correspond to a perfect fluid with energy density $\rho_{m}$
and pressure $p_{m}$, and their ratio gives the matter equation-of-state
parameter
\begin{equation}\label{utop1}
 w=\frac{p_{m}}{\rho_{m}}.
\end{equation} $\nabla_{\alpha}$ is the covariant derivative associated with
the Levi-Civita connection of the metric and $\Box \equiv
\nabla^{\beta}\nabla_{\beta}$. 
Taking the trace of equation \eqref{EinstNew} we obtain  ``trace-equation'' 
\begin{equation}\label{10ANew}
 f'(R)R+3\Box f'(R)-2f(R)= T,
\end{equation}
where $T=T^{\alpha}_{\alpha}$ is the trace of the energy-momentum tensor of
ordinary matter.

Our main objective is to investigate anisotropic cosmologies. Let us assume,
as usual, an anisotropic metric of form \cite{E2003}:
\begin{equation}\label{metrica}
ds^{2}=-N(t)^2
dt^{2}+[e^{1}_{1}(t)]^{-2}dr^{2}+[e^{2}_{2}(t)]^{-2}[d\theta^{2}+S(\theta)^{2
}d\varphi^{2}],
\end{equation}
where $N(t)$ is the lapse function, that we will set $N=1$, $e^{1}_{1}(t)$
and $e^{2}_{2}(t)$  are the expansion scale factors, which in principle can
evolve differently. 

Notice that the metric \eqref{metrica} can describe three geometric families,
that is:
\begin{displaymath}
 S(\theta) = \left\{ \begin{array}{lll}
\theta & \textrm{ for $k=0$} & \textrm{(Bianchi I models),}\\
\sinh\theta & \textrm{for $k=-1$} & \textrm{(Bianchi III models),}\\
\sin\theta & \textrm{for $k=1$} & \textrm{(Kantowski-Sachs (KS) models),}
\end{array} \right.
\end{displaymath}
where $k$ is the spatial curvature parameter.

From the expansion scale factors are defined the kinematic variables 
\begin{subequations}\label{jdfgoig}
\begin{eqnarray}
 H&=&-\frac{1}{3}\frac{d}{dt} {\ln
\left[\;e^{1}_{1}\;(e^{2}_{2})^{2}\right]},\\
 \sigma&=&\frac{1}{3} \frac{d}{dt}{\ln
\left[e^{1}_{1}\;(e^{2}_{2})^{-1}\right]}.
\end{eqnarray}
\end{subequations}

Furthermore, ${}^2\!K=(e^{2}_{2})^{2}$ is the Gauss curvature of the
3-spheres \cite{Henk1996} and their evolution equation is given by
\cite{Genly2008}
\begin{eqnarray}\label{curvaturaK}
\dot{{}^{2}\!K}&=&-2(\sigma+H)\;({}^{2}\!K).
\end{eqnarray}
Additionally, the evolution equation for $e^{1}_{1}$ reads (see equation (42)
in section 4.1 in \cite{Genly2008})
\begin{eqnarray}\label{evolucione1}
\dot{e}^{1}_{1}= \left(-H+2\sigma\right){e}^{1}_{1}.
\end{eqnarray}

From the trace equation \eqref{10ANew} for the Kantowski-Sachs geometry
($k=1$) and assuming the matter content described as a perfect fluid we
obtain:
\begin{equation}\label{TrazaNew}
 -3\frac{d^{2}}{dt^{2}}f'(R)-9H\frac{d}{dt}f'(R)+f'(R)R-2f(R)=
-\rho_{m}+3p_{m}.
\end{equation} 
The Ricci scalar is written as
\begin{equation}\label{RicciNew}
 R= 12H^{2}+6\sigma^{2}+6\dot{H}+2 \;{}^{2}\!K.
\end{equation}

Now, it is straightforward to write the equation \eqref{EinstNew} as
\cite{DeFelice:2010aj} \footnote{Alternatively, we can write the field
equations \eqref{EfEqs} as $G_{\alpha \beta}= T_{\alpha \beta}+T^X_{\alpha
\beta}$, where  $T^X_{\alpha \beta}=G_{\alpha
\beta}\left(1-f'(R)\right)+f''(R) \nabla_\alpha \nabla_\beta
R+f'''(R)\left(\nabla_\alpha R\right)\left(\nabla_\beta R\right) -g_{\alpha
\beta}\left[\frac{1}{2}(R f'(R)-f(R))+f''(R) \Box R+f'''(R)g^{\mu \nu}
\left(\nabla_\mu R\right)\left(\nabla_\nu R\right)\right]$, or $T^X_{\alpha
\beta}=G_{\alpha \beta}\left(1-f'(R)\right)+f''(R) \nabla_\alpha \nabla_\beta
R+f'''(R)\left(\nabla_\alpha R\right)\left(\nabla_\beta R\right)
-\frac{g_{\alpha \beta}}{6}\left[2 T+f(R)+R f'(R)\right]$ (if used the trace
equation $\Box R=\frac{1}{3 f''(R)} \left[T-3 f'''(R)g^{\mu \nu}
\left(\nabla_\mu R\right)\left(\nabla_\nu R\right)+2 f-R f'(R)\right]$ for
eliminating second order derivatives of $R$ with respect to $t$), i.e., the
recipe I in \cite{Jaime:2012gc,Jaime:2010kn}. The above expression for
$T^X_{\alpha \beta}$ studied in \cite{Jaime:2012gc,Jaime:2010kn} corresponds
exactly to the energy momentum tensor (EMT) of
geometric dark energy given by $T^X_{\alpha \beta}=(1-f'(R))R_{\alpha
\beta}+\frac{1}{2}g_{\alpha \beta}(f(R)-R)+\left(\nabla_\alpha
\nabla_\beta-g_{\alpha \beta}\Box\right)f'(R)$ discussed in the references
\cite{Starobinsky:2007hu, Motohashi:2010tb, Motohashi:2010zz,
Motohashi:2011wy}.}:
\begin{subequations}
\label{TmunuEff}
\begin{align}
& A G_{\alpha \beta}= T_{\alpha \beta}+T^{DE}_{\alpha \beta},\label{EfEqs}\\
& T^{DE}_{\alpha \beta}=(A-f'(R))R_{\alpha \beta}+\frac{1}{2}g_{\alpha
\beta}(f(R)-A R)+\left(\nabla_\alpha \nabla_\beta-g_{\alpha
\beta}\Box\right)f'(R), \label{TXEff}
\end{align}
\end{subequations}
where A is some constant. In order to reproduce the standard matter era
($3H^2 \simeq \rho_m$) for $z \gg 1$,
we can choose $A = 1$. An alternative possible choice is $A = F_0$, where
$F_0$ is
the present value of $f'(R)$. This choice may be suitable if the deviation of
$F_0$ from 1 is small (as
in scalar-tensor theory with a nearly massless scalar field
\cite{Boisseau:2000pr,Torres:2002pe}). 
 
Substituting the Kantowski-Sachs metric into the equations \eqref{EfEqs} with
the definition for $T^{DE}_{\alpha \beta}$ given by \eqref{TXEff},  and after
some algebraic manipulations, we obtain  
\begin{subequations}\label{EERROP}
\begin{eqnarray}
 A\left(3H^{2}-3\sigma^{2}+{}^{2}\!K\right)&=&\rho_{m} +\rho_{DE},
\label{EERROPa}\\
 A\left(-3(\sigma+H)^{2}-2\dot{\sigma}-2\dot{H}-{}^{2}\!K\right)&=&p_{
m}+p_{DE}-2\pi_{DE}, \label{EERROPb}\\
 A\left(-3\sigma^{2}+3\sigma
H-3H^{2}+\dot{\sigma}-2\dot{H}\right)&=&p_{m}+p_{DE}+\pi_{DE}.
\label{EERROPc}
\end{eqnarray}
\end{subequations}
where \begin{subequations}\label{mnvcdf}
\begin{eqnarray}
\rho_{DE}&=& -3 H \frac{d}{dt}f'(R)+\frac{1}{2}(R
f'(R)-f(R))+\left(3H^2-3\sigma^2+{}^{2}\!K\right)(A-f'(R)), \label{mnvcdfa}\\
  p_{DE}&=& \frac{d^2}{dt^2}f'(R)+2H\frac{d}{dt}
f'(R)+\frac{1}{2}(f(R)-R f'(R))+\nonumber\\ && -\left(3H^2+2\dot
H+3\sigma^2+\frac{1}{3}\,{}{}^{2}\!K\right)(A-f'(R)),\label{mnvcdb}\\
  \pi_{DE}&=&-\frac{\frac{d}{dt}f'(R)}{f'(R)}A\sigma,\label{mnvcdfc}
\end{eqnarray}
\end{subequations}
denote, respectively, the isotropic energy density and pressure and the
anisotropic pressure of the effective energy-momentum tensor for Dark Energy
in modified gravity. Note that for the choice of an FRW background, where
$\sigma={}^{2}\!K=0$, we recover the equations (4.94) and (4.95) in the
review \cite{DeFelice:2010aj}.    

The advantage of using the expression \eqref{TXEff} for the definition of the
effective energy-momentum tensor for Dark Energy in modified gravity, instead
of using the alternative ``curvature-fluid'' energy-momentum tensor
\cite{Goswami:2008fs,Goheer:2008tn}
$$T_{\mu \nu}^{eff}=\frac{1}{f'(R)}\left[\frac{1}{2} g_{\mu
\nu}\left(f(R)-R f'(R)\right) + \nabla_\mu \nabla_\nu
f'(R)-g_{\mu \nu}\Box f'(R)\right],$$   is that by construction
\eqref{TXEff} is always conserved, i.e., $\nabla_\nu {T^{DE}}_{\mu}^{\nu}=0$,
leading to the conservation equation for the effective Dark Energy, which in
anisotropic modified gravity is given by \cite{Genly2008}:
\begin{equation}
\dot \rho_{DE}+3 H(\rho_{DE}+p_{DE})+6\sigma\pi_{DE}=0.
\end{equation} On the other hand, $T_{\mu \nu}^{eff}$ is not conserved in
presence of mater (it is conserved for vacuum solutions only). 

Now, combining the equations \eqref{EERROPa} and \eqref{mnvcdfa}, we obtain 
\begin{equation}
f'(R)\rho_{tot}=-3 A H\frac{d}{dt}f'(R)+\frac{1}{2}A \left[R
f'(R)-f(R)\right]+A \rho_m,
\end{equation} where we have defined the total (effective) energy density
$\rho_{tot}=\rho_{DE}+\rho_m.$  
Eliminating $\pi_{DE}$ between \eqref{EERROPb} and \eqref{EERROPc}, and using
\eqref{RicciNew},
we acquire 
\begin{align}
p_{tot}&=-\frac{A}{3}\left[9(H^2+\sigma^2)+6\dot H+ {}^2\!K\right]\nonumber\\
       &=\frac{A}{3}\left[3(H^2-\sigma^2)-R+{}^2\!K\right],
\end{align}
where we have defined the total (effective) isotropic pressure
$p_{tot}=p+p_{DE}.$ 
Thus, we can define the effective equation of state parameter 
\begin{equation}
w_{tot}\equiv \frac{p_{tot}}{\rho_{tot}}=\frac{2 f'(R) \left(-{}^2\!K-3
H^2+R+3 \sigma^2\right)}{3 \left[6 H \dot R f''(R)-R f'(R)+f(R)-2
\rho_m\right]}
\end{equation}
 
Additionally,  we define the observational density parameters:
\begin{itemize}
 \item the spatial curvature:
    \begin{equation}\label{hgt12}
     \Omega_{k}\equiv -\frac{^{2}K}{3H^{2}},
    \end{equation}
  \item the matter energy density:
    \begin{equation}
      \tilde{\Omega}_{m}\equiv\frac{\rho_{m}}{3 A H^{2}},
    \end{equation}
  \item the ``effective Dark Energy'' density:
   \begin{equation}
      \tilde{\Omega}_{DE}\equiv \frac{\rho_{DE}}{3A H^{2}},
    \end{equation}
  \item the shear density: 
    \begin{equation}\label{hgt123}
     \Omega_{\sigma}\equiv\left(\frac{\sigma}{H}\right)^{2},
    \end{equation}
\end{itemize}
satisfying
$\Omega_{k}+\tilde{\Omega}_{m}+\tilde{\Omega}_{DE}+\Omega_{\sigma}=1$.

Now, for the homogeneous but anisotropic Kantowski-Sachs metric, the
Einstein's equations \eqref{EinstNew} along with \eqref{TrazaNew} can be
reduced with respect to the time derivatives of $\dot{\sigma}$, \,${}^{2}\!K$
and $\dot{H}$, leading to:
 the equation for the shear evolution:
\begin{equation}\label{shear}
 \dot{\sigma}=
-\sigma^{2}-3H\sigma+H^{2}-\frac{\rho_{m}}{3f'(R)}-\frac{1}{6}\left[R-\frac{
f(R)}{f'(R)}\right]+\frac{(H-\sigma)}{f'(R)}\frac{d}{dt}f'(R),
\end{equation}
the Gauss constraint:
\begin{equation}\label{restriccion}
^{2}K=3\sigma^{2}-3H^{2}-\frac{\rho_{m}}{f'(R)}+\frac{1}{2}\left[R-\frac{f(R)
}{f'(R)}\right]-\frac{3H}{f'(R)}\frac{d}{dt}f'(R),
\end{equation}
and the Raychaudhuri equation:
\begin{equation}\label{Raychaudhuri}
 \dot{H}=
-H^{2}-2\sigma^{2}-\frac{1}{6f'(R)}\left[\rho_{m}+3p_{m}\right]-\frac{H}{
2f'(R)}\frac{d}{dt}f'(R)+\frac{1}{6}\left[R-\frac{f(R)}{f'(R)}-\frac{3}{f'(R)
}\frac{d^{2}}{dt^{2}}f'(R)\right].
\end{equation}
Using the trace equation \eqref{TrazaNew} it is possible to eliminate the
derivative $\frac{d^{2}}{dt^{2}}f'(R)$ in \eqref{Raychaudhuri}, obtaining a
simpler form of the Raychaudhuri equation:
\begin{equation}\label{Raychaudhurinew}
 \dot{H}=
-H^{2}-2\sigma^{2}-\frac{\rho_{m}}{3f'(R)}+\frac{f(R)}{6f'(R)}+\frac{1}{f'(R)
}H\frac{d}{dt}f'(R).
\end{equation}
Furthermore, the Gauss constraint \eqref{restriccion} can alternatively be
expressed as
\begin{equation}\label{restriccionnew}
\left[H+\frac{1}{2}\frac{\frac{d}{dt}f'(R)}{f'(R)}\right]^{2}+\frac{1}{3}\;^{
2}K=\sigma^{2}+\frac{\rho_{m}}{3f'(R)}+\frac{1}{6}\left[R-\frac{f(R)}{f'(R)}
\right]+\frac{1}{4}\left[\frac{\frac{d}{dt}f'(R)}{f'(R)}\right]^{2}.
\end{equation}
Finally, the evolution of matter conservation equation is:
\begin{equation}\label{ConsM}
\dot{\rho}_m=-3\gamma H\rho_m,
\end{equation}
where the perfect fluid, with equation of state $p_m=(\gamma-1)\rho_m$,
satisfies the standard energy conditions which implies $1\leq\gamma\leq2$.

In summary, the cosmological equations of $f(R)$-gravity in the
Kantowski-Sachs background are the ``Raychaudhuri equation''
\eqref{Raychaudhurinew}, the shear evolution \eqref{shear}, the trace
equation \eqref{TrazaNew}, the Gauss constraint \eqref{restriccionnew}, the
evolution equation for the 2-curvature ${}^2\!K$ \eqref{curvaturaK} and the
evolution equation for $e^{1}_{1}$ \eqref{evolucione1}. Finally, these
equations should be completed by considering the evolution equation for
matter source \eqref{ConsM}.  These equations contains, as a particular case,
the model $f(R)=R^n$ investigated in \cite{Leon:2010pu}.

\section{The dynamical system}\label{Sect:3}

In the previous section we have formulated the $f(R)$-gravity for the
homogeneous and anisotropic Kantowski-Sachs geometry. In this section we
investigate, from the dynamical systems perspective, the cosmological model
without considering an explicit form for the function $f(R)$. Instead we
leave the function $f(R)$ as a free function and use a parametrization that
allows for the treatment of arbitrary (unspecified) $f(R)$ anzatzes. 

\subsection{Parametrization of arbitrary $f(R)$ functions}\label{arb}

For the treatment of arbitrary $f(R)$ models, we introduce, following the
idea in \cite{Amendola:2006we}, the functions 
\begin{subequations}
\begin{eqnarray}
m=\frac{R f''(R)}{f'(R)}=\frac{d \ln{f'(R)}}{d \ln{R}} \label{funcm},\\
r=-\frac{R f'(R)}{f(R)}=-\frac{d \ln{f(R)}}{d \ln(R)}\label{funcr}.
\end{eqnarray} 
\end{subequations}
Now assuming that $m$ is a singled-valued function of $r,$ say $m=m(r)$, and
leaving the function $f(R)$ still arbitrary, it is possible to obtain a
closed dynamical system for $r$ and for a set of normalized  variables. On
the other hand, given $m(r)$ or 
\begin{eqnarray}\label{FuncionMM}
M(r)=\frac{r(1+r+m(r))}{m(r)},
\end{eqnarray} as input, it is possible to reconstructing the original $f(R)$
function as follows. First, deriving in both sides of \eqref{funcr} with
respect to $R$, and using the definitions \eqref{funcm} and \eqref{funcr}, is
deduced that
\begin{eqnarray}
\frac{dr}{dR}=\frac{r(1+m(r)+r)}{R}.
\end{eqnarray}
Separating variables and integrating the resulting equation, we obtain the
quadrature
\begin{eqnarray}\label{funcionRRRRR}
R(r)=R_{0}\exp{\left[\int\frac{dr}{r(1+m(r)+r)}\right]}.
\end{eqnarray}
Second, using the definition of $r$ we obtain:
\begin{eqnarray}
r=-\frac{d\ln{f(R)}}{d\ln{R}}.\nonumber
\end{eqnarray}
Reordering the terms at convenience are deduced the expressions 
\begin{eqnarray}\label{funcionfRR}
-r d\ln{R}&=&d\ln{f(R)},\nonumber\\
-\int r d\ln{R}&=&\ln{\left[\frac{f(R)}{f_{0}}\right]},\nonumber\\
f(R)&=&f_{0} \exp{\left(-\int r d\ln{R}\right)}.
\end{eqnarray}
Substituting \eqref{funcionRRRRR} in \eqref{funcionfRR} we obtain 
\begin{eqnarray}\label{frnewrrt}
f(r)=f_{0} \exp{\left(-\int \frac{1}{1+m(r)+r}\;dr\right)}.
\end{eqnarray}
Finally, from the equations \eqref{funcionfRR} and \eqref{funcionRRRRR} we
obtain $f(R)$ by eliminating the parameter $r$. 
In the table \ref{Funcionesfr} are shown the functions $m(r)$ and $M(r)$ for
some $f(R)$-models.

\begin{table*}[h]
\begin{center}
  \footnotesize  %cambia a tama�o de letra m'as peque�o
\begin{tabular}{ccc}
\hline
\hline\\[-0.3cm]
$f(R)$ & $m(r)$ &$M(r)$ \\[0.2cm]
\hline\\[-0.2cm]
%%%%%%%%
$\alpha R^n$ & $-r-1$ & $0$ \\[0.3cm]
$R+\alpha R^n$ & $\frac{n(1+r)}{r}$ & $\frac{r(n+r)}{n}$  \\[0.3cm]
$R^p\left(\ln \alpha R\right)^q$ & $\frac{(p+r)^{2}-r(r+1)q}{q r} $ &
$\frac{r(p+r)^{2}}{(p+r)^{2}-r(r+1)q}$  \\[0.3cm]
$R^p \exp \left(q R\right)$ & $\frac{p-r^2}{r}$ & $-\frac{r (p+r)}{r^2-p}$ 
\\[0.3cm]
$R^p \exp \left(\frac{q}{R}\right)$ & $-\frac{r^2+2 r+p}{r}$ & $\frac{r
(p+r)}{r^2+2 r+p}$  \\[0.3cm]\hline \hline
\end{tabular}\end{center}
\caption[crit]{\label{Funcionesfr} Some examples of $f(R)$-models}
\end{table*}

Therefore, following the above procedure, we can transform our cosmological
system into a closed dynamical system for a set of normalized, auxiliary,
variables and $r$. Such a procedure is possible for arbitrary $f(R)$ models,
and for the usual ansatzes of the cosmological literature it results to very
simple forms of $M(r)$, as can be seen in Table \ref{Funcionesfr}.   In
summary, with the introduction of the variables $M$ and $r$, one adds an
extra direction in the phase-space, whose neighboring points correspond to
``neighboring'' $f(R)$-functions. Therefore, after the
general analysis has been completed, the substitution of the specific $M(r)$
for the desired function $f(R)$ gives immediately the specific results. Is
this crucial aspect of the method the one that make it very powerful,
enforcing its applicability.

To end this subsection let us comment  about anisotropic curvature (strong)- and weak-singularities.

It is well-known that the violation of the stability condition $f'(R)>0$ during the evolution of a FRW background results in the immediate lost of homogeneity and isotropy, and thus, an anisotropic curvature singularity is generically formed \cite{Nariai:1973eg,Gurovich:1979xg}. On other hand, at the instance where $f''(R)$ becomes zero for a finite $R=R_c,$ also an undesirable weak singularity forms \cite{Appleby:2009uf}. In this section we want  to discuss more about this kind of singularities.

\begin{itemize}

\item 
At the regime where $f'(R)$ reaches zero for finite $R$, an anisotropic curvature (strong) singularity with power-law behavior of the metric coefficients is generically formed \cite{Nariai:1973eg,Gurovich:1979xg}. This singularity prevents a transition to the region where the effective gravitational constant in negative in a generic, non-degenerate solution. 
Now let us characterize this singularity in terms of our dynamical variables. 
Observe that 
\begin{equation}
f'(R)=-\frac{r f(R)}{R}.
\end{equation}
Hence, for $f(R)= {\mathcal O}(R),$ and $R\neq 0$, the singularity corresponds to the value $r_c=0.$ 
\item 
At the regime where $f''(R)$ becomes zero, for a non-zero $f'(R)$, a weak singularity develops generically \cite{Appleby:2009uf}. 
Now let us characterize this singularity in terms of our dynamical variables. 
Observe that 
\begin{equation}
f''(R)=-\frac{r m(r) f(R)}{R^2}.
\end{equation}
Hence, for $f(R)= {\mathcal O}(R^2),$ the singularity corresponds to the values $r_c$ such that $m(r_c)=0.$ \footnote{$r$ must be different to zero since it is required $f'(R) \neq 0$ at the singularity.}

\end{itemize}

A more complete analysis of all possible singularities requires further investigation and is left for future research.

\subsection{Normalization and Phase-space}\label{NorM}

In order to do a systematic analysis of the phase-space, as well as doing the
stability analysis  of cosmological models, it is convenient to transform the
cosmological equations into their autonomous form
\cite{Copeland1998,Leon:2010ai,Abdelwahab:2011dk}. This will be achieved by
introducing the normalized variables:
\begin{subequations}\label{VRDN}
\begin{align}
& Q=\frac{H}{D},\; \Sigma=\frac{\sigma}{D},\; x=\frac{1}{2 D
f'(R)}\frac{df'(R)}{dt},
y=\frac{1}{6 D^2}\left[R-\frac{f(R)}{f'(R)}\right],\label{3.7a}\\ &
z=\frac{\rho_m}{3 D^2 f'(R)},\;  K=\frac{\;^2K}{3 D^2},
\end{align}
\end{subequations}
where we have defined the normalization factor:
\begin{eqnarray}\label{DDnew}
D&=&\sqrt{\frac{(^2K)}{3}+\left(H+\frac{1}{2
f'(R)}\frac{df'(R)}{dt}\right)^2}\equiv\sqrt{\frac{(^2K)}{3}+\left(H+\frac{
f''(R)}{2 f'(R)}\frac{dR}{dt}\right)^2}.
\end{eqnarray}
From the definitions \eqref{VRDN} we obtain the bounds $y\geq 0,\, z\geq 0$
and $K\geq 0$. However, $r$ can in principle take values over the whole real
line. 

Now, from the Gauss constraint \eqref{restriccionnew} and the equation
\eqref{DDnew} follows the algebraic relations
\begin{subequations}\label{Eliminacion}
\begin{align}
& x^2+y+z+\Sigma^2=1,   \\
& K+(Q+x)^2=1,
\end{align}
\end{subequations}
that allow to express to $z$ and $K$ in terms of the other variables. Thus,
our relevant phase space variables will be $r$ and the variables
\eqref{3.7a}. 

Using the variables \eqref{3.7a}, and the new time variable $\tau$ defined as
\[d\tau=D dt,\]  we obtain the autonomous system
\begin{subequations}\label{Sistema00}
\begin{align}
&r'=2  x M(r),\label{main_eq}\\
&Q'=-\frac{1}{2(1+r)}
\{2(1+r)-2(1+r)x^{2}-2ry+2Q^{3}(1+r)\Sigma+2(1+r)\Sigma^{2}+ \nonumber\\
& \ \ \ \ \ \ \ \ \ \ \ \ +
Q^{2}(1+r)(2+3x^{2}(-2+\gamma)+4x\Sigma-6\Sigma^{2}+3\gamma(-1+y+\Sigma^{2}
))+ \nonumber\\
&\ \ \ \ \ \ \ \ \ \ \ \ +
Q(1+r)(-2\Sigma+x(-2+3x^{2}(-2+\gamma)+2x\Sigma-6\Sigma^{2}+\nonumber\\
&\ \ \ \ \ \ \ \ \ \ \ \ +3\gamma(-1+y+\Sigma^{2})))\},
\\
&\Sigma'=\frac{1}{2}\{
-2+2(Q+x)^2-3(Q+x)(2+x^2(-2+\gamma)+(-1+y)\gamma)\Sigma- \nonumber\\
&\ \ \ \ \ \ \ \ \ \ \ \
-2(-1+Q+x)(1+Q+x)\Sigma^2-3(Q+x)(-2+\gamma)\Sigma^3\},
\\
&x'=-\frac{1}{2(1+r)}\;\{-4-4r+6Qx+6Qrx+10x^{2}+10rx^{2}-6Qx^{3}- \nonumber\\
&\ \ \ \ \ \ \ \ \ \ \ \
-6Qrx^{3}-6x^{4}-6rx^{4}+2ry+3\gamma+3r\gamma-3Qx\gamma- \nonumber\\
&\ \ \ \ \ \ \ \ \ \ \ \
-3Qrx\gamma-6x^{2}\gamma-6rx^{2}\gamma+3Qx^{3}\gamma+3Qrx^{3}\gamma+        
\nonumber\\
&\ \ \ \ \ \ \ \ \ \ \ \
+3x^{4}\gamma+3rx^{4}\gamma-3y\gamma-3ry\gamma+3Qxy\gamma+3Qrxy\gamma+       
\nonumber\\
&\ \ \ \ \ \ \ \ \ \ \ \ +
3x^{2}y\gamma+3rx^{2}y\gamma+2(1+r)x(-1+Q+x)(1+Q+x)\Sigma+        
\nonumber\\
&\ \ \ \ \ \ \ \ \ \ \ \ +(1+r)(4+3x(Q+x)(-2+\gamma)-3\gamma)\Sigma^{2}\},
\\
&y'=-\frac{1}{1+r}\; y\;
\{3(1+r)x^3(-2+\gamma)+(1+r)x^{2}(3Q(-2+\gamma)+2\Sigma)+       \nonumber\\
&\ \ \ \ \ \ \ \ \ \ \ \
+x(4+2r-3\gamma-3r\gamma+3y\gamma+3ry\gamma+4Q(1+r)\Sigma+   
  \nonumber\\
&\ \ \ \ \ \ \ \ \ \ \ \
+3(1+r)(-2+\gamma)\Sigma^{2})+(1+r)(-2\Sigma+Q(2(Q-3\Sigma)\Sigma+ 
 \nonumber\\
&\ \ \ \ \ \ \ \ \ \ \ \ +3\gamma(-1+y+\Sigma^{2})))\}
\end{align}
\end{subequations}
where ( ${}'$ ) denotes derivative with respect to the new time variable
$\tau$,  defined in the phase space:
\begin{eqnarray}\label{DOm11}
 \Psi&=& \left\{(r,Q,\Sigma,x,y)| r\in \mathbb{R},
Q\in[-2,2],\Sigma\in[-1,1], |Q+x|\leq1, x\in[-1,1],  \right. \nonumber \\ &&
\left. \ \ \ \ \ \ \ \ \ \ \ \ \ \ \ \ \ \ \ \ y\in[0,1], 
x^2+y+\Sigma^2\leq1 \right\}.
\end{eqnarray}
Observe that the phase space \eqref{DOm11} is in general non-compact since
$r\in \mathbb{R}$. Additionally, in the cases where $M(r)$ has poles, i.e.,
given $M(r)=P(r)/Q(r),$ there are r-values, $r_i,$ $r_i<r_{i+1},$ such that
$Q(r_i)=0,$ then $r$ cannot take values on the whole real line, but in the
disconnected region ${\cal R}=\cup_i (r_{i},r_{i+1})$ where $r_i$ runs over
the set of poles of $M(r).$ In these cases, due the fact that $Q(r)=0$
defines a singular surface in the phase space, the dynamics of the system is
heavily constrained. In particular, it implies that do not exist global
attractor, so it is not possible to obtain general conclusions on the
behavior of the orbits without first providing information about
the initial conditions and the functional form of $M(r)$
\cite{Carloni:2007br}. The study is more difficult to address if $P(r)$ and
$Q(r)$ vanish simultaneously, since this would imply the existence of
infinite eigenvalues. For example, as discussed in \cite{Carloni:2007br}, for
the model $R^p \exp \left(q R\right)$, some of the eigenvalues
diverge for $p = 0, 1$. This is a consequence of the fact that for these two
values of the parameter the cosmological equations assume a special form and
is not a pathology of the system. 

It is worth to mention that our variables are related with those introduce in
the reference \cite{Amendola:2006we} by the relation:
\begin{align}\label{comparison}
Q d\tau=d N,\; r=\frac{x_3}{x_2}, \; D Q=H, \; x=-\frac{1}{2}Q x_1,\; y= Q^2
(x_2+x_3). 
\end{align}
So, in principle, we can recover their results in the absence of radiation
($x_4=0$), by taking the limit $|Q+x|\rightarrow 1, \Sigma\rightarrow 0.$

Finally, comparing with the results in \cite{Amendola:2006we} we see that our
equation \eqref{main_eq} reduces to the equation 
\begin{equation}\label{equiv_eq}
\frac{d r}{d\ln a}=r(1+r+m)\frac{\dot R}{H R}
\end{equation} presented in \cite{Amendola:2006we} for the choice $\sigma=0,
^2K=0$ (up to a time scaling, which of course preserves the qualitative
properties of the flow). From \eqref{main_eq} it follows that the asymptotic
solutions (corresponding to fixed points in the phase space) satisfy $x=0$ or
$M(r)=0$ or both conditions.  Due to the equivalence of \eqref{main_eq} and
\eqref{equiv_eq} when $\sigma=0, ^2K=0$, it follows that the solutions
having $M(r)=0$ and/or $x=0$ contain as particular cases those solutions
satisfying $r=0$ and/or $\frac{\dot R}{H R}=0$ that where omitted in
\cite{Amendola:2006we} according to \citep{Carloni:2007br}. In fact, $r=0$ is
one of the values that annihilates the function and due the  identity
$\frac{\dot R}{H R}\equiv \frac{2x}{Q m(r)}$ it implies that, for $m(r)$
finite and nonzero,  the points having $x=0$ also satisfy $\frac{\dot R}{H
R}=0$. For the above reason the criticism of \cite{Carloni:2007br} is not
applicable to the present paper.

\subsection{Stability analysis of the (curves of) critical points}

In order to obtain the critical points we need the set the right hand side of
\eqref{Sistema00} equal to the vector zero. From the first of equations
\eqref{Sistema00} are distinguished two cases: the first are the critical
points satisfying $x=0$, and the second one are those corresponding to an
$r$-coordinate such that $M(r)=0.$ We denote de $r$-values where $M(r)$ is
zero by $r=r^{*}$, i.e., $r^{*}=M^{-1}(0)$.
In the tables \ref{tab2} and \ref{tab1}, are presented the critical points
for the case \mbox{($r=r^{*}=M^{-1}(0)$)} and $x=0$ respectively. 

\begin{table*}[h]
\begin{center}\resizebox{\columnwidth}{!}{
\begin{tabular}{@{\hspace{7pt}}l@{\hspace{12pt}}c@{\hspace{12pt}}c@{\hspace{
12pt}}c@{\hspace{12pt}}c@{\hspace{12pt}}c@{\hspace{12pt}}l}
\hline
\hline\\[-0.3cm]
Labels & $r$ & $Q$ & $\Sigma$ & $y$ & $x$ & Existence\\[0.2cm]
\hline\\[-0.2cm]
%%%%%%%%
$A^{\pm}(r^{*})$ & $r^{*}$ & \scriptsize $\pm\;\frac{1+2r^{*}}{3(1+r^{*})}$ &
$0$ & \scriptsize $\frac{(4r^{*}+5)(1+2r^{*})}{9(1+r^{*})^{2}}$ & \scriptsize
$\pm\frac{2+r^{*}}{3(1+r^{*})}$ & $r^{*}\leq -\frac{5}{4}$ or
$r^{*}\geq-\frac{1}{2}$\\[0.3cm]
$B^{\pm}(r^{*})$ & $r^{*}$ & \scriptsize $\pm\;\frac{-2}{3(-2+\gamma)}$ & $0$
& $0$ & \scriptsize $\pm\;\frac{-4+3\gamma}{3(-2+\gamma)}$ &
$1\leq\,\gamma\leq \frac{5}{3}$\\[0.3cm]
$C^{+}(r^{*})$ & $r^{*}$ & \scriptsize $Q$ & \scriptsize $\sqrt{-Q(Q-2)}$ &
$0$ & \scriptsize $-Q+1$ & $0\leq\;Q\;\leq2$ \\[0.3cm]
$C^{-}(r^{*})$ & $r^{*}$ & \scriptsize $Q$ & \scriptsize $\sqrt{-Q(Q+2)}$ &
$0$ & \scriptsize $-Q-1$ & $-2\leq\;Q\;\leq0$ \\[0.3cm]
$N^{\pm}(r^{*})$ & $r^{*}$ & $0$ & $0$ & $0$ & $\pm\;1$ & always \\[0.3cm]
$L^{\pm}(r^{*})$ & $r^{*}$ & $\pm2$ & $0$ & $0$ & $\mp1$& always \\[0.3cm]
$P_{1}^{\pm }(r^{*})$ & $r^{*}$ & $\pm\;1$ & $\pm\;1$ & $0$ & $0$ & always
\\[0.3cm]
$P_{2}^{\pm }(r^{*})$ & $r^{*}$ & $\pm\;1$ & $\mp\;1$ & $0$ & $0$ & always
\\[0.3cm]
$P_{3}^{\pm }(r^{*})$ & $r^{*}$ & \scriptsize $\pm\;\frac{2}{4-3\gamma}$ &
\scriptsize $\pm\;\frac{3\gamma+2r^{*}}{r^{*}(-4+3\gamma)}$ & $y_1(r^{*})$ &
\scriptsize $\pm\;\frac{3\gamma(1+r^{*})}{(-4+3\gamma)r^{*}}$ & $\notin \Psi$
for $\gamma\in [1,2]$ \\[0.3cm]
$P_4^{\pm}(r^{*})$ & $r^{*}$ & \scriptsize
$\pm\frac{2(r^{*})^{2}+5r^{*}+5}{7(r^{*})^{2}+16r^{*}+10}$ & \scriptsize
$\mp\frac{2(r^{*})^{2}+2r^{*}-1}{7(r^{*})^{2}+16r^{*}+10}$ & $y_2(r^{*})$ &
\scriptsize $\pm\frac{3(2+r^{*})(1+r^{*})}{7(r^{*})^{2}+16r^{*}+10}$ &
$r^{*}\leq\frac{1}{2}(-1-\sqrt{3})$; or \\[0.3cm]
&&&&&& $r^{*}\geq\frac{1}{2}(-1+\sqrt{3})$ \\[0.3cm]
$P_{5}^{\pm }(r^{*})$ & $r^{*}$ & \scriptsize $\pm \frac{2}{(-4+3\gamma)}$ &
\scriptsize $\frac{2\sqrt{3\gamma-5}}{-4+3\gamma}$ & $0$ & \scriptsize $\pm
\frac{3(-2+\gamma)}{-4+3\gamma}$ & $\frac{5}{3}\leq\gamma\leq2$ \\[0.3cm]
$P_{6}^{\pm }(r^{*})$ & $r^{*}$ &\scriptsize $\pm\frac{-2r^{*}}{3\gamma
r^{*}+3\gamma-2r^{*}}$ & $0$ & $y_3(r^{*})$ & \scriptsize $\pm
\frac{3\gamma(1+r^{*})}{3\gamma r^{*}+3\gamma-2r^{*}}$ & \scriptsize
$1\leq\gamma<\frac{4}{3}, b^{-}\leq r^{*}\leq -1$\\[0.3cm]
&&&&&& \scriptsize $1\leq\gamma<\frac{4}{3}, -\frac{3\gamma}{4}\leq r^{*}\leq
b^{+}$ or \\[0.2cm]
&&&&&& \scriptsize $\frac{4}{3}<\gamma<\frac{5}{3}, b^{-}\leq r^{*}\leq
-\frac{3}{4}$ or \\[0.2cm]
&&&&&& \scriptsize $\frac{4}{3}<\gamma<\frac{5}{3}, -1\leq r^{*}\leq b^{+}$
or \\[0.2cm]
&&&&&& \scriptsize $\frac{5}{3}<\gamma<2, -1\leq r^{*}\leq b^{+}$ or
\\[0.2cm]
&&&&&& \scriptsize $\gamma=\frac{4}{3}, -\frac{4}{5}-\frac{\sqrt{6}}{5}\leq
r^{*}\leq -\frac{4}{5}+\frac{\sqrt{6}}{5}$ or \\[0.2cm]
&&&&&& \scriptsize $\gamma=\frac{5}{3}, -1\leq r^{*}\leq -\frac{1}{3}$ or
\\[0.2cm]
&&&&&& \scriptsize $\gamma=\frac{5}{3}, r^{*}=-\frac{5}{4}$
\\[0.3cm]\hline \hline
\end{tabular}}\end{center}
\caption[crit]{\label{tab2} The critical points of the system
\eqref{Sistema00} for the case \mbox{($r=r^{*}=M^{-1}(0)$)}. We use the
notations
 $y_1(r^*)=\frac{-6(1+r^{*})(-4r^{*}+2\gamma
r^{*}+3\gamma^{2}-5\gamma)}{(r^{*})^{2}(-4+3\gamma)^{2}}$,
$y_2(r^*)=\frac{9(4(r^{*})^{2}+10r^{*}+7)(1+r^{*})^{2}}{(7(r^{*})^{2}+16r^{*}
+10)^{2}}$,\;\;
 $y_3(r^*)=\frac{2(1+r^{*})(4r^{*}+3\gamma)}{(3\gamma
r^{*}+3\gamma-2r^{*})^{2}}$ and
$b^{\pm}=\frac{-4-9\gamma}{4(1+3\gamma)}\;\pm\;\frac{1}{4}\sqrt{\frac{
16+48\gamma+9\gamma^{2}}{(1+3\gamma)^{2}}}.$}
\end{table*}

\begin{table*}[h]
   \centering
\begin{tabular}{@{\hspace{7pt}}l@{\hspace{12pt}}c@{\hspace{12pt}}c@{\hspace{
12pt}}c@{\hspace{12pt}}c@{\hspace{12pt}}c@{\hspace{12pt}}l}
\hline
\hline\\[-0.3cm]
Pts. & $r$ & $Q$ & $\Sigma$ & $y$ & $x$ & Existence \\[0.2cm]
\hline\\[-0.2cm]
%%%%%%%%
$P_{7}^{\pm}$ & $r_{c}$ & $\pm\;2$ & $\pm\;1$ & $0$ & $0$ & \small $\notin
\Psi$  \\[0.3cm]
$P_{8}^{\pm}$ & $r_{c}$ & $\pm\;1$ & $\pm\;1$ & $0$ & $0$ & \small always
\\[0.3cm]
$P_{9}^{\pm}$ & $r_{c}$ & $\pm1\;$ & $\mp\;1$ & $0$ & $0$ & \small always
\\[0.3cm]
$P_{10}^{\pm}$ & $-2$ & $\pm\;1$ & $0$ & $1$ & $0$ & \small always \\[0.3cm]
$P_{11}^{\pm}$ & $-2$ & $\pm\;\frac{1}{2}$ & $\mp\;\frac{1}{2}$ &
$\frac{3}{4}$ & $0$ & \small always  \\[0.3cm]\hline \hline
\end{tabular}
\caption[crit]{\label{tab1} The critical points of the system
\eqref{Sistema00} for the case $x=0$.}
\end{table*}

Observe that the points $P_{3}^{\pm }(r^{*})$ do not belong to the phase
space for $\gamma \geq 1$. If we relax this condition, these points indeed
exist (see the reference \cite{Leon:2010pu} for the corresponding points  for
the special case $f(R)=R^{n}$).

Additionally, the system \eqref{Sistema00} admits two circles of critical
points, given by $C^{\pm}(r^{*}): r=r^{*}$, $y=0$,
\mbox{$\Sigma^{2}+Q^{2}\pm(-2 Q)=0$,} $x+Q=\pm1.$ These points correspond to
solutions in the full phase space satisfying $K=y=z=0$. 
%%%%%%%%%%%%%%%%%%%%%%%%%%%%%%%%%%%%%%%%%%%%%%%%%%%%%%%%%%%%%%%%%%%
%%%%%%%%%%%%%%%%%%%%%%%%%%%%%%%%%%%%%%%%%%%%%%%%%%%%%%%%%%

\begin{table*}[h]
\resizebox{\columnwidth}{!}{
\begin{tabular}{@{\hspace{5pt}}l@{\hspace{10pt}}c@{\hspace{10pt}}c@{\hspace{
10pt}}c@{\hspace{10pt}}c@{\hspace{10pt}}c}
\hline
\hline\\[-0.3cm]
Points &$\lambda_{1}$ &$\lambda_{2}$ &$\lambda_{3}$ &$\lambda_{4}$
&$\lambda_{5}$ \\[0.2cm]
\hline\\[-0.2cm]
%%%%%%%%
$L^{\pm}(r^{*})$ & $\pm 4$ & $0$ &\scriptsize $\pm(10-6\gamma)$ &\scriptsize
$\pm(8+\frac{2}{1+r^{*}})$ & \scriptsize $\mp 2M'(r^{*})$ \\[0.3cm]
$N^{\pm}(r^{*})$ & $\pm4$ & $\pm 2$ & $0$ &\scriptsize
$\pm(4-\frac{2}{1+r^{*}})$ &\scriptsize $\pm 2M'(r^{*})$ \\[0.3cm]
$P_{1}^{\pm }(r^{*})$ & $\pm 6$ & $\pm 2$ & $0$ & $0$ &\scriptsize $\mp
3(-2+\gamma)$ \\[0.3cm]
$P_{2}^{\pm }(r^{*})$ & $\pm 6$ & $\pm 6$ & $0$ & $0$ &\scriptsize
$\mp3(-2+\gamma)$ \\[0.3cm]
$A^{+}(r^{*})$ & \scriptsize $-\frac{(1+2r^{*})(5+4r^{*})}{3(1+r^{*})^{2}}$
&\scriptsize $-\frac{(1+2r^{*})(5+4r^{*})}{3(1+r^{*})^{2}}$ &\scriptsize
$\frac{2-4r^{*}(1+r^{*})}{3(1+r^{*})^{2}}$ & $f_1(r^*,\gamma)$ & \scriptsize
$\frac{2(2+r^{*})M'(r^{*})}{3(1+r^{*})}$ \\[0.3cm]
$A^{-}(r^{*})$ &\scriptsize $\frac{-2+4r^{*}(1+r^{*})}{3(1+r^{*})^{2}}$
&\scriptsize $\frac{(1+2r^{*})(5+4r^{*})}{3(1+r^{*})^{2}}$ &\scriptsize
$\frac{(1+2r^{*})(5+4r^{*})}{3(1+r^{*})^{2}}$ & $f_2(r^*,\gamma)$
&\scriptsize $-\frac{2(2+r^{*}M'(r^{*}))}{3(1+r^{*})}$ \\[0.3cm]
$B^{\pm}(r^{*})$ &\scriptsize $\pm\frac{4}{6-3\gamma}$ &\scriptsize
$\pm(-2+\frac{2}{6-3\gamma})$ & \scriptsize $\pm(-2+\frac{2}{6-3\gamma})$
&\scriptsize $\pm\frac{-2(3\gamma+4r^{*})}{3(-2+\gamma)(1+r^{*})}$
&\scriptsize $\pm\frac{2(-4+3\gamma)M'(r^{*})}{3(-2+\gamma)}$ \\[0.3cm]
$P_4^{\pm}(r^{*})$ &\scriptsize
$\pm\frac{-3(7+2r^{*}(5+2r^{*}))}{10+r^{*}(16+7r^{*})}$ &\scriptsize
$\pm\frac{-15\gamma-3r^{*}(4+5\gamma+2(1+\gamma)r^{*})}{10+r^{*}(16+7r^{*})}$
& \scriptsize $f_3^{\pm}(r^*)$ &\scriptsize $f_3^{\mp}(r^*)$ &\scriptsize
$\pm\frac{6(1+r^{*})(2+r^{*})M'(r^{*})}{10+r^{*}(16+7r^{*})}$ \\[0.3cm]
$P_{5}^{\pm }(r^{*})$ & $0$ & $0$ & \scriptsize
$\pm4+\frac{4\sqrt{-5+3\gamma}}{4-3\gamma}$ &\scriptsize
$\pm\frac{6(\gamma+2(-1+\gamma)r^{*})}{(-4+3\gamma)(1+r^{*})}$ &\scriptsize
$\pm\frac{6(-2+\gamma)M'(r^{*})}{-4+3\gamma}$ \\[0.3cm]
$P_{6}^{\pm }(r^{*})$ &\scriptsize
$\pm\frac{6\gamma+4r^{*}}{3\gamma+(-2+3\gamma)r^{*}}$ &\scriptsize $\pm
\frac{-3(\gamma+2(-1+\gamma)r^{*})}{3\gamma+(-2+3\gamma)r^{*}}$ & \scriptsize
$ f_4^{\pm}(r^*,\gamma)$ &\scriptsize $f_4^{\mp}(r^*,\gamma) $ &\scriptsize
$\pm\frac{6\gamma(1+r^{*})M'(r^{*})}{3\gamma+(-2+3\gamma)r^{*}}$
\\[0.3cm]\hline \hline
\end{tabular}}
\caption[crit]{\label{tab4} Eigenvalues for the critical points in Table
\ref{tab2} (Point $P_3^\pm$ is omitted since it does not belongs to the phase space $\Psi$). We use the notations 
\scriptsize{$f_1(r^*,\gamma)=\frac{2}{3}(-1+\frac{1}{(1+r^{*})^{2}}
)+\gamma(-2+\frac{1}{1+r^{*}})$,
$f_2(r^*,\gamma)=\frac{3\gamma+r^{*}(4+9\gamma+(2+6\gamma)r^{*})}{3(1+r^{*})^
{2}}$, \scriptsize
$f_3^{\pm}(r^*)=\frac{-3(7+2r^{*}(5+2r^{*})\pm\sqrt{-1+26r^{*}+20(r^{*})^{2}}
\sqrt{7+2r^{*}(5+2r^{*})})}{20+2r^{*}(16+7r^{*})}$}, and \\
\scriptsize{$f_4^{\pm}(r^*,\gamma)=\frac{-(3\gamma+3r^{*}
(-2+3\gamma+2(-1+\gamma)r^{*})\pm\sqrt{1+r^{*}}\sqrt{81\gamma^{2}+r^{*}
(3\gamma(52+87\gamma)+4r^{*}(41+57\gamma+54\gamma^{2}+(5+3\gamma)^{2}r^{*}))}
)}{(2(1+r^{*})(3\gamma+(-2+3\gamma)r^{*}))}.$}
}
\end{table*}

%%%%%%%%%%%%%%%%%%%%%%%%%%%%%%%%%%%%%%%%%%%%%%%%%%%%%%%%%%%%%%%%%%%
%%%%%%%%%%%%%%%%%%%%%%%%%%%%%%%%%%%%%%%%%%%%%%%%%%%%%%%%%%

\begin{table*}[h]
\begin{tabular}{@{\hspace{7pt}}l@{\hspace{12pt}}c@{\hspace{12pt}}c@{\hspace{
12pt}}c@{\hspace{12pt}}c@{\hspace{12pt}}c@{\hspace{12pt}}l}
\hline
\hline\\[-0.3cm]
Points &$\lambda_{1}$ &$\lambda_{2}$ &$\lambda_{3}$ &$\lambda_{4}$
&$\lambda_{5}$ & Stability \\[0.2cm]
\hline\\[-0.2cm]
%%%%%%%%

$P_{8}^{\pm}$ &$\pm6$ &$\pm2$ & $0$ &$0$ &$\mp 3(-2+\gamma)$ & Non-hyperbolic
\\[0.3cm]
$P_{9}^{\pm}$ &$\pm6$ &$\pm6$ &$0$ &$0$ &$\mp 3(-2+\gamma)$ &  Non-hyperbolic
 \\[0.3cm]
$P_{10}^{+}$ & $-3$ &$-2$ &$-3\gamma$ &$\frac{1}{4}\alpha^{-}$ &
$\frac{1}{4}\alpha^{+}(r^{*})$ & Non-hyperbolic if $M(-2)=0$  \\[0.3cm]
      &&&&&& Stable (attractor) if
$M(-2)>0$ \\[0.3cm]
            &&&&&& Unstable (saddle) if $M(-2)<0$  \\[0.3cm]
$P_{10}^{-}$  &$3$ &$2$ &$3\gamma$ &$-\frac{1}{2}\alpha^{+}(r^{*})$
&$-\frac{1}{2}\alpha^{-}$ & Non-hyperbolic if $M(-2)=0$ \\[0.3cm]
       &&&&&& Unstable if $M(-2)\neq0$ 
\\[0.3cm]
$P_{11}^{+}$ & $-3$ &$\frac{3}{2}$ &$-\frac{3}{2}\gamma$
&$\frac{1}{4}\beta^{-}$ & $\frac{1}{4}\beta^{+}$ & Non-hyperbolic if
$M(-2)=0$  \\[0.3cm]
       &&&&&& Unstable (saddle) if
$M(-2)\neq0$  \\[0.3cm]
$P_{11}^{-}$ & $3$ &$-\frac{3}{2}$ &$\frac{3}{2}\gamma$
&$-\frac{1}{4}\beta^{+}$ & $-\frac{1}{4}\beta^{-}$ & Non-hyperbolic if
$M(-2)=0$  \\[0.3cm]
       &&&&&& Unstable (saddle) if
$M(-2)\neq0$  \\[0.3cm]\hline \hline
\end{tabular}
\caption[crit]{\label{tab3} \centering Eigenvalues and stability for the
critical points $P_{8}$ - $P_{11}$. We use the notations
$\alpha^{\pm}(r^{*})=-3\pm\,\sqrt{9-8M(-2)}$ and
$\beta^{\pm}=-3\pm\,\sqrt{9-24M(-2)}.$}
\end{table*}

In the tables \ref{tab4} and \ref{tab3} are summarized the stability results
for the (curves of) fixed points. The more interesting points are the
following:
\begin{itemize}
\item $L^+(r^{*})$ is a local repulsor if $1\leq \gamma<\frac{5}{3},
r^{*}<-\frac{5}{4}, M'(r^{*})<0;$ or $1\leq\gamma<\frac{5}{3}, r^{*}>-1,
M'(r^{*})<0.$
\item $N^+(r^{*})$ is a local repulsor if $r^{*}<-1, M'(r^{*})>0;$ or
$r^{*}>-\frac{1}{2}, M'(r^{*})>0.$
\item $A^{-}(r^{*})$ is a repulsor if
$-2<r^{*}<\frac{-1-\sqrt{3}}{2},\;M'(r^{*})>0;$ or $r^{*}<-2,\;M'(r^{*})<0;$
or $r^{*}>\frac{1}{2}(-1+\sqrt{3}),\;M'(r^{*})<0.$  
 \item The point $P_{10}^{-}$ es a repulsor for $M(-2)>0.$
\end{itemize}

\begin{itemize}
\item $L^-(r^{*})$ is a local attractor if $1\leq \gamma<\frac{5}{3},
r^{*}<-\frac{5}{4}, M'(r^{*})<0;$ or $1\leq\gamma<\frac{5}{3}, r^{*}>-1,
M'(r^{*})<0.$
\item $N^-(r^{*})$ is a local attractor if $r^{*}<-1, M'(r^{*})>0;$ or
$r^{*}>-\frac{1}{2}, M'(r^{*})>0.$
\item $A^{+}(r^{*})$ is a attractor if
$-2<r^{*}<\frac{-1-\sqrt{3}}{2},\;M'(r^{*})>0;$ or $r^{*}<-2,\;M'(r^{*})<0;$
or $r^{*}>\frac{1}{2}(-1+\sqrt{3}),\;M'(r^{*})<0.$  
 \item The point $P_{10}^{+}$  represent a future attractor for $M(-2)>0$ and
if $M(-2)>\frac{9}{8}$, it is a stable focus. In the  special case 
$M(-2)=0,$ $P_{10}^{+}$ coincides with $A^{+}(-2)$ and becomes
non-hyperbolic. This case cannot be analyzed using the linearization
technique. 
 \end{itemize}

Some saddle points of physical interest with stable manifold 4D  are:
\begin{itemize}
 \item $A^{+}(r^{*})$ is saddle with stable manifold 4D if
%\begin{footnotesize}
   \begin{eqnarray}
   -2<r^{*}&<&\frac{-1-\sqrt{3}}{2},M'(r^{*})<0\;\;\text{or}
\nonumber\\
   r^{*}&<&-2,M'(r^{*})>0 \;\;\text{or}\nonumber\\
   r^{*}&>&\frac{-1+\sqrt{3}}{2},M'(r^{*})>0
\;\;\text{or}\nonumber\\
   \frac{5}{3}<\gamma\leq2,
\frac{1}{2}(-1-\sqrt{3})<r^{*}&<&-\frac{5}{4},M'(r^{*})>0\;\;\text{or}
\nonumber\\
   1\leq\gamma\leq\frac{5}{3},
\frac{1}{2}(-1-\sqrt{3})<r^{*}&<&\frac{-4-9\gamma}{4(1+3\gamma)}-\frac{1}{4}
\sqrt{\frac{16+48\gamma+9\gamma^{2}}{(1+3\gamma)^{2}}},M'(r^{*})>0\;\;\text{
or}\nonumber\\
\frac{-4-9\gamma}{4(1+3\gamma)}+\frac{1}{4}\sqrt{\frac{16+48\gamma+9\gamma^{2
}}{(1+3\gamma)^{2}}}<r^{*}&<&\frac{1}{2}(-1+\sqrt{3}),M'(r^{*})<0.\nonumber
     \end{eqnarray}
%\end{footnotesize}
\item $B^{+}(r^{*})$ is saddle with stable manifold 4D if
%\begin{footnotesize}
 \begin{eqnarray}
 1\leq\gamma&<&\frac{4}{3},
-1<r^{*}<-\frac{3\gamma}{4},M'(r^{*})<0\;\;\text{or} \nonumber\\
   \frac{4}{3}<\gamma&<&\frac{5}{3},
-\frac{3\gamma}{4}<r^{*}<-1,M'(r^{*})<0.\nonumber
 \end{eqnarray}
%\end{footnotesize}
\item $P_{4}^{+}(r^{*})$ is saddle with stable manifold 4D if
%\begin{footnotesize}
   \begin{eqnarray}
   -2<r^{*}&<&\frac{-1-\sqrt{3}}{2},M'(r^{*})>0\;\;\text{or}
\nonumber\\
   r^{*}&<&-2,M'(r^{*})<0 \;\;\text{or}\nonumber\\
   r^{*}&>&\frac{-1+\sqrt{3}}{2},M'(r^{*})<0.\nonumber
   \end{eqnarray}
%\end{footnotesize}
\end{itemize}

The points $P_{1}^{\pm }(r^{*})$, $P_{2}^{\pm }(r^{*})$ and $P_{5}^{\pm
}(r^{*})$ have a 2D center manifold.  Observe that the points $P_{1}^{\pm
}(r^{*})$ and $P_{2}^{\pm }(r^{*})$ are particular cases of $P_{8}^{\pm}$ and
$P_{9}^{\pm}$ respectively, because both coincide when $r=r^{*}$. The points
$P_{1}^{-}(r^{*})$ and $P_{2}^{-}(r^{*})$  have a 3D stable manifold for
$\gamma\neq2$ and the point $P_{5}^{-}(r^{*})$ have a 3D stable manifold if
  \[ \frac{5}{3}\leq \gamma <2,\; r^{*}<\frac{\gamma}{2-2\gamma},\;
M'(r^{*})<0 \;\; \text{or}\;\; \frac{5}{3}\leq \gamma <2,\; r^{*}>-1,\;
M'(r^{*})<0. \]

%%%%%%%%%%%%%%%%%%%%%%%%%%%%%%%%%%%%%%%%%%%%%%%%%%%%%%%%%%%%%%%%%%%
%%%%%%%%%%%%%%%%%%%%%%%%%%%%%%%%%%%%%%%%%%%%%%%%%%%%%%%%%%

%%%%%%%%%%%%%%%%%%%%%%%%%%%%%%%%%%%%%%%%%%%%%%%%%%%%%%%%%%%%%%%%%%%
%%%%%%%%%%%%%%%%%%%%%%%%%%%%%%%%%%%%%%%%%%%%%%%%%%%%%%%%%%

%%%%%%%%%%%%%%%%%%%%%%%%%%%%%%%%%%%%%%%%%%%%%%%%%%%%%%%%%%%%%%%%%%%
%%%%%%%%%%%%%%%%%%%%%%%%%%%%%%%%%%%%%%%%%%%%%%%%%%%%%%%%%%

\section{Formalism for the physical description of the solutions. Connection
with the observables}\label{Formalism}
In this section we present a formalism based on the reference
\cite{Leon:2010pu} for obtaining the physical description of a critical
point, and also connecting with the basic observables, that are relevant for
a physical discussion. 

Firstly, in order obtain first-order evolution rates for $e^{1}_{1}$,
$^{2}K$, $\rho_{m}$ and $R$ as functions of $\tau$ we use the equations
\eqref{evolucione1}, \eqref{curvaturaK}, \eqref{ConsM} and the relation
$d\tau=D dt$, for obtaining the first order differential equations
\begin{subequations}\label{wqert}
\begin{eqnarray}
 {{e}^{1}_{1}}'&=&-(Q^{*}-2\Sigma^{*})e^{1}_{1},\\
 ^{2}{K}'&=&-2\left(Q^{*}+\Sigma^{*}\right)^{2}K,\\
 {\rho}_{m}'&=&-3\, \gamma \,Q^{*}\rho_{m},\\
 {R}'&=&\frac{2x^{*}R}{m(r^{*})} \;.
\end{eqnarray}
\end{subequations}
where $^{*}$ denotes the evaluation at an specific critical point, and $'$ 
denotes derivative with respect to $\tau$. 
The last equation follows from definition of $x$ given by \eqref{VRDN} and
the definition of $m(r)$ given by \eqref{funcm}.

In order to express the functions $e^{1}_{1}(\tau)$, $^{2}K(\tau)$,
$\rho_{m}(\tau)$ and $R(\tau)$ in terms of the co-moving time variable $t,$
we substitute in the solutions of  \eqref{wqert} in terms of the  $\tau$, by
the expression $\tau=\tau(t)$ obtained by inverting the solution of
\begin{equation}\label{mkop}
  \frac{dt}{d\tau}=\frac{1}{D^{*}},
\end{equation}
with $D^{*}$ being the first-order solution of
\begin{equation}\label{mkop1}
 {D}'=D\;\Upsilon^{*}
\end{equation}
where
\begin{small}
 \begin{equation}
 \Upsilon^{*}=\frac{1}{2}\{2(x^{*}-\Sigma^{*})+(Q^{*}+x^{*}
)\left(3((\gamma
-2)(x^{*})^{2}+\gamma(y^{*}-1))+2(Q^{*}+x^{*})\Sigma^{*}+3(\gamma -2)(\Sigma
^*)^{2}\right)\}.\nonumber
 \end{equation}
\end{small}
Solving equations \eqref{mkop}, \eqref{mkop1} for $x^*\neq 0,$  $y^*\neq 0,$ 
$\Upsilon^*\neq 0$, $r^*\neq\{0,1\}$, with initial conditions $D(0)=D_{0}$
and $t(0)=t_{0}$ we obtain
\begin{equation}
 t(\tau)=\frac{1-e^{-\tau\Upsilon^{*}}}{D_{0}\Upsilon^{*}}+t_{0}.
\end{equation}
Thus, inverting the last equation for $\tau$ and substituting in the
solutions of \eqref{wqert}, with initial conditions
${e}^{1}_{1}(0)={e}^{1}_{1\;0}$, $^{2}K(0)=\;^{2}K_{0}$,
$\rho_{m}(0)=\rho_{m\;0}$ and $R(0)= R_{0}$ we obtain 
\begin{subequations}\label{nghu}
\begin{eqnarray}
{e}^{1}_{1}(t)&=&{e}^{1}_{1\;0}\left(\ell_{1}-t\Upsilon^{*}D_{0}\right)^{
\frac{Q^{*}-2\Sigma^{*}}{\Upsilon^{*}}},\\
 ^{2}K(t)&=&\;^{2}K_{0}\left(\ell_{1}-t\Upsilon^{*}D_{0}\right)^{\frac{
2(Q^{*}+\Sigma^{*})}{\Upsilon^{*}}},\\
 \rho_{m}(t)&=&\rho_{m0}\left(\ell_{1}-t\Upsilon^{*}D_{0}\right)^{
\frac{3\gamma Q^{*}}{\Upsilon^{*}}},\\
 R(t)&=&
R_{0}\left(\ell_{1}-t\Upsilon^{*}D_{0}\right)^{-\frac{2x^{*}}{m(r^{*}
)\Upsilon^{*}}}.
\end{eqnarray}
\end{subequations}
where $\ell_{1}=D_{0}t_{0}\Upsilon^{*}+1$. 

Additionally, we can introduce a length scale $\ell$ along the flow lines,
describing the volume expansion (contraction) behavior of the congruence
completely, via the standard relation, namely
\begin{equation}\label{BBB}
 \frac{\dot{\ell}}{\ell}=\frac{1}{3}\Theta\equiv H
\end{equation}
The length scale $\ell$ along the flow lines, defined in \eqref{BBB}, can be
expressed as \cite{Henk1996}:
\begin{equation}\label{nghu1}
\ell(t)=\ell_{0}\left(\ell_{1}-t\Upsilon^{*}D_{0}\right)^{-\frac{Q^{*}}{
\Upsilon^{*}}}.
\end{equation}
where $\ell_{0}=[({e}^{1}_{1\;0})(\;^{2}K_{0})]^{-\frac{1}{3}}$. In summary,
the expressions \eqref{nghu} and \eqref{nghu1} determine the cosmological
solution, that is the evolution of various quantities, at a critical point.
In the particular case $\Upsilon^{*}\rightarrow 0^{+}$,  $\ell_1\rightarrow
1$, and using the algebraic fundamental limit, we obtain
\begin{subequations}
\label{caso1111}
\begin{eqnarray}
{e}^{1}_{1}(t)&=&{e}^{1}_{1\;0}\, e^{-D_{0}\, (Q^{*}-2\,\Sigma^{*})\,t},\\
^{2}K(t)&=&\;^{2}K_{0}\, e^{-2\,D_{0}\,(Q^{*}+\Sigma^{*})\,t},\\
\rho_{m}(t)&=&\rho_{m\;0}\, e^{-3\gamma\, D_{0}\, Q^*\, t},\\
R(t)&=&R_{0}\, e^{\frac{2\, D_{0}\, x^{*}\,t}{m(r^*)}},\\
\ell(t)&=&\ell_{0}\, e^{D_{0}\, Q^{*}\, t}.
\end{eqnarray}
\end{subequations}
In the simple cases $x^{*}=0$ and $y^{*}=0$, we deduce from the definitions
of the dynamical variables and $r$, the relationships:
\begin{itemize}
 \item  If $y^{*}=0$ and $r^{*}\neq\{0,1\}$, $R=0$.
 \item  If $y^{*}=0$ and $r^{*}=1$, $R$ is arbitrary.
 \item  If $x^{*}=0$, then either $f'(R)$ is a constant, say $c$,
which implies that asymptotically $f(R)\rightarrow c R +const.$,  or $\dot
R=0$, which means a spacetime of constant Ricci curvature. In both case GR is
recovered. In the special case for which $R$ becomes asymptotically a
constant, say $R_0$, it must satisfy, in general, a transcendental equation: 
\begin{equation}
 r^{*}=-c\frac{R_0}{f(R_0)}.
\end{equation}
where $c$ is constant. Solving this relationship, the corresponding value of
$R(r^{*})=R_0$ is obtained.
\end{itemize}

Now, let us now come to the observables. Using the above expressions, we can
calculate the deceleration parameter $q$ defined as usual as \cite{Henk1996}:
\begin{equation}
 q=3u^{\mu}\nabla_{\mu}[{\Theta}^{-1}]-1=-\frac{\ell\;\ddot{\ell}}{
(\dot{\ell})^{2}},
\end{equation}
and the effective (total) equation of state parameter of the universe, which
is defined as \eqref{utop1}:
\begin{equation}
 w_{eff}\equiv \frac{p_{m}+p_{DE}}{\rho_{m}+\rho_{DE}},
\end{equation}
where $p_{DE}$ and $\rho_{DE}$ are defined by \eqref{mnvcdf}. Using the
dynamical variables \eqref{VRDN} and the equations \eqref{Raychaudhurinew},
\eqref{TrazaNew} and the state equation for the matter ($p=(\gamma
-1)\rho_{m}$) \cite{Coley1999}, we obtain:
\begin{subequations}
\label{jhkj1}
\begin{eqnarray}
 q&=&\frac{1-x(2Q+x)+\Sigma^{2}-r(-1+2Qx+x^{2}+y-\Sigma^{2})}{Q^{2}
(1+r)},\\
\label{EoS:total}
 w_{eff}&=&\frac{1}{3}+\frac{2ry}{3(1+r)(-1+2Qx+x^{2}+\Sigma^{2})}.
\end{eqnarray}
\end{subequations}

Now, the various density parameters defined in \eqref{hgt12} -
\eqref{hgt123}, in terms of the auxiliary
variables straightforwardly read
\begin{subequations}\label{poper1}
\begin{eqnarray} 
 \Omega_{k}&=& \frac{(-1+Q+x)(1+Q+x)}{Q^{2}},\\
  \tilde{\Omega}_{m}&=&\frac{\left(1-x^{2}-y-\Sigma^{2}\right)f'(R)}{A Q^{2}},\\
  \tilde{\Omega}_{DE}&=& \frac{f'(R)/A \left(\Sigma ^2+x^2+y-1\right)- \left(2 Q
x+\Sigma ^2+x^2-1\right)}{ Q^2},\\
  \Omega_{\sigma}&=& \left(\frac{\Sigma}{Q}\right)^{2}.
\end{eqnarray}
\end{subequations}

Additionally, the equation of state parameter for Dark Energy, defined as $w_{DE}=p_{DE}/\rho_{DE}$, reads:
		\begin{equation}
		\small
		w_{DE}= \frac{A \left(r \left(2 Q x+\Sigma ^2+x^2+2 y-1\right)+2 Q x+\Sigma ^2+x^2-1\right)-3 (\gamma -1) (r+1) f'(R) \left(\Sigma ^2+x^2+y-1\right)}{3 (r+1) \left(A \left(2 Q x+\Sigma
   ^2+x^2-1\right)-f'(R) \left(\Sigma ^2+x^2+y-1\right)\right)},\label{wDE}
    \end{equation}
which depend on the parametrization used for the
$T^{(eff)}_{\alpha \beta}$. 

 Although there is a slight confusion in the terminology in the literature,
the majority of authors use the term {\em phantom Dark Energy} referring to
$w_{DE}<-1$, or the term {\em phantom universe} referring to $w_{tot}\equiv
w_{eff}<-1$. Comparing \eqref{wDE} and \eqref{EoS:total},
it is obvious that when $\tilde{\Omega}_{DE}+\Omega_\sigma=1$,  then $w_{eff}$ and $w_{DE}$
coincide. Therefore, in the discussion of the present manuscript, in the
critical points where  $\tilde{\Omega}_{DE}+\Omega_\sigma=1,$ the physical results obtained
using $w_{eff}$ and $w_{DE}$ will remain the same, independently of the
definition of $T^{(eff)}_{\alpha \beta}$. This is indeed the case in almost
all the obtained interesting stable points. However, we mention that in
general this is not the case, for instance in the present universe where 
the condition $\tilde{\Omega}_{DE}+\Omega_\sigma=1$ is violated, $w_{tot}$ and $w_{DE}$ are different, and in particular
according to observations  $w_{tot}>-1$ while $w_{DE}$ can be a little bit
above or below $-1$. Thus, in the current universe, when people ask whether
we lie in the {\em phantom} regime they mean whether $w_{DE}<-1$ not
$w_{tot}<-1$. In our work, we prefer the term {\em phantom
universe} since the quantity $w_{tot}$ will not change due to changes in the
parametrization of the $T^{(eff)}_{\alpha \beta}$. Additionally, it is
straightforward to additionally use $w_{DE}$ in order to examine whether DE
is in the phantom regime. In this case in general, with the new
parametrization where $w_{DE}$ changes, although $w_{tot}$ remains the same,
the DE-sector results could change.

Finally, introducing the auxiliary variable 
\be u=f'(R)/A,
\ee we obtain the closed form  
\begin{align}
\tilde{\Omega}_{m}&=\frac{\left(1-x^{2}-y-\Sigma^{2}\right)u}{Q^{2}},\\
\tilde{\Omega}_{DE}&= \frac{u \left(\Sigma ^2+x^2+y-1\right)- \left(2 Q
x+\Sigma ^2+x^2-1\right)}{ Q^2},\\
w_{DE}&=\frac{ \left(r \left(2 Q x+\Sigma ^2+x^2+2 y-1\right)+2 Q x+\Sigma ^2+x^2-1\right)-3 (\gamma -1) (r+1) u \left(\Sigma ^2+x^2+y-1\right)}{3 (r+1) \left( \left(2 Q x+\Sigma
   ^2+x^2-1\right)-u \left(\Sigma ^2+x^2+y-1\right)\right)}
\end{align}
where the resulting evolution equation for $u$ is 
\begin{equation}
\label{equ}
u'= 2 u x.
\end{equation}
Observe that  equation \eqref{equ} decouples from the equations
\eqref{Sistema00}. 

Let us make an important comment here. From \eqref{equ} it follows that at
equilibrium we have two choices, $x=0$ or $u=0.$
\begin{itemize}
\item

For the equilibrium points having $x=0$ it appears an additional zero
eigenvalue, and thus $u$, that is $f'(R)$, acquires a constant value. 

\item 

For the equilibrium points having $x\neq 0$, since they have   $u=0$  they
correspond to $f'(R)=0.$ In this case the
additional eigenvalue along the $u$-direction is $2 x^{*}$, where $x^{*}$ is
the value of the $x$-coordinate at the fixed point. Thus, perturbations along
the $u$-axis are stable in the extended phase space only if $x^{*}<0$.

\end{itemize}

Now, following a similar approach as in the references
\cite{Amendola:2006we,Leon:2010pu}, it proves more convenient to define the
dimensionless matter density 
\begin{equation}\label{Omega_m_new}
 \Omega_{m}\equiv\frac{A \rho_{m}}{3f'(R)H^{2}},
  \end{equation} 
and the expression for $\Omega_{DE}\equiv\frac{A \rho_{DE}}{3f'(R)H^{2}}$ is
given implicitly through  
$\Omega_{k}+\left({\Omega}_{m}+{\Omega}_{DE}\right)f'(R)/A+\Omega_{\sigma}
=1$.
Observe that by choosing $A=F_0=f'(R)|_{today}$ in \eqref{Omega_m_new} we
recover the usual dimensionless energy density $\frac{\rho_{m}}{3H^2}$ that
we observe today. In Table \ref{Paramm23} we display the values of
$\Omega_{m}$ evaluated at the equilibrium points.

Finally, it is interesting to notice that in Kantowski-Sachs geometry, the
geometry becomes isotropic if  $\sigma$ becomes zero, as can be seen from
\eqref{metrica} and the equation \eqref{jdfgoig}. Thus, critical points with
$\Sigma=0$ (or more physically $\Omega_{\sigma}=0$) correspond to usual
isotropic Friedmann points. In case that such an isotropic point attracts the
universe,  then we obtain future asymptotic isotropization
\cite{Goheer:2007wu,Leon:2010pu}.

%%%%%%%%%%%%%%%%%%%%%%%%%%%%%%%%%%%%%%%%%%%%%%%%%%%%%%%%%%%%%%%%%%%
%%%%%%%%%%%%%%%%%%%%%%%%%%%%%%%%%%%%%%%%%%%%%%%%%%%%%%%%%%
\newpage

\section{Cosmological implications}\label{Sect:5}

In this section we  discuss the physical implications for a generic $f(R)$ in
a universe with a Kantowski-Sachs geometry.

Since $Q$ is the Hubble scalar divided by a positive constant we have that
$Q>0$ ($Q<0$) at a fixed point, corresponds to an expanding (contracting)
universe. The case $Q=0$ corresponds to an static universe. 
Combining the above information with the information obtained for
deceleration parameter, $q$, is is possible to characterize the cosmological
solutions according to:
\begin{itemize}
 \item  If $Q>0$ and $q<0$ ($q>0$),  the critical point represents a
universe in accelerating (decelerating) expansion,
 \item  If $Q<0$ and $q>0$ ($q<0$), the critical point represents a
universe in accelerating (decelerating) contraction.
\end{itemize}
Additionally, if $w_{eff}<-1$, then the total EoS parameter of the universe
exhibits phantom behavior. The values $w_{eff}$ evaluated at critical points
are presented in the second column of table \ref{Obser}.

%%%%%%%%%%%%%%%%%%%%%%%%%%%%%%%%%%%%
\begin{table*}[h]
\begin{center} 
\begin{tabular}{@{\hspace{5pt}}l@{\hspace{10pt}}c@{\hspace{10pt}}c}
\hline
\hline\\[-0.3cm]
Points &  $w_{eff}$  & $q$ \\[0.2cm]
\hline\\[-0.2cm]
%%%%%%%%
$A^{\pm}(r^{*})$  & $\frac{1-7r^{*}-6(r^{*})^{2}}{3(1+r^{*})(1+2 r^{*})}$ &
$\frac{1-2r^{*}-2(r^{*})^{2}}{1+3r^{*}+2(r^{*})^{2}}$ \\[0.3cm] 
$B^{\pm}(r^{*})$, $L^{\pm}(r^{*})$ &  $\frac{1}{3}$ & $1$\\[0.3cm]
$P_4^{\pm}(r^{*})$  &
$\frac{1-r^{*}(21+4r^{*}(9+4r^{*}))}{3+3r^{*}(21+8r^{*}(3+r^{*}))}$ &
$\frac{1-2r^{*}-2(r^{*})^{2}}{5+5r^{*}+2(r^{*})^{2}}$ \\[0.3cm]
$P_{5}^{\pm }(r^{*})$  & $\frac{1}{3}$ & $-4+3\gamma$ \\[0.3cm]
$P_{6}^{\pm }(r^{*})$  & $-\frac{\gamma+r^{*}}{r^{*}}$ &
$-\frac{3\gamma+2r^{*}}{2r^{*}}$ \\[0.3cm]
$P_{1}^{\pm }(r^{*})$, $P_{2}^{\pm }(r^{*})$, $P_{8}^{\pm}$, $P_{9}^{\pm}$  &
$\frac{1}{3}$ & $2$ \\[0.3cm]
$C^{\pm}(r^{*})$  & $\frac{1}{3}$ & $\frac{2}{1\pm\,\sin \, u}$ \\[0.3cm]
$P_{10}^{\pm}$, $P_{11}^{\pm}$  & $-1$ & $-1$ \\[0.3cm]\hline \hline
\end{tabular}\end{center}
\caption[crit]{\label{Obser} Basic observables $q$ and $w_{eff}$ for each
critical point.}
\end{table*}

%%%%%%%%%%%%%%%%%%%%%%%%%%%%%%%%%%%%%%%%%%%%

\begin{table*}[h]
\begin{center}
\resizebox{\columnwidth}{!}{ %cambia a tama�o de letra m'as peque�o
\begin{tabular}{@{\hspace{5pt}}l@{\hspace{10pt}}c@{\hspace{10pt}}c@{\hspace{
10pt}}c}
\hline
\hline\\[-0.3cm]
Points & $\Omega_{k}$ &$\Omega_{m}$ & $\Omega_{\sigma}$\\[0.2cm]
\hline\\[-0.2cm]
%%%%%%%5
$A^{\pm}(r^{*})$, $L^{\pm}(r^{*})$, $P_{10}^{\pm}$ & $0$ & $0$ & $0$
\\[0.3cm]
$B^{\pm}(r^{*})$ & $0$ & $5-3\gamma$ &  $0$ \\[0.3cm]
$P_4^{\pm}(r^{*})$ &
$-\frac{3(-1+2r^{*}(1+r^{*}))(7+2r^{*}(5+2r^{*}))}{(5+r^{*}(5+2r^{*}))^{2}}$
& $0$  & $\frac{(1-2r^{*}(1+r^{*}))^{2}}{(5+r^{*}(5+2r^{*}))^{2}}$ \\[0.5cm]
$P_{5}^{\pm }(r^{*})$ & $0$ & $0$  & $-5+3\gamma$ \\[0.3cm]
$P_{6}^{\pm }(r^{*})$ & $0$ &
$-\frac{3\gamma+r^{*}(4+9\gamma+(2+6\gamma)r^{*})}{2(r^{*})^{2}}$  & $0$
\\[0.3cm]
$P_{1}^{\pm }(r^{*})$, $P_{2}^{\pm }(r^{*})$, $P_{8}^{\pm}$, $P_{9}^{\pm}$  &
$0$ & $0$  & $1$ \\[0.3cm]
$C^{\pm}$ & $0$ & $0$  & $\frac{\cos^{2}\,u}{(\pm1+\sin\,u)^{2}}$ \\[0.3cm]
$P_{11}^{\pm}$ & $-3$ & $0$  & $1$ \\[0.3cm]\hline \hline
\end{tabular}}\end{center}
\caption[crit]{\label{Paramm23} Energy density parameters $\Omega_{k}$,
$\Omega_{m}$, and $\Omega_{\sigma}$ for each critical point.}
\end{table*}

\begin{table*}[h]
\begin{center}
   %cambia a tama�o de letra m'as peque�o
 \resizebox{\columnwidth}{!}{
\begin{tabular}{@{\hspace{5pt}}l@{\hspace{10pt}}c@{\hspace{10pt}}l}
\hline
\hline\\[-0.3cm]
Points & $\Upsilon^{*}$ & Solution \\[0.2cm]
\hline\\[-0.2cm]
%%%%%%%%
$A^{+}(r^{*})$ & $-\frac{2+r^{*}}{3(1+r^{*})^{2}}$ &
\begin{math}
 \ell(t)= \left\{ \begin{array}{ll}
\ell_{0}\left(\ell_{1}-D_{0}\,t\Upsilon^{*}\right)^{s_{1}} & \textrm{for
$r^{*}\neq-2$}\\
\ell_{0}\,e^{D_{0}\,t} & \textrm{for $r^{*}=-2$}  
\end{array} \right., 
\; 
\rho_{m}(t)=\rho_{m0}\left[\frac{\ell(t)}{\ell_{0}}\right]^{-3\gamma}, 
\end{math}
\\[0.2cm]
          && \begin{math}
 R(t)= \left\{ \begin{array}{ll}
R_{0}\left(\ell_{1}- D_{0}\,t\Upsilon^{*}\right)^{\frac{2(1+r^{*})}{m(r)}} &
\textrm{for $r^{*}\neq \{-2,-\frac{5}{4},-\frac{1}{2}\}$} \\
 R_{0} & \textrm{for $r^{*}=-2$} \\
 0 & \textrm{for $r^{*}=\{-\frac{5}{4},-\frac{1}{2}\}$} 
\end{array}. \right.
\end{math} \\[0.3cm]
&& Isotropic. Expanding. Decelerating for  $-1.366 \lesssim r^{*} \leq
-\tfrac{5}{4}$ or $-0.5<r^{*} \lesssim 0.366.$  \\[0.2cm]
&& Accelerating for $r^{*}\lesssim -1.366$ or $r^{*}\gtrsim 0.366$. Phantom
behavior for $r^{*}<-2.$ \\[0.2cm]
&&  \textsl{de Sitter}, constant Ricci curvature, for $r^{*}\,=\,-2$.
\\[0.3cm]
$A^{-}(r^{*})$ & $\frac{2+r^{*}}{3(1+r^{*})^{2}}$ & \begin{math}
 \ell(t)= \left\{ \begin{array}{ll}
\ell_{0}\left(\ell_{1}-D_{0}\,t\Upsilon^{*}\right)^{s_{1}} & \textrm{for
$r^{*}\neq-2$} \\
\ell_{0}\,e^{-D_{0}\,t} & \textrm{for $r^{*}=-2$} 
\end{array}, \right.
\;
\rho_{m}(t)=\rho_{m0}\left[\frac{\ell(t)}{\ell_{0}}\right]^{-3\gamma} 
\end{math}
\\[0.2cm]
          && \begin{math}
 R(t)= \left\{ \begin{array}{ll}
R_{0}\left(\ell_{1}- D_{0}\,t\Upsilon^{*}\right)^{\frac{2(1+r^{*})}{m(r)}} &
\textrm{for $r^{*}\neq \{-2,-\frac{5}{4},-\frac{1}{2}\}$} \\
 R_{0} & \textrm{for $r^{*}=-2$} \\
 0 & \textrm{for $r^{*}=\{-\frac{5}{4},-\frac{1}{2}\}$} 
\end{array}. \right.
\end{math} \\[0.3cm]
&& Isotropic. Contracting. Accelerating for  $-1.366 \lesssim r^{*} \leq
-\tfrac{5}{4}$ or $-0.5<r^{*} \lesssim 0.366.$  \\[0.2cm]
&& Decelerating for $r^{*}\lesssim -1.366$ or $r^{*}\gtrsim 0.366$. Phantom
behavior for $r^{*}<-2.$ \\[0.2cm]
&&  Exponential collapse, constant Ricci curvature, for $r^{*}\,=\,-2$.
\\[0.3cm]
$B^{+}(r^{*})$ & $\frac{4}{3(-2+\gamma)}$ &
$\ell(t)=\ell_{0}\sqrt{\ell_{1}-D_{0}t\Upsilon^{*}}$,
$\;\;\rho_{m}(t)=\rho_{m0}(\ell_{1}-D_{0}t\Upsilon^{*})^{-3\gamma/2}$,
$R(t)=0.$ \\[0.2cm]
&& Isotropic. Expanding. Decelerating. Total matter/enegy mimics radiation.
\\[0.3cm]
$B^{-}(r^{*})$ & $\frac{4}{3(2-\gamma)}$ &
$\ell(t)=\ell_{0}\sqrt{\ell_{1}-D_{0}t\Upsilon^{*}}$,
$\;\;\rho_{m}(t)=\rho_{m0}(\ell_{1}-D_{0}t\Upsilon^{*})^{-3\gamma/2}$,
$R(t)=0.$ \\[0.2cm]
&& Isotropic. Contracting. Accelerating. Total matter/enegy mimics radiation.
\\[0.3cm]
$P_{4}^{+}(r^{*})$ & $-\frac{3(2+r^{*})}{10+r^{*}(16+7r^{*})}$ & \begin{math}
 \ell(t)= \left\{ \begin{array}{ll}
\ell_{0}\left(\ell_{1}-D_{0}\,t\Upsilon^{*}\right)^{s_{2}} & \textrm{for
$r^{*}\neq-2$} \\
\ell_{0}\,e^{\frac{D_{0}\,t}{2}} & \textrm{for $r^{*}=-2$}
\end{array}, \right.  \; 
\rho_{m}(t)=\rho_{m0}\left[\frac{\ell(t)}{\ell_{0}}\right]^{-3\gamma},
\end{math}
\\[0.2cm]
          && \begin{math}
 R(t)= \left\{ \begin{array}{ll}
R_{0}\left(\ell_{1}- D_{0}\,t\Upsilon^{*}\right)^{\frac{2(1+r^{*})}{m(r)}} &
\textrm{for $r^{*}\neq -2$} \\
 R_{0} & \textrm{for $r^{*}=-2$}  
\end{array}. \right.
\end{math} \\[0.3cm] && non-flat universe ($\Omega_{k}\neq0$). Accelerating
expansion for $r^{*}<-\tfrac{1}{2}(1+\sqrt{3})$ or
$r^{*}>-\tfrac{1}{2}(1-\sqrt{3}).$
\\[0.2cm] && Phantom solutions for $-2.395 \lesssim \frac{1}{4}
\left(-5-\sqrt{21}\right)<r^{*}<-2, M'(r^{*})<0.$ 
\\[0.2cm] && de Sitter solutions for
$r^{*}=\tfrac14\left(-5-\sqrt{21}\right)$. \\[0.3cm]
$P_{4}^{-}(r^{*})$ & $\frac{3(2+r^{*})}{10+r^{*}(16+7r^{*})}$ & \begin{math}
 \ell(t)= \left\{ \begin{array}{ll}
\ell_{0}\left(\ell_{1}-D_{0}\,t\Upsilon^{*}\right)^{s_{2}} & \textrm{for
$r^{*}\neq-2$} \\
\ell_{0}\,e^{-\frac{D_{0}\,t}{2}} & \textrm{for $r^{*}=-2$}  
\end{array}, \right. \; 
\rho_{m}(t)=\rho_{m0}\left[\frac{\ell(t)}{\ell_{0}}\right]^{-3\gamma}, 
\end{math}
\\[0.2cm]
          && \begin{math}
 R(t)= \left\{ \begin{array}{ll}
R_{0}\left(\ell_{1}- D_{0}\,t\Upsilon^{*}\right)^{\frac{2(1+r^{*})}{m(r)}} &
\textrm{for $r^{*}\neq -2$} \\
 R_{0} & \textrm{for $r^{*}=-2$} 
\end{array}. \right.
\end{math}\\[0.2cm] && non-flat universe ($\Omega_{k}\neq0$). Decelerating
contraction for $r^{*}<-\tfrac{1}{2}(1+\sqrt{3})$ or
$r^{*}>-\tfrac{1}{2}(1-\sqrt{3}).$
\\[0.3cm] && Phantom solutions for $-2.395 \lesssim \frac{1}{4}
\left(-5-\sqrt{21}\right)<r^{*}<-2, M'(r^{*})<0.$ 
\\[0.3cm] && Exponential collapse for
$r^{*}=\tfrac14\left(-5-\sqrt{21}\right)$. \\[0.3cm]
$P_{5}^{+}(r^{*})$ & $-\frac{6(-1+\gamma)}{3\gamma-4}$ &
$\ell(t)=\ell_{0}(\ell_{1}-D_{0}\Upsilon^{*} t)^\frac{1}{3(-1+\gamma)}$,
$\;\;\rho_{m}(t)=\rho_{m0}(\ell_{1}- D_{0}\Upsilon^{*}
t)^\frac{\gamma}{1-\gamma}$, $R(t)=0.$ \\[0.2cm]
&& Anisotropic. Expanding. Decelerating. Total matter/energy mimics
radiation. \\[0.3cm]
$P_{5}^{-}(r^{*})$ & $\frac{6(-1+\gamma)}{3\gamma-4}$ &
$\ell(t)=\ell_{0}(\ell_{1}-D_{0}\Upsilon^{*} t)^\frac{1}{3(-1+\gamma)}$,
$\;\;\rho_{m}(t)=\rho_{m0}(\ell_{1}- D_{0}\Upsilon^{*}
t)^\frac{\gamma}{1-\gamma}$, $R(t)=0$ \\[0.2cm]
&& Anisotropic. Contracting. Accelerating. Total matter/energy mimics
radiation. \\[0.3cm]
\hline \hline
\end{tabular}}\end{center}\caption[crit]{\label{Paramm3}\centering Equations
that determine the behavior of $\ell(t)$, $\rho_{m}(t)$ y $R(t)$ for the
critical points $A^{\pm}(r^{*})$, $B^{\pm}(r^{*})$, $P_4^{\pm}(r^{*})$ and
$P_5^{\pm}(r^{*})$. We use the notations $s_{1}=-1+2r^{*}+\frac{3}{2+r^{*}}$
and $s_{2}=\frac{5+5r^{*}+2(r^{*})^{2}}{3(2+r^{*})}.$
}
\end{table*}

\begin{table*}[h]
\begin{center}
 \resizebox{\columnwidth}{!}{
    %cambia a tama�o de letra m'as peque�o
\begin{tabular}{@{\hspace{5pt}}l@{\hspace{10pt}}c@{\hspace{10pt}}l}
\hline
\hline\\[-0.3cm]
Points & $\Upsilon^{*}$ & Solution \\[0.2cm]
\hline\\[-0.2cm]
%%%%%%%%
$P_{6}^{+}(r^{*})$ & $-\frac{3\gamma}{3\gamma+(-2+3\gamma)r^{*}}$ &
 $\ell(t)=\ell_{0}\left(\ell_{1}-D_{0}\Upsilon^{*}\,t\right)^{-\frac{
2r^{*}}{3\gamma}}$, \;\;\;\;
$\rho_{m}(t)=\rho_{m0}\left[\frac{\ell(t)}{\ell_{0}}\right]^{-3\gamma},$
\\[0.3cm]
          && \begin{math}
  R(t)= \left\{ \begin{array}{ll}
R_{0}\left(\ell_{1}- D_{0}\Upsilon^{*}\,t\right)^{\frac{2(1+r^{*})}{m(r)}} &
\textrm{for $r^{*}\neq \{-\frac{3\gamma}{4},-1\}$} \\
 0 & \textrm{for $r^{*}=-\frac{3\gamma}{4}$} \\
 \textrm{arbitrary} & \textrm{for $r^{*}=-1$} 
\end{array}. \right.
\end{math} \\[0.2cm]
&& Isotropic. Decelerated expansion. Unstable. \\[0.2cm]
&& It reduces to $P_3^+$ investigated in \cite{Leon:2010pu} for
$R^n$-gravity, $r^*=-n, \gamma=\tfrac23 n$. 
 \\[0.3cm]
$P_{6}^{-}(r^{*})$ & $\frac{3\gamma}{3\gamma+(-2+3\gamma)r^{*}}$ &
 $\ell(t)=\ell_{0}\left(\ell_{1}-D_{0}\Upsilon^{*}\,t\right)^{-\frac{
2r^{*}}{3\gamma}}$, \;\;\;\;
$\rho_{m}(t)=\rho_{m0}\left[\frac{\ell(t)}{\ell_{0}}\right]^{-3\gamma},$\\[
0.3cm]
          && \begin{math}
  R(t)= \left\{ \begin{array}{ll}
R_{0}\left(\ell_{1}- D_{0}\Upsilon^{*}\,t\right)^{\frac{2(1+r^{*})}{m(r)}} &
\textrm{for $r^{*}\neq \{-\frac{3\gamma}{4},-1\}$} \\
 0 & \textrm{for $r^{*}=-\frac{3\gamma}{4}$} \\
 \textrm{arbitrary} & \textrm{for $r^{*}=-1$} 
\end{array}. \right.
\end{math} \\[0.2cm]
&& Isotropic. Accelerated contraction. Unstable. \\[0.2cm]
&&  It reduces to $P_3^-$ investigated in \cite{Leon:2010pu} for
$R^n$-gravity, $r^*=-n, \gamma=\tfrac23 n$. 
 \\[0.3cm]
$P_{8}^{+}$ & $-3$ & $\ell(t)=\ell_{0}(\ell_{1}+3 D_{0} t)^\frac{1}{3}$,
$\;\;\rho_{m}(t)=\rho_{m0}(\ell_{1}+ 3 D_{0} t)^{-\gamma}$, $R(t)=0.$
\\[0.2cm]
&& Dominated by anisotropy. Decelerated expansion. Total matter/energy mimics
radiation. 
 \\[0.3cm]
$P_{8}^{-}$ & $3$ & $\ell(t)=\ell_{0}(\ell_{1}- 3 D_{0} t)^\frac{1}{3}$,
$\;\;\rho_{m}(t)=\rho_{m0}(\ell_{1}- 3 D_{0} t)^{-\gamma}$, $R(t)=0.$
\\[0.2cm]
&& Dominated by anisotropy. Accelerated contraction. Total matter/energy
mimics radiation. 
 \\[0.3cm]
$P_{9}^{+}$ & $-3$ & $\ell(t)=\ell_{0}(\ell_{1}+ 3 D_{0} t)^\frac{1}{3}$,
$\;\;\rho_{m}(t)=\rho_{m0}(\ell_{1}+ D_{0} t)^{-\gamma}$, $R(t)=0.$ \\[0.2cm]
&& Dominated by anisotropy. Decelerated expansion. Total matter/energy mimics
radiation. 
 \\[0.3cm]
$P_{9}^{-}$ & $3$ & $\ell(t)=\ell_{0}(\ell_{1}- 3 D_{0} t)^\frac{1}{3}$,
$\;\;\rho_{m}(t)=\rho_{m0}(\ell_{1}- 3 D_{0} t)^{-\gamma}$, $R(t)=0.$
\\[0.2cm]
&& Dominated by anisotropy. Accelerated expansion. Total matter/energy mimics
radiation. 
 \\[0.3cm]
$P_{10}^{+}$ & $0$ & $\ell(t)=\ell_{0}\,e^{D_{0}\,t}$,
$\;\;\rho_{m}(t)=\rho_{m0}\,e^{-3 D_{0}\,t\,\gamma}$, $R(t)=R_{0}.$ \\[0.2cm]
&& Isotropic. Accelerated de Sitter expansion. constant Ricci curvature. 
\\[0.3cm]
$P_{10}^{-}$ & $0$ & $\ell(t)=\ell_{0}\,e^{-D_{0}\,t}$,
$\;\;\rho_{m}(t)=\rho_{m0}\,e^{3 D_{0}\,t\,\gamma}$, $R(t)=R_{0}.$ \\[0.2cm]
&& Isotropic. Exponential collapse. constant Ricci curvature. \\[0.3cm]
$P_{11}^{+}$ & $0$ & $\ell(t)=\ell_{0}\,e^{\frac{D_{0}\,t}{2}}$,
$\;\;\rho_{m}(t)=\rho_{m0}\,e^{-\frac{3}{2} D_{0}\,t\,\gamma}$, $R(t)=R_{0}$
\\[0.2cm]
&& Non-flat ($\Omega_k=-3$). Anisotropic. Accelerating de Sitter expansion.
constant Ricci curvature. \\[0.3cm]
$P_{11}^{-}$ & $0$ & $\ell(t)=\ell_{0}\,e^{-\frac{D_{0}\,t}{2}}$,
$\;\;\rho_{m}(t)=\rho_{m0}\,e^{\frac{3}{2} D_{0}\,t\,\gamma}$,
$R(t)=R_{0}$\\[0.2cm]
&& Non-flat ($\Omega_k=-3$). Anisotropic. Exponential collapse. constant
Ricci curvature. \\[0.3cm]\hline \hline
\end{tabular}}\end{center}\caption[crit]{\label{Paramm1}\centering Evolution
rates for $\ell(t)$, $\rho_{m}(t)$ and $R(t)$ evaluated at the critical
points  $P_{6}^{\pm }(r^{*})$ - $P_{11}^{\pm}$.}
\end{table*}

Furthermore, it is necessary to extract the behavior of the physically
important quantities  $\ell(t)$, $\rho_m(t)$ and $R(t)$
at the critical point. The quantity $\ell(t)$ is the length scale along the
flow lines and in the case of zero
anisotropy (for instance in FRW cosmology) it is just the usual scale factor.
Additionally, $\rho_m(t)$ is the matter energy density and $R(t)$ is the
Ricci scalar. These solutions are presented in the last column of Tables
\ref{Paramm3}, \ref{Paramm1} and \ref{Paramm2} \footnote{The mathematical
details for obtaining the evolution rates of $\ell(t)$, $\rho_m(t)$ and
$R(t)$ are presented in the section \ref{Formalism}.}. Lastly, critical
points with $\Sigma^{*}=0$ correspond to isotropic universe. 
 
Let us analyze the physical behavior in more details. For $A^+(r^{*}),$ the
effective EoS parameter and  the deceleration parameter are given by
$w_{eff}=\frac{1-7r^{*}-6(r^{*})^{2}}{3+9r^{*}+6(r^{*})^{2}}$ and
$q=\frac{1-2r^{*}-2(r^{*})^{2}}{1+3r^{*}+2(r^{*})^{2}}$ respectively. Thus,
the condition for an accelerating expanding universe ($w_{eff}<-\frac{1}{3},
q<0$) are reduced to $r^{*}<-\frac{1}{2}(1+\sqrt{3})\approx -1.366$ or
$r^{*}>\frac{1}{2}(-1+\sqrt{3})\approx 0.366.$ Henceforth, the critical point
$A^{+}(r^{*})$ represents an isotropic universe in decelerating (resp.
accelerating) expansion for  $-1.366 \lesssim r^{*} \leq -\tfrac{5}{4}$ or
$-0.5<r^{*} \lesssim 0.366$ (resp. $r^{*}\lesssim -1.366$ or $r^{*}\gtrsim
0.366$). For $r^{*}<-2$ the total EoS parameter of the universe exhibits
phantom behavior. In the case $r^{*}\,=\,-2$, the solution $A^{+}(r^{*})$
have a constant Ricci curvature \mbox{($R(t)=R_{0}$)} (see table
\ref{Paramm3}) and the corresponding cosmological solutions are of
\textsl{de 
Sitter} type. Furthermore, as the curvature energy density goes to zero (see
table \ref{Paramm23}), as the critical point is approached, we obtain an
asymptotically flat universe. 
Summarizing, $A^+(r^{*})$ is an attractor representing a universe in
accelerating expansion provided:
\begin{itemize}
\item $r^{*}<-2,\, M'(r^{*})<0.$ $A^+(r^{*})$. In this case the effective EoS
parameter satisfies $w_{eff}<-1.$ That is, a phantom solution. 
\item $r^{*}\gtrsim 0.366,\, M'(r^{*})<0.$ $A^+(r^{*})$. It adopts the
``appearance'' of a quintessence field, i.e., $-1<w_{eff}<-\frac{1}{3}.$ 
\item If $M'(r^{*})<0$, then $$\lim_{r^{*}\rightarrow \infty} w_{eff}=-1.$$
In this case $A^+(r^{*})$ represents a \textsl{de Sitter}  solution.
\item If $-2<r^{*}\lesssim -1.366, \, M'(r^{*})>0,$  the effective EoS
parameter satisfies $w_{eff}>-1.$
\item In the limit $r^{*}\rightarrow -2,$ $A^+(r^{*})$ reduces to the
non-hyperbolic point $(Q, \Sigma, y, x, r)=(1,0,1,0,-2)$. The stability of
this point depends on the particular choice of $M(r).$ Specially, for the
case $f(R)=R+\alpha R^2$,  $A^+(-2)$ is locally asymptotically unstable
(saddle) \footnote{In the appendix \ref{stabilityAplus} it is presented the
full center manifold analysis to prove that \emph{de Sitter} solution is
locally asymptotically unstable (saddle) for Quadratic Gravity $f(R)=R+\alpha
R^2$.   This result also is consistent with the result obtained in
\cite{Leon:2010ai} for such models but in the conformal formulation as
scalar-tensor theory in Einstein frame.}.  This is of great physical
significance since such a behavior can describe the inflationary epoch of
universe \cite{Leon:2010pu}. 
\item Finally, if $M(r^{*})=M'(r^{*})=0,$  the critical point $A^+(r^{*})$
have at most a 4D stable manifold. 
\end{itemize}

For $A^+(r^{*})$ we obtain the first order asymptotic solutions:
\begin{subequations}
\begin{align}
& \ell(t)= \left\{ \begin{array}{ll}
\ell_{0}\left(\ell_{1}-D_{0}\,t\Upsilon^{*}\right)^{s_{1}} & \textrm{for
$r^{*}\neq-2$} \\
\ell_{0}\,e^{D_{0}\,t} & \textrm{for $r^{*}=-2$}, 
\end{array} \right.
\\ 
& \rho_{m}(t)=\rho_{m0}\left[\frac{\ell(t)}{\ell_{0}}\right]^{-3\gamma},\\
& R(t)= \left\{ \begin{array}{ll}
R_{0}\left(\ell_{1}-
D_{0}\,t\Upsilon^{*}\right)^{\frac{2(1+r^{*})}{m(r^{*})}} & \textrm{for
$r^{*}\neq \{-2,-\frac{5}{4},-\frac{1}{2}\}$} \\
 R_{0} & \textrm{for $r^{*}=-2$} \\
 0 & \textrm{for $r^{*}=\{-\frac{5}{4},-\frac{1}{2}\}$}
\end{array}, \right.
\end{align}
\end{subequations}
where $s_{1}=-1+2r^{*}+\frac{3}{2+r^{*}}$ and
$\Upsilon^{*}=-\frac{2+r^{*}}{3(1+r^{*})^{2}}$.

\begin{table*}[h]
\begin{center}
\resizebox{\columnwidth}{!}{
   %cambia a tama�o de letra m'as peque�o
\begin{tabular}{@{\hspace{5pt}}l@{\hspace{10pt}}c@{\hspace{10pt}}l}
\hline
\hline\\[-0.3cm]
Points & $\Upsilon^{*}$ & Solution \\[0.2cm]
\hline\\[-0.2cm]
%%%%%%%%
$L^{+}(r^{*})$ & $-4$ & $\ell(t)=\ell_{0}\sqrt{\ell_{1}+ 4 D_{0} t}$,
$\;\;\rho_{m}(t)=\rho_{m0}(\ell_{1}+4 D_{0}t)^{-3\gamma/2}$, $R(t)=0.$
\\[0.2cm]
&& Expanding. Decelerated. Total matter/energy mimics radiation. \\[0.3cm]
$L^{-}(r^{*})$ & $4$ & $\ell(t)=\ell_{0}\sqrt{\ell_{1}- 4 D_{0} t}$,
$\;\;\rho_{m}(t)=\rho_{m0}(\ell_{1}-4 D_{0}t)^{-3\gamma/2}$, $R(t)=0.$
\\[0.2cm]
&& Contracting. Accelerated. Total matter/energy mimics radiation. \\[0.3cm]
$P_{1}^{+}(r^{*})$,$P_{2}^{+}(r^{*})$ & $-3$ & $\ell(t)=\ell_{0}(\ell_{1}+ 3
D_{0} t)^{1/3}$, $\;\;\rho_{m}(t)=\rho_{m0}(\ell_{1}+3 D_{0}t)^{-\gamma}$,
$R(t)=0$ \\[0.2cm]
&& Dominated by anisotropy. Decelerated expansion. \\[0.2cm]
&& Total matter/energy mimics radiation. \\[0.3cm]
$P_{1}^{-}(r^{*})$,$P_{2}^{-}(r^{*})$, & $3$ & $\ell(t)=\ell_{0}(\ell_{1}- 3
D_{0} t)^{1/3}$, $\;\;\rho_{m}(t)=\rho_{m0}(\ell_{1}-3 D_{0}t)^{-\gamma}$,
$R(t)=0$ \\[0.2cm]
&& Dominated by anisotropy. Contracting. Accelerating. \\[0.2cm]
&& Total matter/energy mimics radiation. \\[0.3cm]
$C^{+}(r^{*})$ & $-3-\sin u$ & $\ell(t)=\ell_{0}(\ell_{1}- D_{0}\Upsilon^{*}
t)^{-\frac{1+\sin u}{\Upsilon^{*}}}$, $\;\rho_{m}(t)=\rho_{m0}(\ell_{1}-
D_{0}\Upsilon^{*} t)^{\frac{3\gamma(1+\sin u)}{\Upsilon^{*}}}$, $R(t)=0.$
\\[0.2cm]
&& Expanding. Decelerating. Total matter/energy mimics radiation.  \\[0.3cm]
$C^{-}(r^{*})$ & $3-\sin u$ & $\ell(t)=\ell_{0}(\ell_{1}- D_{0}\Upsilon^{*}
t)^{\frac{1-\sin u}{\Upsilon^{*}}}$, $\;\rho_{m}(t)=\rho_{m0}(\ell_{1}-
D_{0}\Upsilon^{*} t)^{\frac{3\gamma(-1+\sin u)}{\Upsilon^{*}}}$,
$R(t)=0.$\\[0.2cm]
&& Contracting. Accelerating. Total matter/energy mimics radiation. 
\\[0.3cm]\hline \hline
\end{tabular}}\end{center}\caption[crit]{\label{Paramm2}  Evolution rates for
$\ell(t)$, $\rho_{m}(t)$ and $R(t)$, evaluated a the curves of critical
points $C^{\pm}$ and the particular cases $L^{\pm}$, $P_{1}^{\pm}$ and
$P_{2}^{\pm}$.}
\end{table*}

The critical points $A^{-}(r^{*})$ and $B^{-}(r^{*})$ represent contracting
isotropic universes, which are unstable and thus they cannot be the late-time
state of the universe. On the other hand, $B^{+}(r^{*})$ have large
probability to represent a decelerating  zero curvature ($R(t)=0$)
future-attractor if $1\leq \gamma < \frac{5}{3} \approx 1.67, r^{*}\neq-1$
since in such a case the critical point have a stable manifold 4D. 

For $B^+(r^{*})$ we obtain the first order asymptotic solutions:
\begin{subequations}
\begin{align}
&\ell(t)=\ell_{0}\sqrt{\ell_{1}-D_{0}t\Upsilon^{*}} ,\\
&\rho_{m}(t)=\rho_{m0}(\ell_{1}-D_{0}t\Upsilon^{*})^{-3\gamma/2},\\ 
&R(t)=0,
\end{align}
\end{subequations}
where $\Upsilon^{*}=\frac{4}{3(-2+\gamma)}$.  For the choice $\gamma=1,$ this curve is a
saddle with a 4D stable manifold  for $-1<r^*<\tfrac34, M'(r^*)<0$.   

The points $N^{\pm}(r^{*})$  represent static universes ($Q=0$), with $N^-$
($N^+$) stable (unstable). The points $L^{-}(r^{*})$ correspond to an
isotropic asymptotically flat universe with accelerated contraction for
$r^{*}\neq -1.$ It is an attractor for $1\leq \gamma < \frac{5}{3},
r^{*}<-\frac{5}{4}, M'(r^{*})<0$ or $1\leq \gamma < \frac{5}{3}, r^{*}>-1,
M'(r^{*})<0$. The critical point $L^{+}(r^*)$ correspond to a universe in
decelerating expansion for $r^{*}\neq -1$ and it is unstable (local
past-attractor) for $1\leq \gamma < \frac{5}{3}, r^{*}<-\frac{5}{4},
M'(r^{*})<0$ or $1\leq \gamma < \frac{5}{3}, r^{*}>-1, M'(r^{*})<0$.  

The critical points $P_{1}^{-}(r^{*})$ and $P_{2}^{-}(r^{*})$ possesses a 3D
stable manifold and a 2D center manifold. Both represent an asymptotically
flat universe in accelerating contraction. In contrast, the non-hyperbolic
points $P_{1}^{+}(r^{*})$ and $P_{2}^{+}(r^{*})$ correspond to a universe in
decelerating expansion, not representing the late-time universe, because they
possesses a 3D unstable manifold.

The points $P_{4}^{+}(r^{*})$, have a stable manifold 4D and correspond to a
non-flat universe ($\Omega_{k}\neq0$). The values of $w_{eff}$ and $q$ at the
corresponding cosmological solutions are given by 
$w_{eff}=\frac{1-r^{*}(21+4r^{*}(9+4r^{*}))}{3+3r^{*}(21+8r^{*}(3+r^{*}))}$
and $q=\frac{1-2r^{*}-2(r^{*})^{2}}{5+5r^{*}+2(r^{*})^{2}},$ respectively.
\begin{itemize}
\item  For $-2.395 \lesssim \frac{1}{4} \left(-5-\sqrt{21}\right)<r^{*}<-2,
M'(r^{*})<0,$ $P_4^+(r^{*})$ represent phantom solutions ($w_{eff}<-1$).
\item For $r^{*}\lesssim -2.395$ they are non-phantom  accelerating solution.
\item For $r^{*}=\left(-5-\sqrt{21}\right)$, they are de Sitter solutions.
\item For $r^{*}>\frac{1}{2}(-1+\sqrt{3}), M'(r^{*})<0,$ $P_4^+$ represent an
accelerating solution with $-\frac{2}{3}<w_{eff}<-\frac{1}{3}$. In the limit
$r^{*}\rightarrow \infty$ we obtain $w_{eff}=-\frac{2}{3}.$
\item For $-2<r^{*}<\frac{1}{2}(-1-\sqrt{3}), M'(r^{*})>0$ the EoS parameter
satisfies $w_{eff}>-1.$
\item For $r^{*}=-2,$ $P_4^+(r^{*})$ is non-hyperbolic with a 3D stable
manifold. 
\end{itemize}

For $P_4^+(r^{*})$ we have the first order solutions:\\
\begin{subequations}
\begin{align}
& \ell(t)= \left\{ \begin{array}{ll}
\ell_{0}\left(\ell_{1}-D_{0}\,t\Upsilon^{*}\right)^{s_{2}} & \textrm{for
$r^{*}\neq-2$} \\
\ell_{0}\,e^{\frac{D_{0}\,t}{2}} & \textrm{for $r^{*}=-2$}, 
\end{array} \right.
\\
&\rho_{m}(t)=\rho_{m0}\left[\frac{\ell(t)}{\ell_{0}}\right]^{-3\gamma}, \\
&R(t)= \left\{ \begin{array}{ll}
R_{0}\left(\ell_{1}- D_{0}\,t\Upsilon^{*}\right)^{\frac{2(1+r^{*})}{m(r)}} &
\textrm{for $r^{*}\neq -2$} \\
 R_{0} & \textrm{for $r^{*}=-2$}
\end{array}, \right.
\end{align}
\end{subequations} 
where $\Upsilon^{*}=-\frac{3(2+r^{*})}{10+r^{*}(16+7r^{*})}.$

The critical point $P_{4}^{-}(r^{*})$ represents a decelerated contracting
universe which is unstable and thus it cannot be a late-time solution. The
non-hyperbolic point $P_{5}^{+}(r^{*})$ (resp. $P_{5}^{-}(r^{*})$) correspond
for $5/3\leq\gamma\leq2,\,r^{*}\neq-1$ to a zero-curvature cosmological
solution  in decelerating expansion (resp. accelerating contraction).
$P_{5}^{+}(r^{*})$ have a 2D center manifold and a 3D unstable manifold.
Thus, it cannot represent the late-time universe. On the other hand
$P_{5}^{-}(r^{*})$ have a 2D center manifold and a 3D stable manifold.
Evaluating to first order and assuming $\gamma\neq 0,$ we get the asymptotic
solutions:
\begin{subequations} 
\begin{align}
&\ell(t)\propto t^\frac{1}{3(-1+\gamma)}, \\
&\rho_{m}(t)\propto t^\frac{\gamma}{1-\gamma},\\ 
&R(t)=0.
\end{align}
\end{subequations}

The critical points $P_{6}^{+ }(r^{*})$ (resp. $P_{6}^{-}(r^{*})$)
corresponding to a universe in decelerated expansion (resp. accelerated
contraction) are of saddle type, thus they cannot be a late-time solution of
universe.  

For $P_6^+(r^{*})$ we have the first order asymptotic solutions:
\begin{subequations}
\begin{align}
&\ell(t)=\ell_{0}\left(\ell_{1}-D_{0}\Upsilon^{*}\,t\right)^{-\frac{2r^{*}}{
3\gamma}}, \\
&\rho_{m}(t)=\rho_{m0}\left[\frac{\ell(t)}{\ell_{0}}\right]^{-3\gamma},\\
&R(t)= \left\{ \begin{array}{ll}
R_{0}\left(\ell_{1}- D_{0}\Upsilon^{*}\,t\right)^{\frac{2(1+r^{*})}{m(r)}} &
\textrm{for $r^{*}\neq \{-\frac{3\gamma}{4},-1\}$} \\
 0 & \textrm{for $r^{*}=-\frac{3\gamma}{4}$} \\
 \textrm{arbitrary} & \textrm{for $r^{*}=-1$} 
\end{array}, \right.
\end{align}
\end{subequations}
where $\Upsilon^{*}=-\frac{3\gamma}{3\gamma+(-2+3\gamma)r^{*}}.$   

$P_6^+(r^{*})$ represent a matter dominated solution if $1\leq \gamma \leq
\frac{5}{48} \left(5+3 \sqrt{17}\right)\lesssim 1.8093$ and $r^*=-\frac{3
\gamma }{6 \gamma +4}$ with $M'(r^{*})\neq 0;$  or $r^{*}\rightarrow -1,$
$M(r^*)=0.$ In this limit appears two eigenvalues  $\lambda_1\rightarrow
+\infty$ y $\lambda_2\rightarrow -\infty.$ Thus, it behaves as a saddle point
since at least two eigenvalues are of different signs. In this case the
solutions don't remain a long period of time near these solutions, because
they have a strongly unstable direction (associated to the positive infinite
eigenvalue).

For the equilibrium curves $P_{8}^{\pm}$ and $P_{9}^{\pm}$ we have the first
order solutions:
\begin{subequations}
\begin{align} 
&\ell(t)=\ell_{0}(\ell_{1}\pm 3 D_{0} t)^\frac{1}{3}, \\
&\rho_{m}(t)=\rho_{m0}(\ell_{1}\pm 3 D_{0} t)^{-\gamma},\\
&R(t)=0.
\end{align}
\end{subequations}
These curves contains as particular cases the critical points $P_{1}^{\pm
}(r^{*})$ and $P_{2}^{\pm }(r^{*})$ when $r_c=r^{*}$. Since these points
don't describe accurately the current universe, we won't discuss in details
the physical properties of the corresponding cosmological solutions.

The critical point $P_{10}^{+}$ describe a flat isotropic universe in
accelerated expansion of \textsl{de Sitter} type because $w_{eff}=-1.$ This
point always exists. If $M(-2)>0$, $P_{10}^{+}$ is a future-attractor (see
table \ref{tab3}). Evaluating to first order in $P_{10}^{+}$ the asymptotic
solutions are
\begin{subequations}
\begin{align}
&\ell(t)=\ell_{0}\,e^{D_{0}\,t},\\
&\rho_{m}(t)=\rho_{m0}\,e^{-3 D_{0}\,t\,\gamma},\\
& R(t)=R_{0}.
\end{align}
\end{subequations}
For $M(-2)=0$ this critical point is non-hyperbolic, having a 4D stable
manifold. Observe that $A^+$ and $P_{10}^{+}$ coincides if the function
$M(r)$ vanishes at $r=-2$. The point $P_{10}^{-}$ represents a flat isotropic
universe in decelerating contraction, and thus, it cannot represents
accurately the late-time universe.

The critical point $P_{11}^{-}$ corresponds to contracting decelerated
universe which is unstable. Thus, it cannot reproduce the late-time state of
the universe. On the other hand $P_{11}^{+}$ have a 4D stable manifold. Thus
it have a large probability to represents the late-time universe.
Additionally, it is a  \textsl{de Sitter} solution provided $r=-2,\,
0<M(-2)\leq \frac{3}{8}$. Evaluating to first order at $P_{11}^{+}$ we obtain
the first order asymptotic solutions 
\begin{subequations}
\begin{align}
&\ell(t)=\ell_{0}\,e^{\frac{D_{0}\,t}{2}},\\
&\rho_{m}(t)=\rho_{m0}\,e^{-\frac{3}{2} D_{0}\,t\,\gamma},\\ 
&R(t)=R_{0}.
\end{align}
\end{subequations}

Finally, for $u\,\in [0,2\pi],\, u\neq\pi/2$ the curve of critical points
$C^{-}(r^{*})$ correspond to flat universes in accelerating contraction which
correspond to an attractor. For $u\,\in [0,2\pi],\, u\neq3\pi/2$, the
critical points in $C^{+}(r^{*})$ are unstable, thus they cannot attract the
universe at late times. 

\subsection{Comparison with previously investigated models}

Having discussed in the previous section the stability conditions and
physical relevance of equilibrium points, in this subsection we will
proceeded to the comparison with previously investigated models.

Firstly, as we observe, for the choice $f(R)=R^n$ and for Kantowski-Sachs geometry studied in
\cite{Leon:2010pu}, $r$ is
always equal to the constant $-n,$ since $M$ is identically to zero. Then,
for the choice $r^*=-n,$ the point $A^+(r^*)$ coincides with the fixed point
$A^+$ discussed in \cite{Leon:2010pu}. In this special case it is not
required to consider the stability along the $r$-direction, and thus $A^+$ is
stable for $n>\frac{1}{2}(1+\sqrt{3})$ and a saddle otherwise
\cite{Leon:2010pu}. The upper bound $n<3$ appearing in the table 2 of
\cite{Leon:2010pu} is not required here, since we relax the link between $n$
and the EoS of matter ($w$) given by $n=\tfrac32(1+w),$ introduced in
\cite{Goswami:2008fs,Goheer:2008tn} as a requirement for the existence of an
Einstein static universe. As commented in \cite{Leon:2010pu}, when the
$n$-$w$ relation in $R^n$ cosmology is relaxed, in general we cannot obtain
the epoch sequence of the universe where the Einstein static solution (in
practice as a saddle-unstable one) allows to the transition from the 
Friedmann matter-dominated phase to the late-time accelerating phase. The
special point $A^+(0)$ corresponds to the point $P_4$ studied in the context of arbitrary $f(R)$- FRW models
\cite{Amendola:2006we}, which satisfies $\Omega_m=0 $ and $w_{eff}=\tfrac13,$
i.e., it is a solution for which the total matter/energy mimics radiation.
The eigenvalues of the linearization around $A^+(0)$ are
$\left\{-\frac{5}{3},-\frac{5}{3},\frac{2}{3},-\frac{1}{3} (\gamma
+2),\frac{4 M'(0)}{3}\right\},$ thus, it is always a saddle. This is a
difference comparing to the result in \cite{Amendola:2006we} where it  can
be stable for some interval in the parameter space.     
 
Concerning $B^+(r^{*})$, in the special case of $\gamma=\frac{4}{3}$ (radiation), it
corresponds to a matter-dominated solution for which the total matter/energy
mimics radiation. In this case the critical point is a saddle point in the
same way as it is the analogous point $B^+$ discussed in \cite{Leon:2010pu}.
Since $B^+$ behaves as saddle, its neighboring cosmological solutions
  abandon the matter-dominated epoch and the orbits are attracted by
an accelerated solution. This result has a great physical significance since
a cosmological viable universe possesses a standard 
matter epoch followed by an accelerating expansion epoch
\cite{Amendola:2006we}. 

The point $B^+(r^*)$ for the
special case $\gamma=1$ is the analogous of $P_2$ studied in
\cite{Amendola:2006we}. In fact the values $(x_1,x_2,x_3)=(-1,0,0)$
correspond for $H>0$ to $(Q,\Sigma,x,y)=(\tfrac23,0,\tfrac13,0)$ and since
$x_2=x_3=0$ and $x\neq 0,$ follows that $r$ is an arbitrary value that
satisfies $M(r)=0,$ that is, $r=r^*$. This point was referred in
\cite{Amendola:2006we} as the $\phi$-matter dominated epoch (or wrong matter
epoch since $\Omega_m=2>1$).

$L^{+}(r^*)$  is the analogous to $P_3$ investigated in
\cite{Amendola:2006we} since $(x_1,x_2,x_3)=(1,0,0)$ correspond to
$(Q,\Sigma,x,y)=(2,0,-1,0)$ through the coordinate transformation
\eqref{comparison} and the condition $x_2=x_3=0,x\neq 0,$ implies that $r$ is
an arbitrary value that satisfies $M(r)=0,$ that is, $r=r^*$. This point was
referred in \cite{Amendola:2006we} as the purely kinetic points and in our
context, and for the
choice $\gamma=1$, it is a local source for $r^{*}<-\frac{5}{4}, M'(r^{*})<0$
or $r^{*}>-1, M'(r^{*})<0$. The condition $r^{*}<-\frac{5}{4}$ (respectively
$r^{*}>-1$) implies, under the assumption that $r^*/m(r^*)$ remains bounded,
the condition $m(r^*)>1/4$ (respectively $m(r^*)<0$), which are the
instability
condition for $P_3$ discussed in \cite{Amendola:2006we}. It is worthy to
mention that there the authors do not refer explicitly to the additional
condition $M'(r^{*})<0$ obtained here, since there the authors do not
investigate the stability along the $r$-axis.  

The curves of fixed points $P_1^+(r^*)$ and $P_2^+(r^*)$ are analogous points
$P_1^\pm$ and $P_2^\pm$ studied in  \cite{Leon:2010pu} for the special case
of $R^n$-gravity under the
identifications $r^*=-n$ and $\gamma=\tfrac23 n$.

For the point $P_4^+(r^*)$, and using the identification $r^*=-n,$ and
without  considering the stability
conditions due to the extra $r$-coordinate (that is not required in the
analysis of $R^n$-gravity) we recover the stability conditions for $P_4^+$
and the physical behavior discussed in the reference \cite{Leon:2010pu}. 

The points $P_{5}^{\pm}(r^{*})$ reduce to the points $P_5^\pm$ studied in
\cite{Leon:2010pu} when are considered the restrictions $r^*=-n$ and
$\gamma=\tfrac23 n.$ As commented before, the additional variable $r$ is
irrelevant for the stability analysis of $R^n$-gravity since it is constant
and does not evolves with time. 

The points $P_{6}^{\pm}(r^{*})$ reduce to $P_3^\pm$ for the especial choice
$r^*=-n, \gamma=\tfrac23 n$ in the particular case of $R^n$-gravity
investigated in \cite{Leon:2010pu}. 
The points $P_6^+(r^*)$ reduce to the point $P_5$ studied in
\cite{Amendola:2006we} for the choice $\gamma=1.$  

The points $P_{10}^\pm$ exist in
$R^n$-gravity only for $n=2$ and in this case they correspond to the points
${\cal
A}^\pm$ studied in \cite{Leon:2010pu}. Using the coordinate transformation
\eqref{comparison} we obtain that the point  $P_{10}^{+}$ is the analogous to
$P_1$ investigated in the reference \cite{Amendola:2006we}. In this case the
stability condition $M(-2)>0$ implies the condition $0\leq m(-2)< 1$ which is
the stability condition (in the dynamical systems approach) for the de Sitter
solution presented in \cite{Amendola:2006we}. The special case $m(-2)=0$
corresponds to $M(-2)\rightarrow \infty$ which is compatible with $M(-2)>0.$ 
It is worthy to mention that, apart from the dynamical systems stability,
there is another kind of stability condition for the de Sitter solution in
$f(R)$ gravity during expansion, which it is well know it takes the form
$0<R_0 f''(R_0)<f'(R_0),$ where $R_0$ satisfies the equation $R_0 f'(R_0)=2
f(R_0).$ This condition was first derived in \cite{Muller:1987hp}.   

The points $P_{11}^\pm$ exist in $R^n$-gravity only for $n=2$ and in this
case correspond to the points $P_4^\pm$ studied in \cite{Leon:2010pu} which
are saddle points.  

Summarizing, as displayed in tables \eqref{Paramm3}, \eqref{Paramm1}, \eqref{Paramm2} we have found a very rich behavior, and amongst others  the universe can result in isotropized solutions with observables in agreement with observations, such as de Sitter, quintessence-like, or phantom solutions depending on the intervals where values $r=r^*$ such that $M(r^*)=0$ belong. These kind of solutions are permitted for $f(R)$-FRW cosmologies as well.  

Now, as differences  with respect FRW models we have that:
\begin{itemize}
\item There is a large probability to have expanding decelerating zero-curvature future attractors like $B^+(r^*)$ for $1< \gamma <\frac{5}{3}$; 
\item some static kinetic-dominated contracting solutions like $N^-(r^*)$ can be stable; 
\item isotropic, asymptotically flat universes with accelerated contraction like $L^-(r^*)$ can be stable for $1< \gamma <\frac{5}{3}$; 
\item additionally we have solutions which are: non-flat, anisotropic, accelerating de Sitter expansion, with constant Ricci curvature, like $P_{11}^+$, which can attract an open set of orbits;
\item more interesting features are the existence of curves of solutions corresponding to flat universes in accelerated contraction like $C^-(r^*)$ which corresponds to attractors and are typically kinetic-anisotropic scaling solutions. 
\end{itemize}

Besides, as discussed above, our results are generalization of
previously investigated models models both in FRW and KS metrics. Finally, as
we will show in our numerical elaborations in the section \ref{Sect:6} a
cosmological bounce and turnaround are realized in a part of the
parameter-space as a consequence
of the metric choice.

\subsection{Feasibility of the Kantowski-Sachs $f(R)$ models from the
cosmological point of view.}\label{regiones}

In this subsection we briefly discuss the feasibility of the Kantowski-Sachs
$f(R)$ models from the cosmological point of view.
According to the previous discussion, a Kantowski-Sachs geometry in
$f(R)$-gravity leads to a great variety  of physically motivated cosmological
solutions. We have obtained not only acceleration of the expansion, but also
phantom behavior, for example, the critical points $A^{+}(r^{*})$,
$P_{4}^{+}(r^{*})$. Also, we have attracting de Sitter solutions, say
$P_{10}^{+}$ and $P_{11}^{+}.$ Additionally, the points $B^+(r^{*})$ and
$P_6^+(r^{*})$ can represent matter dominated universes. So, a good test for
a cosmologically viable scenario, is the ability of the model to reproduce 
the different stages through which the universe has evolved.

A viable cosmological model must have a phase with an accelerated expansion 
preceded by a matter domain phase \cite{Amendola:2006we}. Actually, since it
is required to have a sufficient growth of density perturbation in the matter
component including baryons, a viable cosmological model should have a matter
dominated stage driven by dust matter ($\gamma=1$), and not by the effective
matter density, after the radiation dominated one (and before an accelerated
stage). Now, since we are more interested on the late-time dynamics and on
the possibility of $f(R)$-gravity to mimic Dark Energy, we have not
considered radiation explicitly in our model. Nevertheless, we can mimic
radiation dominance by setting $\gamma=4/3$ (e.g., the region VII in the
following discussion), but our model cannot reproduce both the dust and
radiation dominated phases of the universe. Just for completeness, we want to
show conditions for the existence of a matter (in a broader sense, not just
restricted to dust) dominated phase 
preceding a Dark-matter dominated one. A complete, and more accurate analysis
deserves additional research.

Following this reasoning, in our scenario, it is possible to define the
following regions where the future-attractor represents cosmological
solutions in accelerating expansion (we assume $1\leq \gamma\leq 2$):
\begin{itemize}
\item[I:] $r^*<-2,\; M'(r^*)<0$. In this region $A^+(r^*)$ is a phantom
dominated ($w_{eff}<-1$) attractor.
\item[II:] $M(-2)>0.$ In this region the attractor is $P_{10}^+$, which is a 
de Sitter solution. 
\item[III:] $-2<r^*<-\frac{1}{2}(1+\sqrt{3}),\; M'(r^*)>0.$ In this region
$A^+(r^*)$ is a quintessence dominated attractor.
\item[IV:] $r^*>\frac{1}{2}(-1+\sqrt{3}),\; M'(r^*)<0.$ In this region
$A^+(r^*)$ is quintessence dominated attractor.
\end{itemize}
On the other hand, we found the critical points  $P_6^+$ and $B^+$
representing cosmological  matter dominated solutions ($\Omega_{m}=1$). As
can be expected these points behaves as saddle points. Going to the parameter
space, it is possible to define three regions for the existence of a
transient matter dominated phase:
\begin{itemize}
\item[V:] $1\leq \gamma\leq 2 ,\; r^* \rightarrow -1.$ In this case
$P_6^+(r^*)$ represents a matter-dominated universe (for the particular case
in that $M(-1)=-1$ this point corresponds to the matter point $P_{5}$ found
in \cite{Amendola:2006we}).
\item[VI:] $1\leq \gamma\leq\frac{5}{48}(5+3\sqrt{17}) ,\; r^*=-\frac{3
\gamma}{4+6 \gamma},\; M'(r^*)\neq 0.$ In this case $P_6^+(r^*)$ represent a
matter-dominated universe.
\item[VII:] $\gamma=\frac{4}{3}, M(r^*)=0.$ In this case $B^+(r^*)$
represents a matter-dominated universe.
\end{itemize}
Summarizing we have determined regions in the  space $(\gamma, r, M'(r))$
containing matter dominated solutions (regions V-VII) and containing
accelerated expanding solutions (regions I-IV). Thus, a $f(R)$ theory with a
curve $M(r)$ connecting in the plane $(r, M'(r))$ a matter dominated phase
prior to an accelerated expanding one can be considered as cosmologically
viable. In this regard, our results  generalize the results presented in
\cite{Amendola:2006we}. It is worthy to mention that even if the above
connection is permitted, these qualitative analysis don't imply that these
models are totally cosmologically viable, because in the analysis is not
defined if the matter era is too short or too long to be compatible with the
observations.  

Nevertheless, the more interesting result here is that introducing an input
function $M(r)$ constructed in terms of the auxiliary quantities  $m=\frac{R
f''(R)}{f'(R)}=\frac{d \ln{f'(R)}}{d \ln{R}}$ and $r=-\frac{R
f'(R)}{f(R)}=-\frac{d \ln{f(R)}}{d \ln(R)},$ and considering very general
mathematical properties such as differentiability, existence of minima,
monotony intervals, etc, we have obtained cosmological solutions compatible
with the modern cosmological paradigm. Furthermore, with the introduction of
the variables $M$ and $r$, one adds an extra direction in the phase-space,
whose neighboring points correspond to ``neighboring'' $f(R)$-functions.
Therefore, after the general analysis has been completed, the substitution of
the specific $M(r)$ for the desired function $f(R)$ gives immediately the
specific results. Is this crucial aspect of the method the one that make it
very powerful, enforcing its applicability.

\section{Examples}\label{Sect:6}

In this section we illustrate our analytical results for a number of
$f(R)$-theories for which it is possible to obtain an explicit  $m(r)$
($M(r)$) functions. Our main purpose is to illustrate the possibility to
realize the matter era followed by a late-time acceleration phase.
Additionally are presented several heteroclinic sequences showing the
transition from an expanding phase to a contracting one, and viceversa.
Moreover, if the universe start from an expanding initial conditions and
result in a contracting phase then we have the realization of a cosmological
turnaround, while if it start from contracting initial conditions and result
in an expanding phase then we have the realization of a cosmological bounce.
Note that the following models, apart from the model 
$f(R)=R^p\exp\left( \frac{q}{R}\right)$, were investigated in
\cite{Ivanov:2011np}. In this reference the authors considered
super-inflating solutions in modified gravity for several well-studied
families
of $f(R)$ functions. Using scalar field reformulation of f(R)-gravity they
describe how the
form of effective scalar field potential can be used for explaining the
existence of stable
super-inflation solutions in the theory under consideration.

\subsection{Model $f(R)=R+\alpha R^2$.}

In this case $M(r)=\frac{1}{2}r(r+2).$ The function $M(r)$ vanishes at $r=0$
and $r=-2,$ that is to say $r^{*}\in\{0,-2\},$ also in this case $M'(0)=1$
and $M'(-2)=-1.$
\begin{itemize}
\item The sufficient conditions for the existence of past-attractors
(future-attractors) are:
   \begin{itemize}
     \item $L^+(-2)$ ($L^-(-2)$) is a local  past (future)-attractor for
$1\leq \gamma<\frac{5}{3}.$
     \item $N^+(0)$ ($N^-(0)$) is always a local  past (future)-attractor.
   \end{itemize}
\item Some  physically interesting saddle points are:
   \begin{itemize}
     \item $A^{+}(-2)$ is a quasi-de Sitter solution ($\ell(t)\propto e^{D_0
t}$), with $H$ changing linear and slowly for $|\alpha| H^2\gg 1$
\cite{Starobinsky:1980te}.  As proved in the Appendix \ref{stabilityAplus},
this solution is locally asymptotically unstable (saddle point) . This result
is of great physical importance because this cosmological solution could
represents early-time acceleration associated with the primordial inflation.
This result is a generalization of the result in \cite{Leon:2010ai} where it
is investigated the stability of these solution in the conformal formulation
of the theory $f(R) = R + \alpha R^2$ as a Scalar-tensor theory for the FRW
metric.  
     \item $A^{+}(0)$ and $P_{4}^{+}(-2)$  are saddle points with a 3D stable
manifold and a 2D unstable manifold. Then it is a saddle point. 
     \item $A^{-}(-2)$ is non-hyperbolic with a 4D unstable manifold when
$r\rightarrow (-2)^-$.  In an analogous way as for $A^{+}(-2)$ it can be
proven that its center manifold is stable. Then it is a  saddle point. 
   \end{itemize}
\end{itemize}

This case corresponds to the Starobinsky inflationary model
\cite{Starobinsky:1980te} and the accelerated phase exists in the asymptotic
past rather than in the future.
Besides, the curve $M(r)$ cannot connect the region VII (defined in section
\ref{regiones}) with a region with accelerated expansion. For this reason
this type of $f(R)$ function  is not cosmologically viable in the sense
discussed in this paper. However we have obtained several heteroclinic
sequences that have physical interest since they show the transition from an
expanding phase to a contracting one, and viceversa which are relevant for
the bounce and the turnaround. These behaviors were known to be possible in
Kantowski-Sachs geometry \cite{Leon:2010pu,Fadragas:2013ina}. 

Some heteroclinic orbits for this example are:
\begin{subequations}
\label{heteroclinicsA}
\begin{align}
& P_{1}^{+}(-2)\longrightarrow \left\{\begin{array}{c}
                          P_{2}^{-}(-2) \\ 
                          P_{10}^{+} 
                         \end{array},    \right. \label{4.9a}
\\
& P_{10}^{-}\longrightarrow \left\{\begin{array}{c}
                          P_{10}^{+} \\ 
                          L^{-}(-2)\\     
      P_{2}^{-}(-2)\\ 
   P_{1}^{-}(-2)
                         \end{array},    \right. \label{4.9b}
\\
& N^+(0)\longrightarrow \left\{\begin{array}{c}
                          N^+(-2)\longrightarrow P_{10}^+ \\ 
                          A^+(0) 
                         \end{array},    \right. \label{4.9c}
\\
& N^{+}(-2)\longrightarrow  P_{10}^{+}, \label{4.9d}\\
& L^{+}(-2)\longrightarrow  P_{10}^{+}, \label{4.9e}\\
& P_{2}^{+}(-2)\longrightarrow \left\{\begin{array}{c}
                          P_{2}^{+}(0)\longrightarrow P_{1}^{-}(0) \\ 
                          P_{1}^{-}(-2)\\    
        P_{10}^{+} 
                         \end{array},    \right. \label{4.9f}  
      \\  
& A^{-}(0)\longrightarrow \left\{\begin{array}{c}
                          P_{1}^{-}(0)\\ 
                          N^{-}(0) 
                         \end{array},    \right. \label{4.9g}
\\
& P_{1}^{+}(0)\longrightarrow \left\{\begin{array}{c}
                               P_{2}^{-}(0)\longrightarrow P_{2}^{-}(-2)\\
                               A^{+}(0)
                             \end{array}.    \right. \label{4.9h}    
\end{align}
\end{subequations} 

\begin{figure}[h]
\centering
\subfigure[]{
\includegraphics[height=5cm,width=7cm]{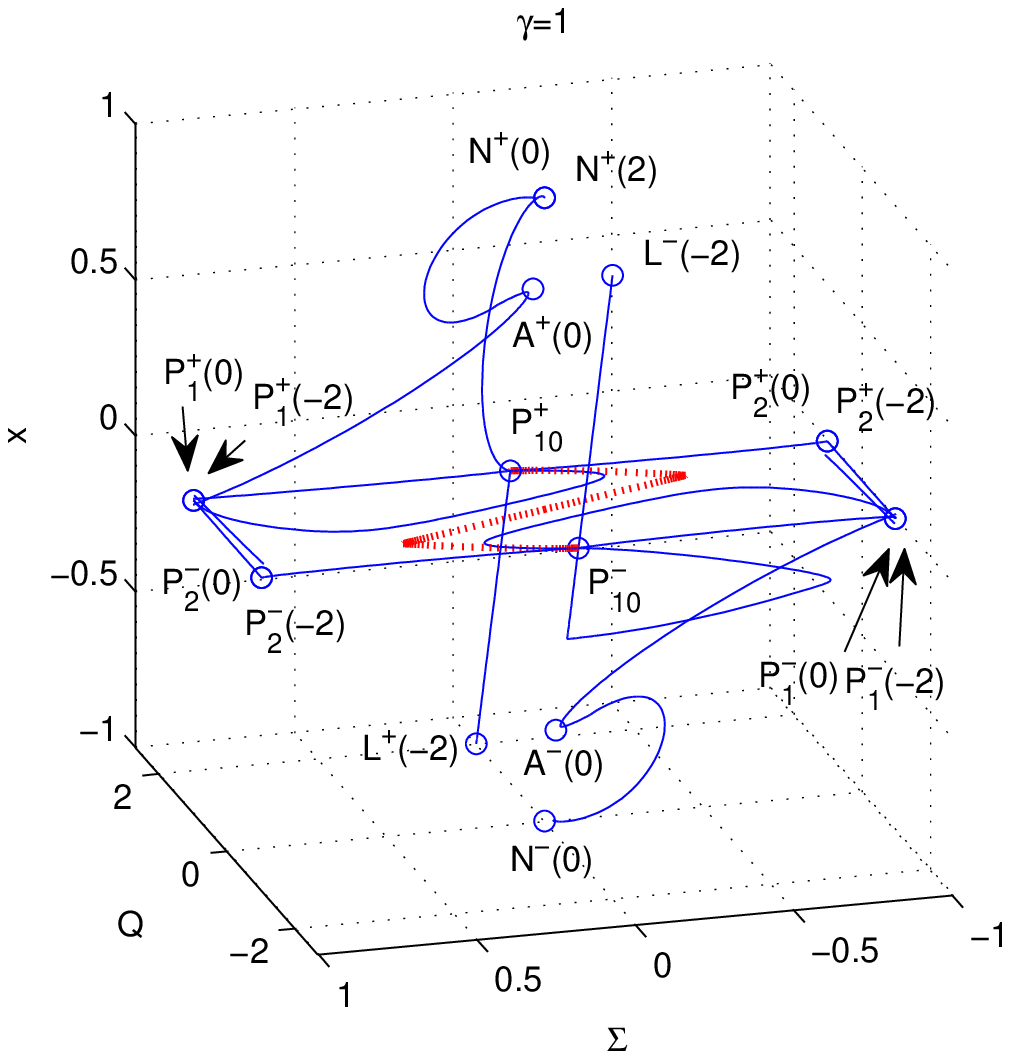}}
\subfigure[]{
\includegraphics[height=5cm,width=7cm]{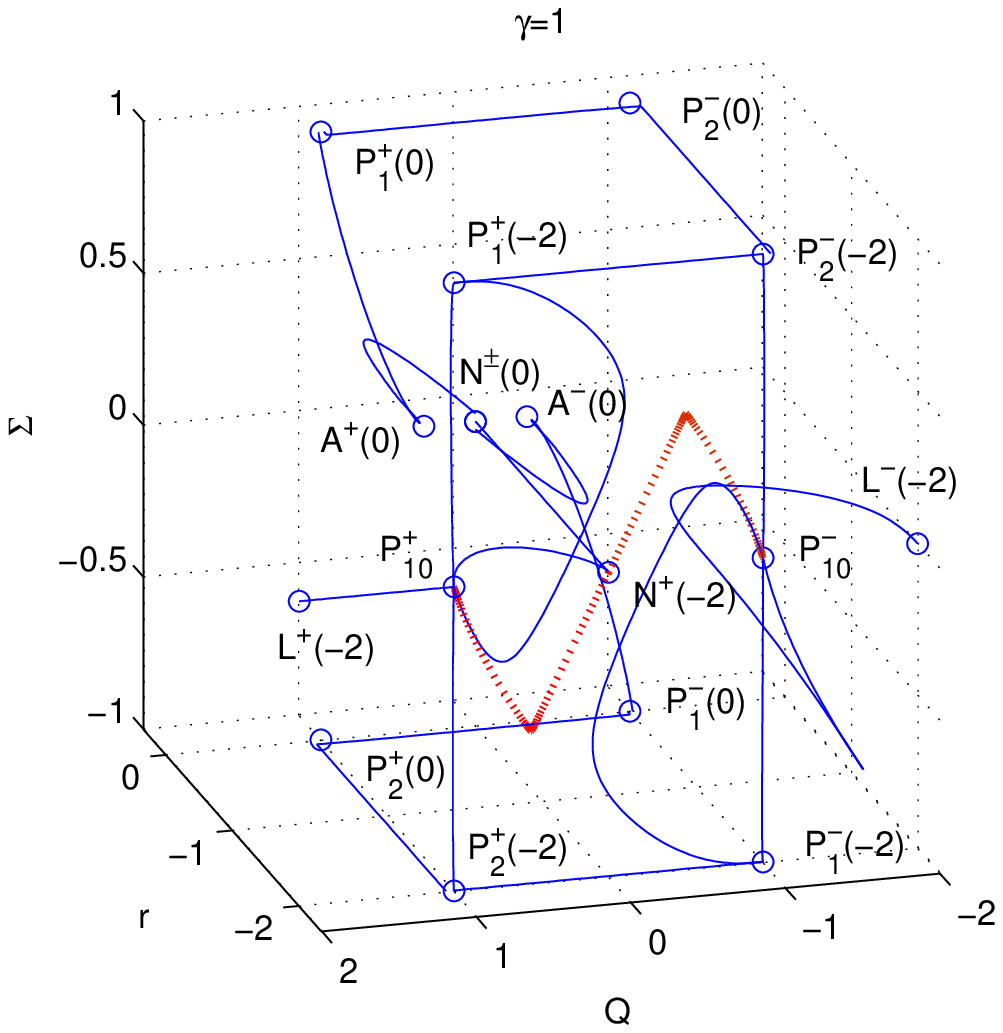}}
\subfigure[]{
\includegraphics[height=5cm,width=7cm]{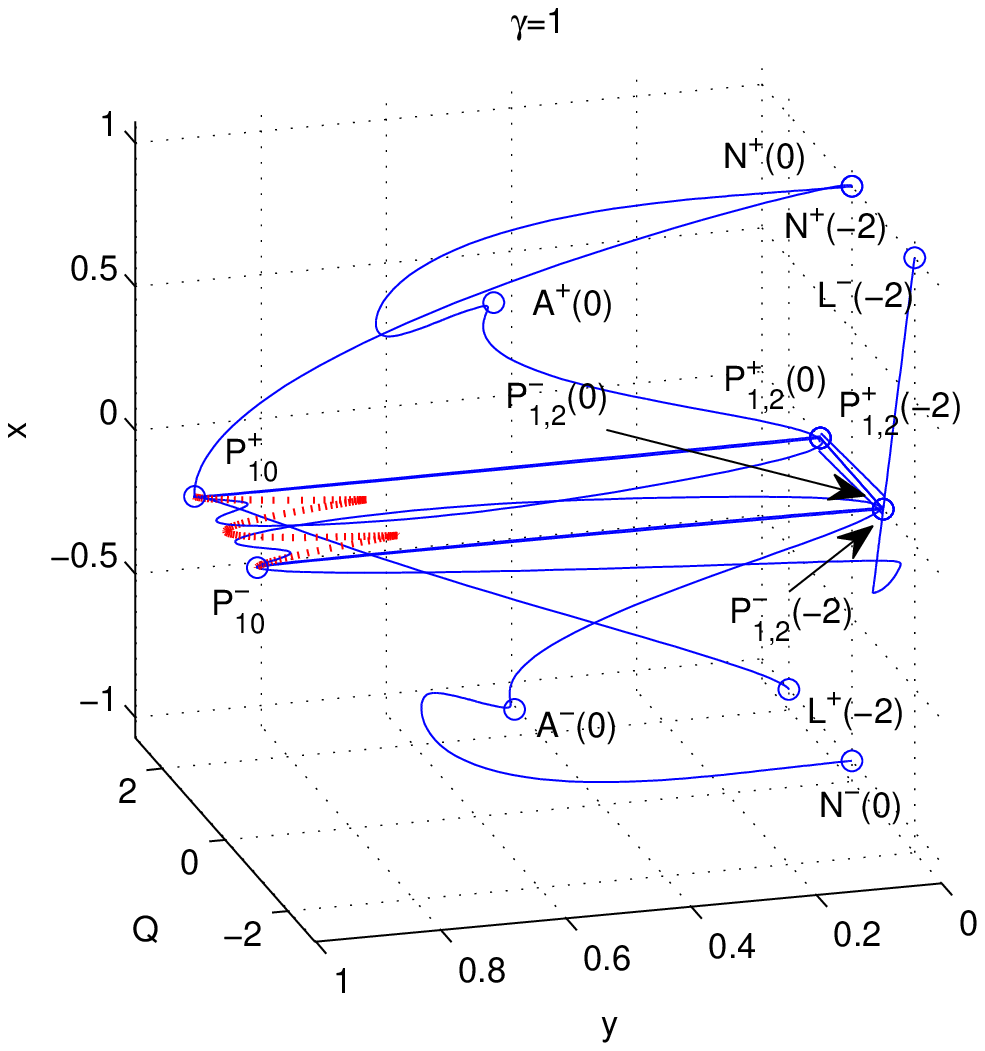}}
\subfigure[]{
\includegraphics[height=5cm,width=7cm]{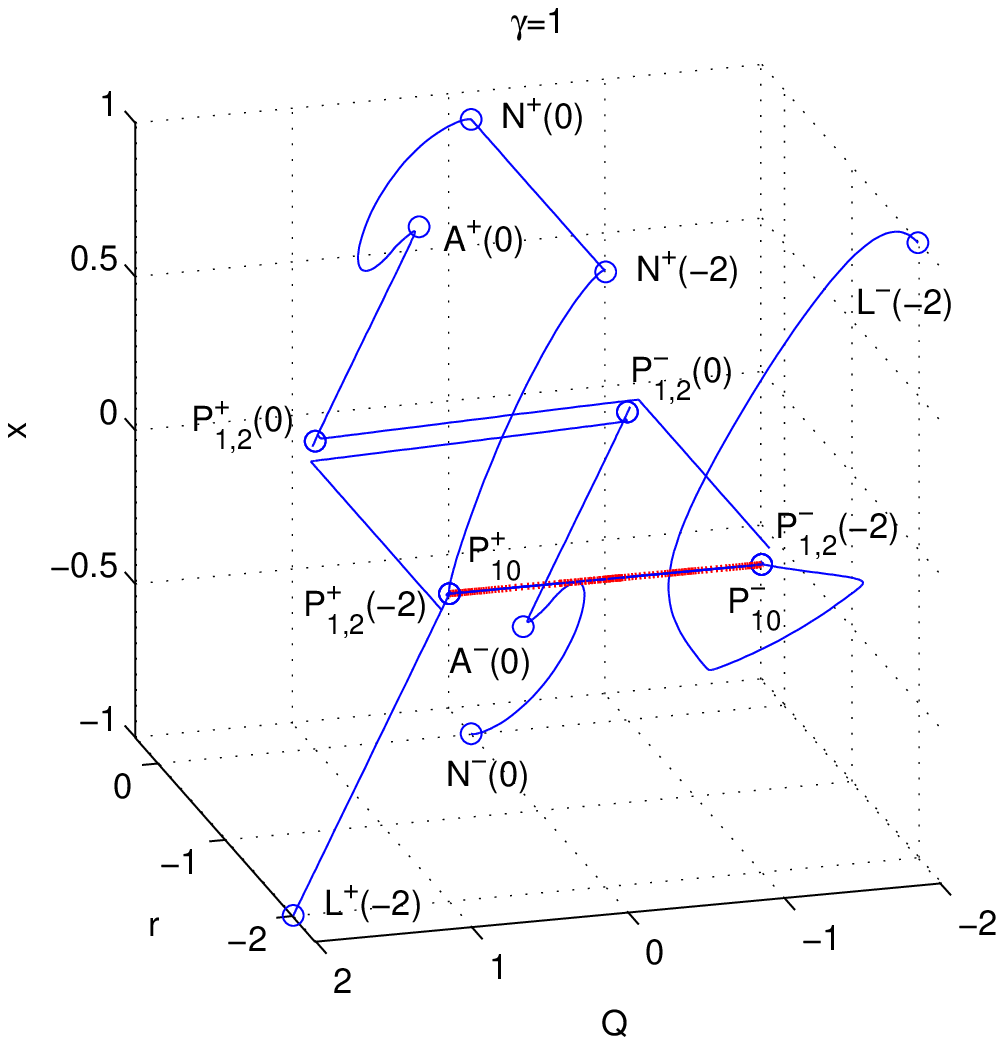}}
\subfigure[]{
\includegraphics[height=5cm,width=7cm]{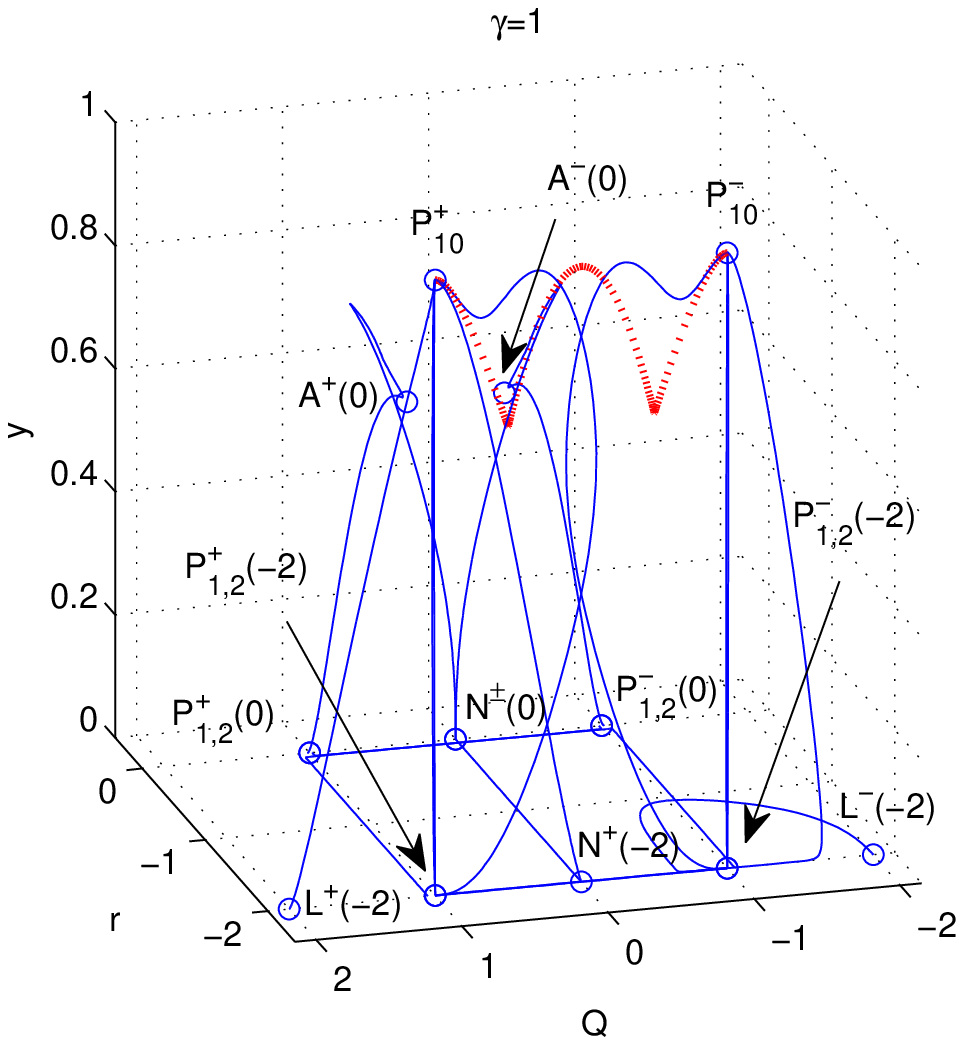}}
\subfigure[]{
\includegraphics[height=5cm,width=7cm]{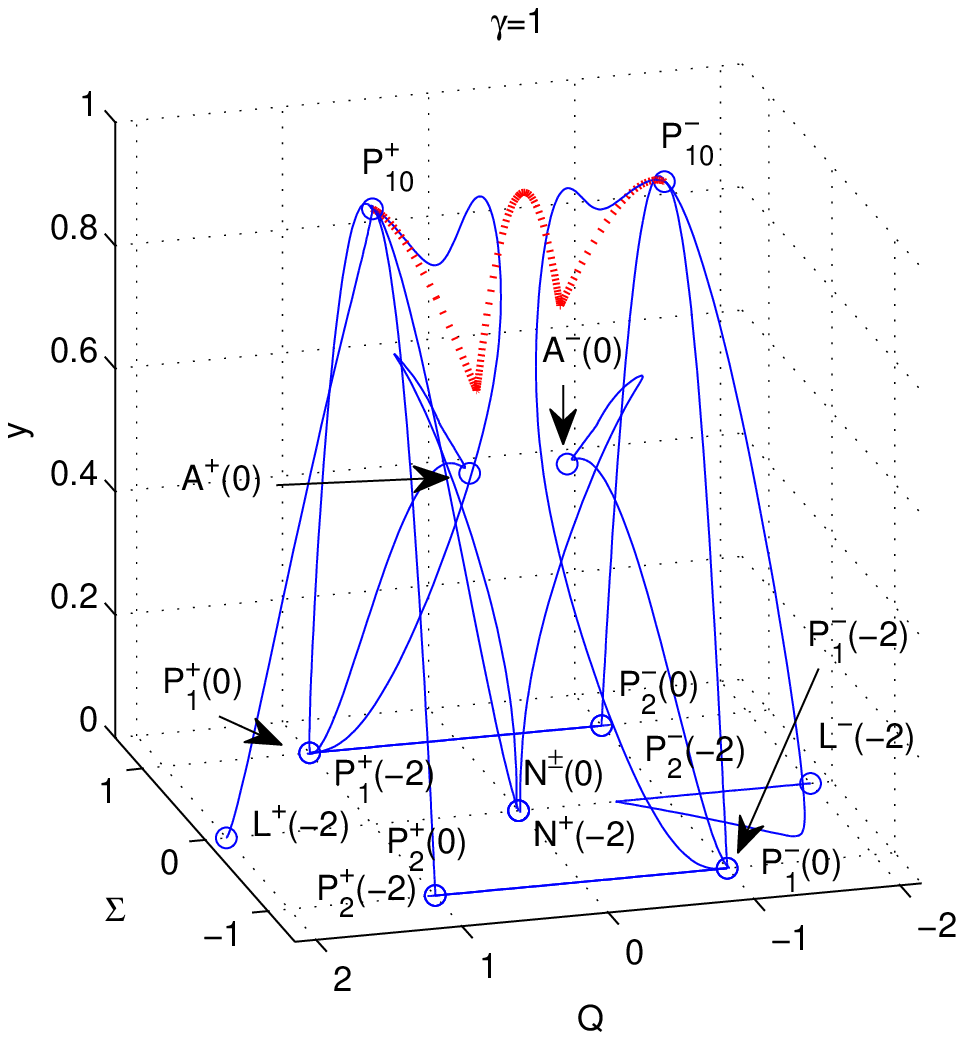}}
\caption{\label{fig1} Some heteroclinic orbits given by
\eqref{heteroclinicsA} for the model $f(R)=R+\alpha R^2$ and dust matter
($\gamma=1$). It is illustrated the transition from an expanding phase to a
contracting one, and viceversa. The dotted (red) line corresponds to the
orbit joining directly the contracting de Sitter solution $P_{10}^-$ with
the expanding one  $P_{10}^+$.}
\end{figure}

In order to present the aforementioned results in a transparent way, we
proceed to several numerical simulations showed in figure \ref{fig1}.   
There are presented the heteroclinic orbits given by \eqref{heteroclinicsA}
for the model $f(R)=R+\alpha R^2$ and dust matter ($\gamma=1$). It is
illustrated the transition from an expanding phase to a contracting one (see
the sequences at first and second lines of \eqref{4.9f} and at the first line
of \eqref{4.9h}), and viceversa (see sequence at first line of \eqref{4.9b}).
The dotted (red) line corresponds to the orbit joining directly the
contracting de Sitter solution $P_{10}^-$ with the expanding one  $P_{10}^+$.

%%%%%%%%%%%%%%%%%%%%%%%%%%%%%%%%%%%%%
%%%%%%%%%%%%%%%%%%%%%%%%%%%%%%%%%%%%%%
%%%%%%%%%%%%%%%%%%%%%%%%%%%%%%%%%%%%%
\subsection{Model $f(R)=R^p\exp\left( q R\right)$.} 

In this case $M(r)=\frac{r(p+r)}{p-r^{2}},$ $r^*\in\left\{0,-p\right\}$, 
$M'(0)=1$ and $M'(-p)=\frac{1}{-1+p}$.

\begin{itemize}
\item The sufficient conditions for the existence of past-attractors
(future-attractors) are:
   \begin{itemize}
     \item $L^{+}(-p)$ ($L^-(-p)$) is a local past (future)-attractor for
$1\leq \gamma<\frac{5}{3}, p<1.$
     \item $N^{+}(0)$ ($N^-(0)$) is always a local past (future)-attractor 
  \item $N^{+}(-p)$ ($N^{-}(-p)$) is a local past
(future)-attractor for $p>1$
      \item $A^{-}(-p)$ ($A^{+}(-p)$) is a past (future)-attractor for
$\frac{1+\sqrt{3}}{2}<p<2$ or $p<\frac{(1-\sqrt{3})}{2}$.
      \item The point $P^{-}_{10}$ ($P_{10}^{+}$) is a past
(future)-attractor if $2<p<4,$ specially if $2.72<p<4$, the critical point
$P_{10}^{+}$ is a stable focus.
   \end{itemize}
\item Some saddle points with physical interest are:
   \begin{itemize}
     \item $A^{+}(-p)$ is  a saddle point with a 4D stable manifold for $p>2$
or $\frac{5}{3}<\gamma\leq2,\; \frac{5}{4}<p<\frac{1+\sqrt{3}}{2}$  or  
$1\leq\gamma\leq\frac{5}{3},\;\frac{1}{4}\left(\frac{4+9\gamma}{1+3\gamma}
+\sqrt{\frac{16+48\gamma+9\gamma^{2}}{(1+3\gamma)^{2}}}\right)<p<\frac{
1+\sqrt{3}}{2}$. 
     \item $B^{+}(-p)$ is a saddle point with a 4D stable manifold for
$p<1,\;1\leq\gamma<\frac{4p}{3}$.
     \item $P_{4}^{+}(-p)$ is a saddle point with a 4D stable manifold for
$\frac{1+\sqrt{3}}{2}<p<2$ or $p<\frac{1-\sqrt{3}}{2}$. 
   \end{itemize}
\end{itemize}

\begin{figure}[h]
\centering
\subfigure[]{
\includegraphics[height=5.25cm,width=6cm]{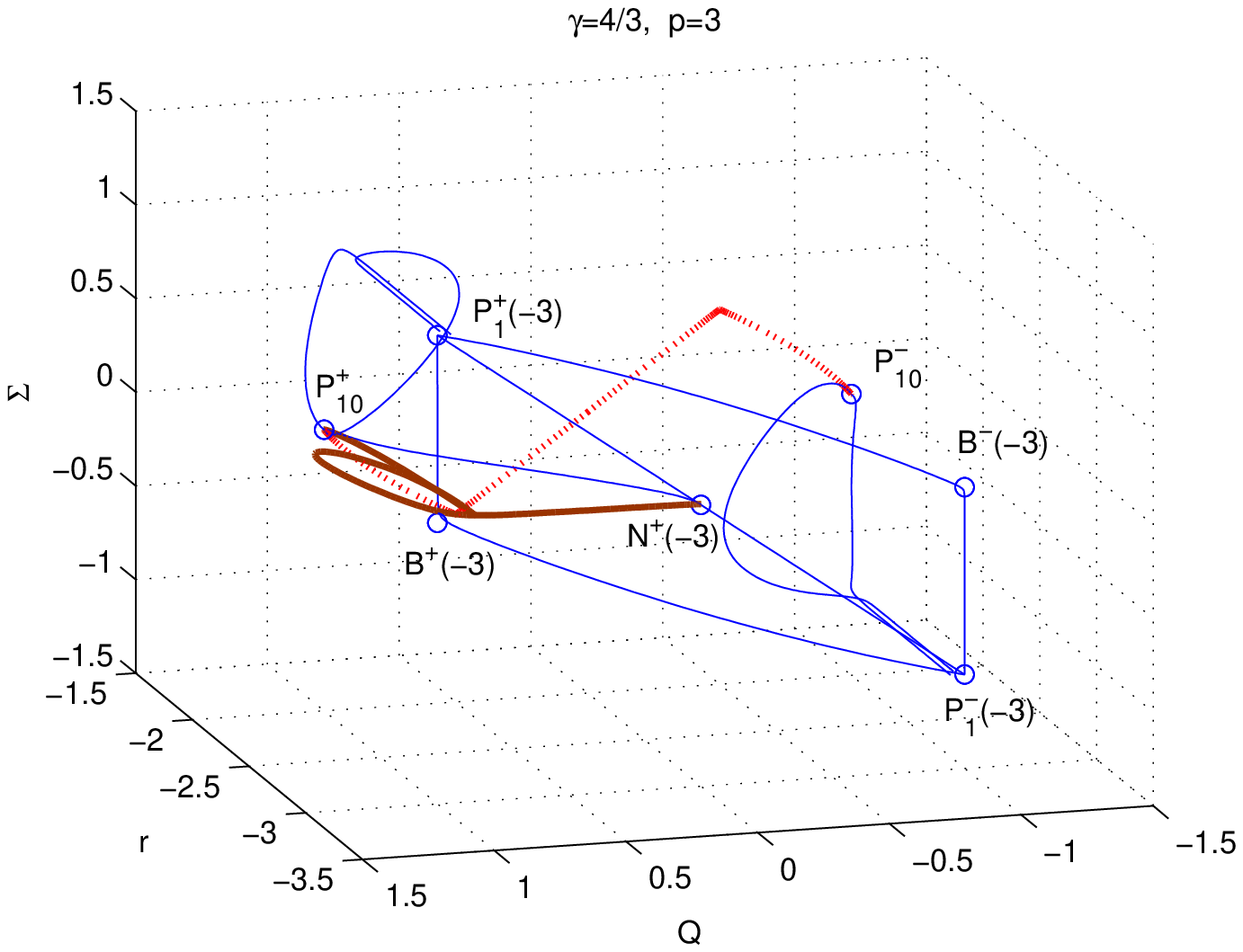}}
\subfigure[]{
\includegraphics[height=5.25cm,width=6cm]{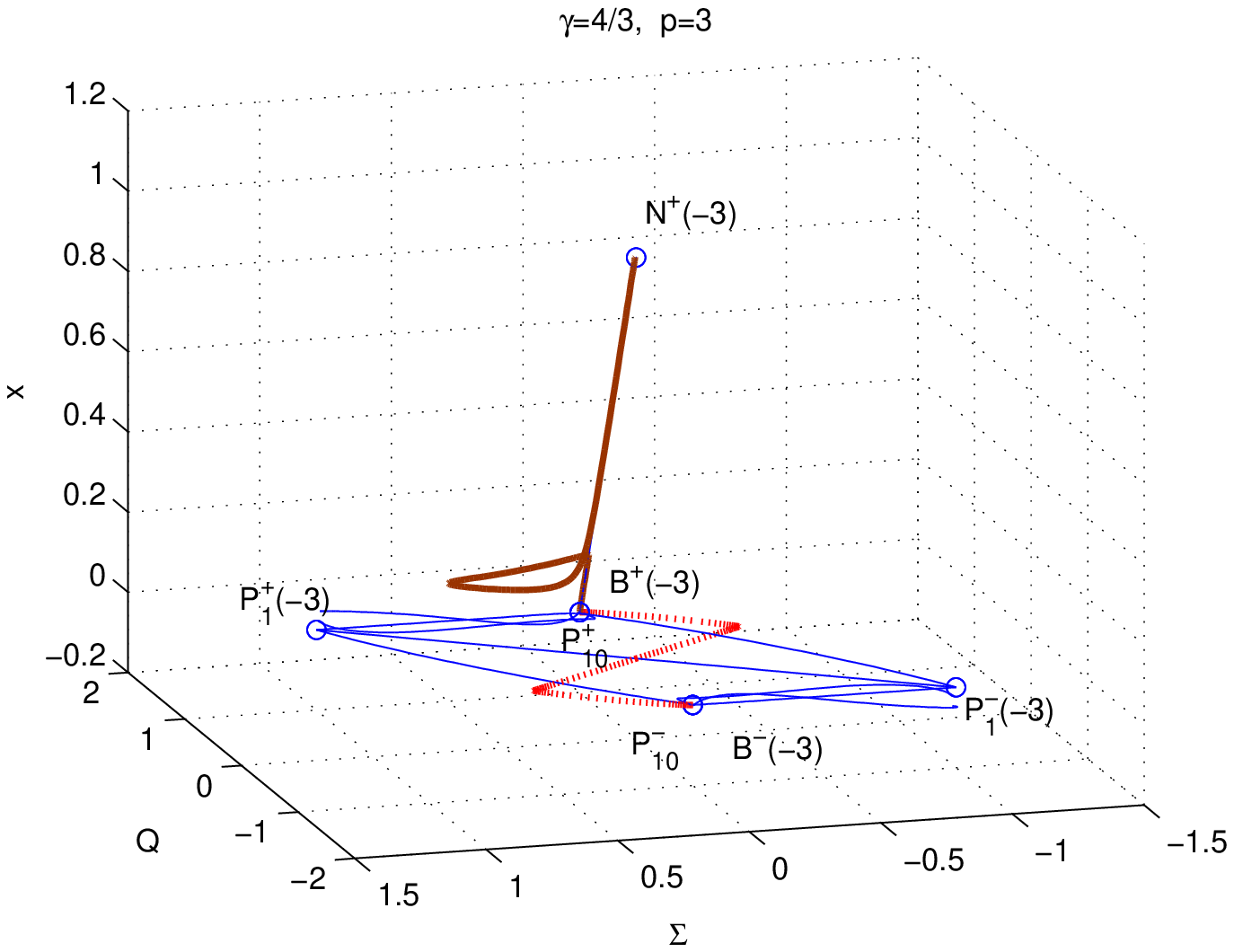}}
\subfigure[]{
\includegraphics[height=5.25cm,width=6cm]{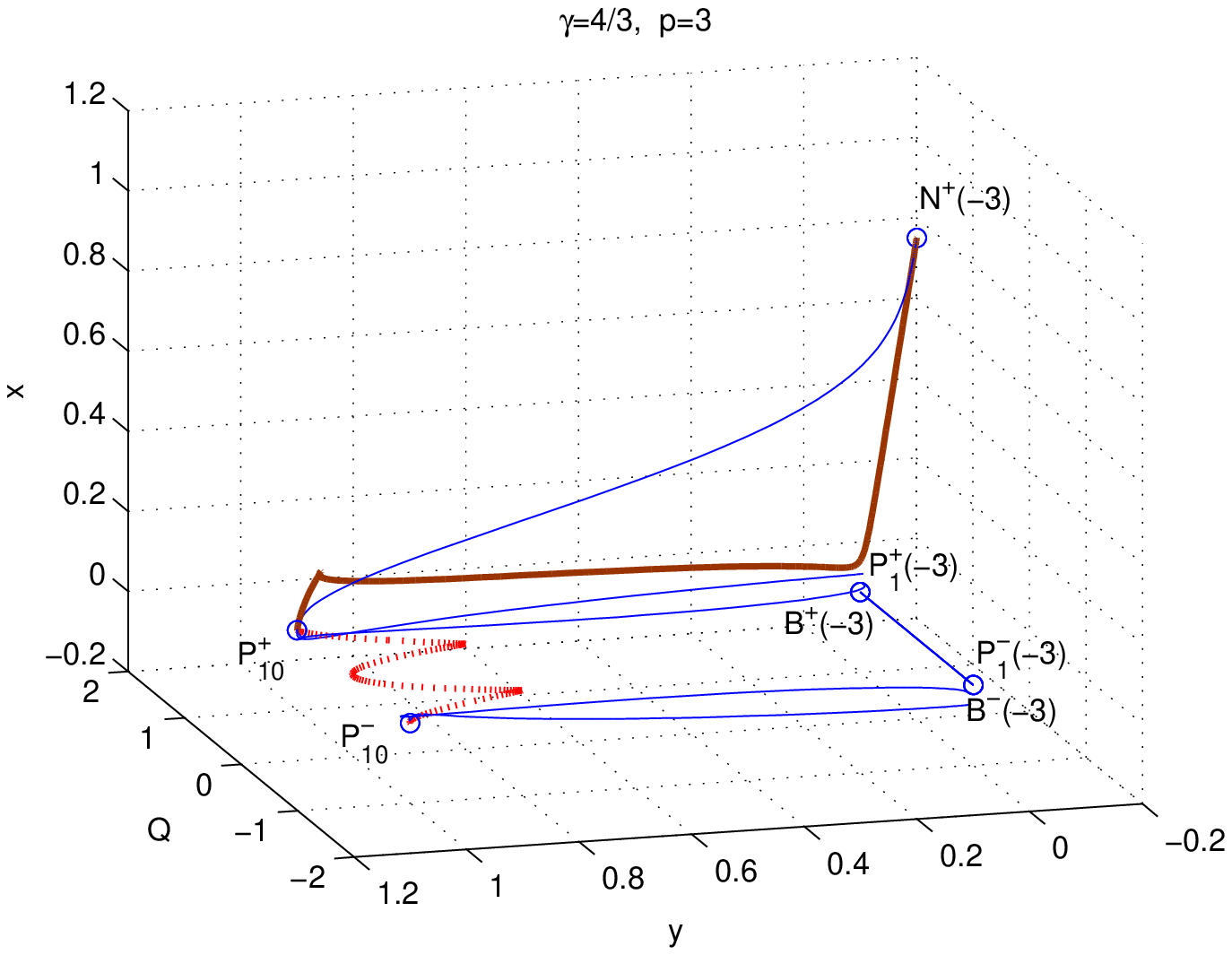}}
\subfigure[]{
\includegraphics[height=5.25cm,width=6cm]{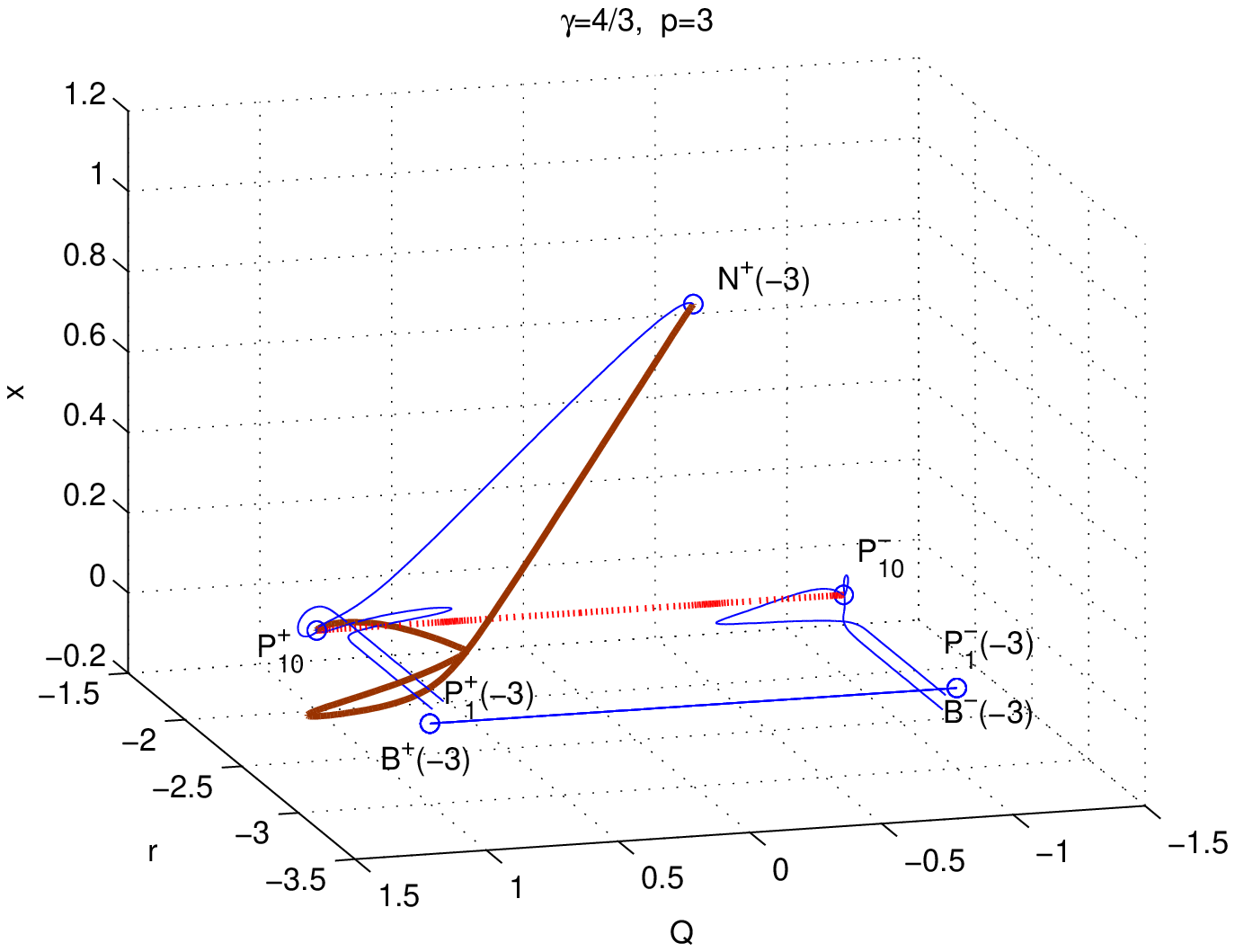}}
\subfigure[]{
\includegraphics[height=5.25cm,width=6cm]{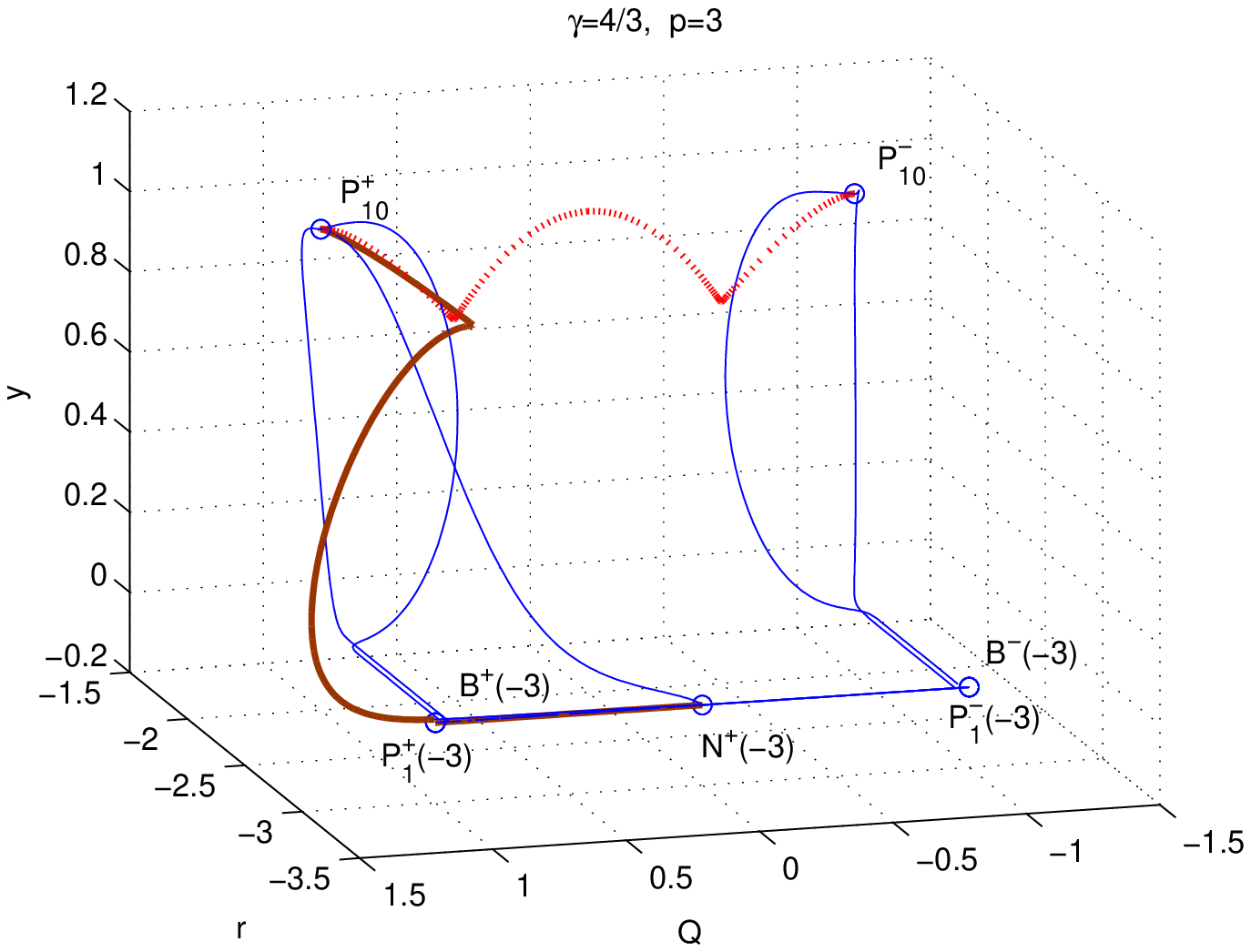}}
\subfigure[]{
\includegraphics[height=5.25cm,width=6cm]{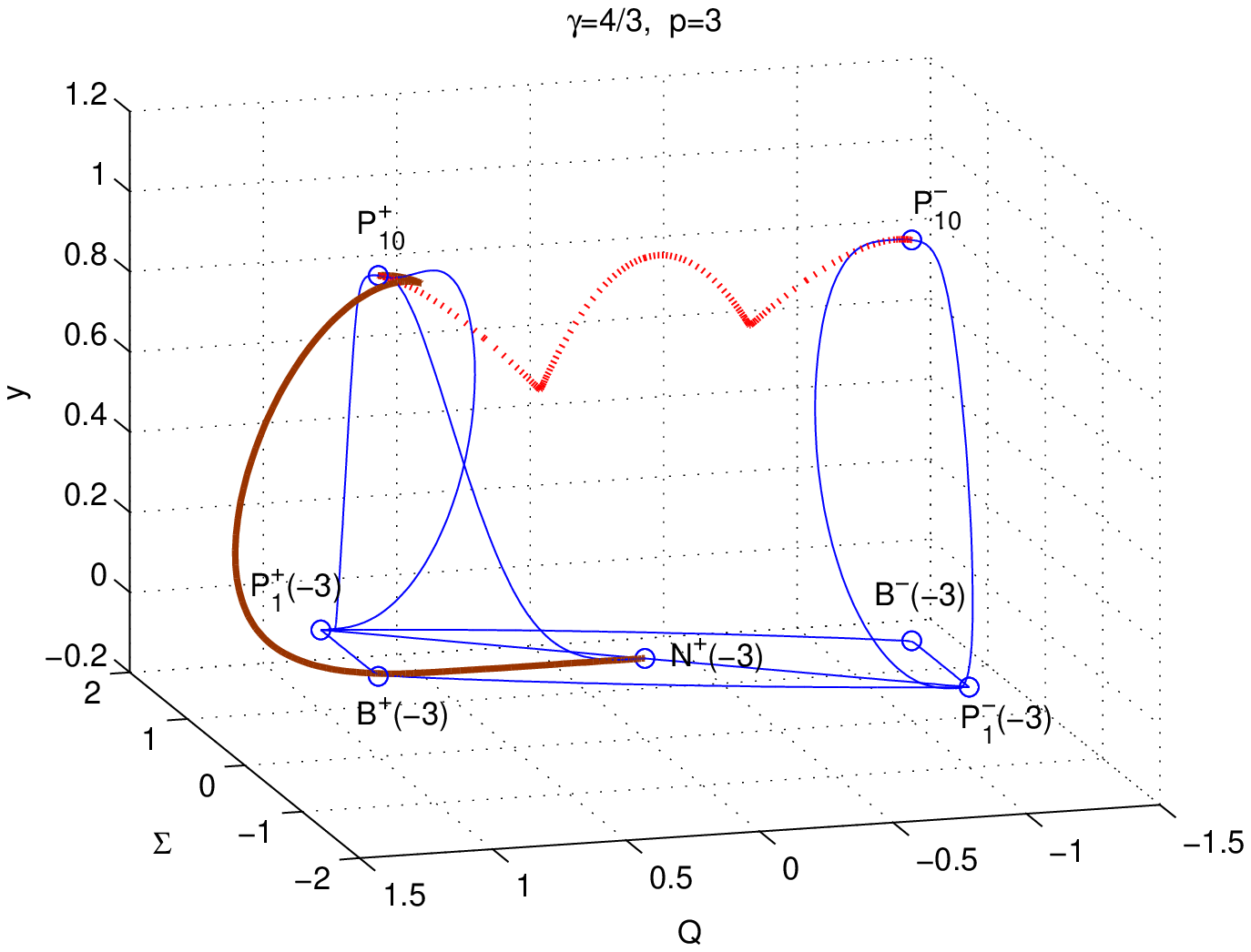}}
\caption{\label{fig2} Some heteroclinic orbits given by
\eqref{heteroclinicsC.1} for the model $f(R)=R^p\exp\left( q R\right)$ with
radiation ($\gamma=4/3$) for $p=3$ and $q$ arbitrary. It is illustrated the
transition from an expanding phase to a contracting one, and viceversa. The
dotted (red) line corresponds to the orbit joining directly the contracting
de Sitter solution $P_{10}^-$ with the expanding one  $P_{10}^+$ (first line
of \eqref{4.10c}). The thick (brown) line denotes the orbit that connects
the matter-dominated (radiation like, $\gamma=\frac{4}{3}$) solution
$B^+(-3)$ (region VII) to the accelerated de Sitter phase $P_{10}^+$ (region
II). This orbit, corresponding to the first line in \eqref{4.10b} is past
asymptotic to the static solution $N^+(-3)$.}
\end{figure}

\begin{figure}[h]
\centering
\subfigure[]{
\includegraphics[height=5.25cm,width=6cm]{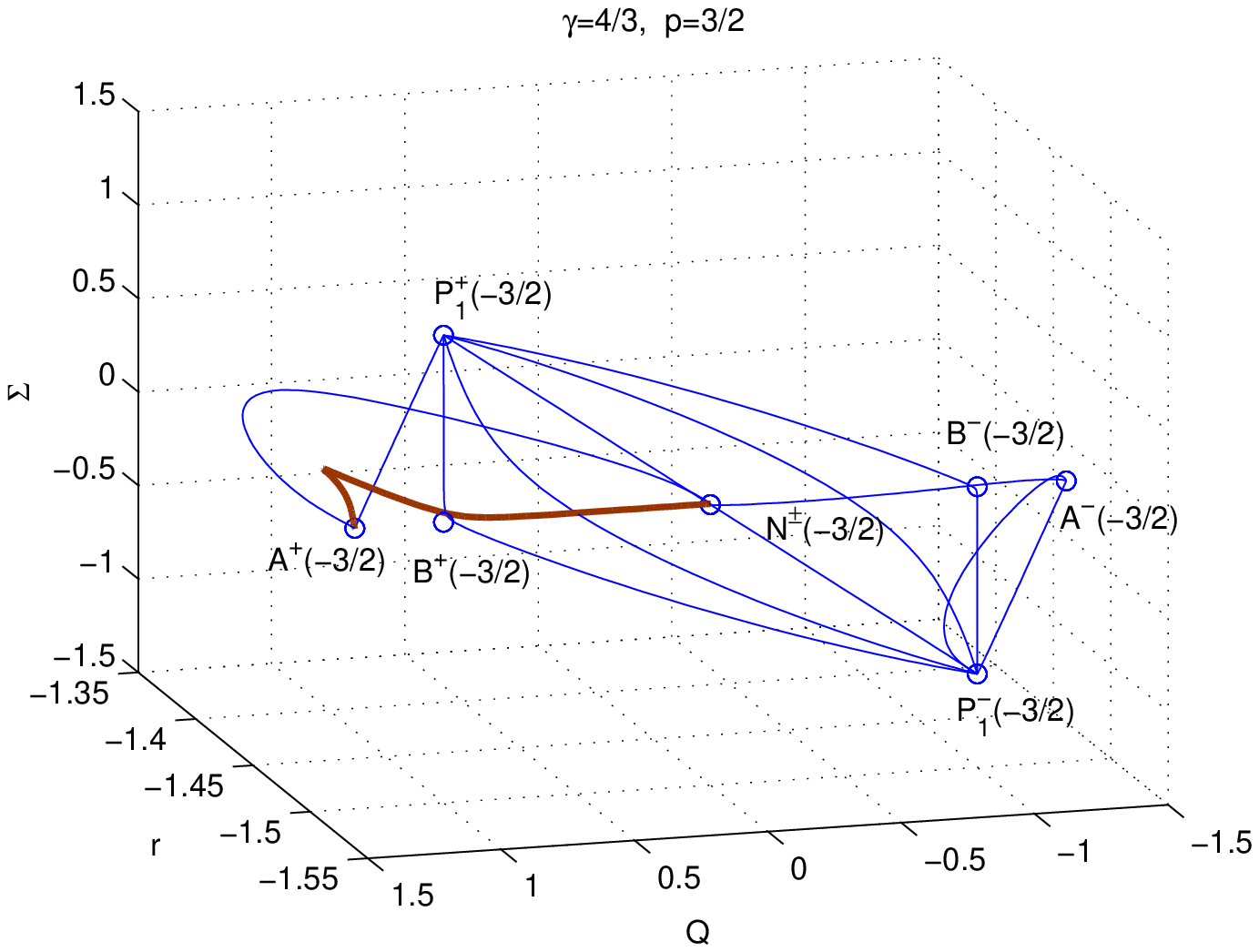}}
\subfigure[]{
\includegraphics[height=5.25cm,width=6cm]{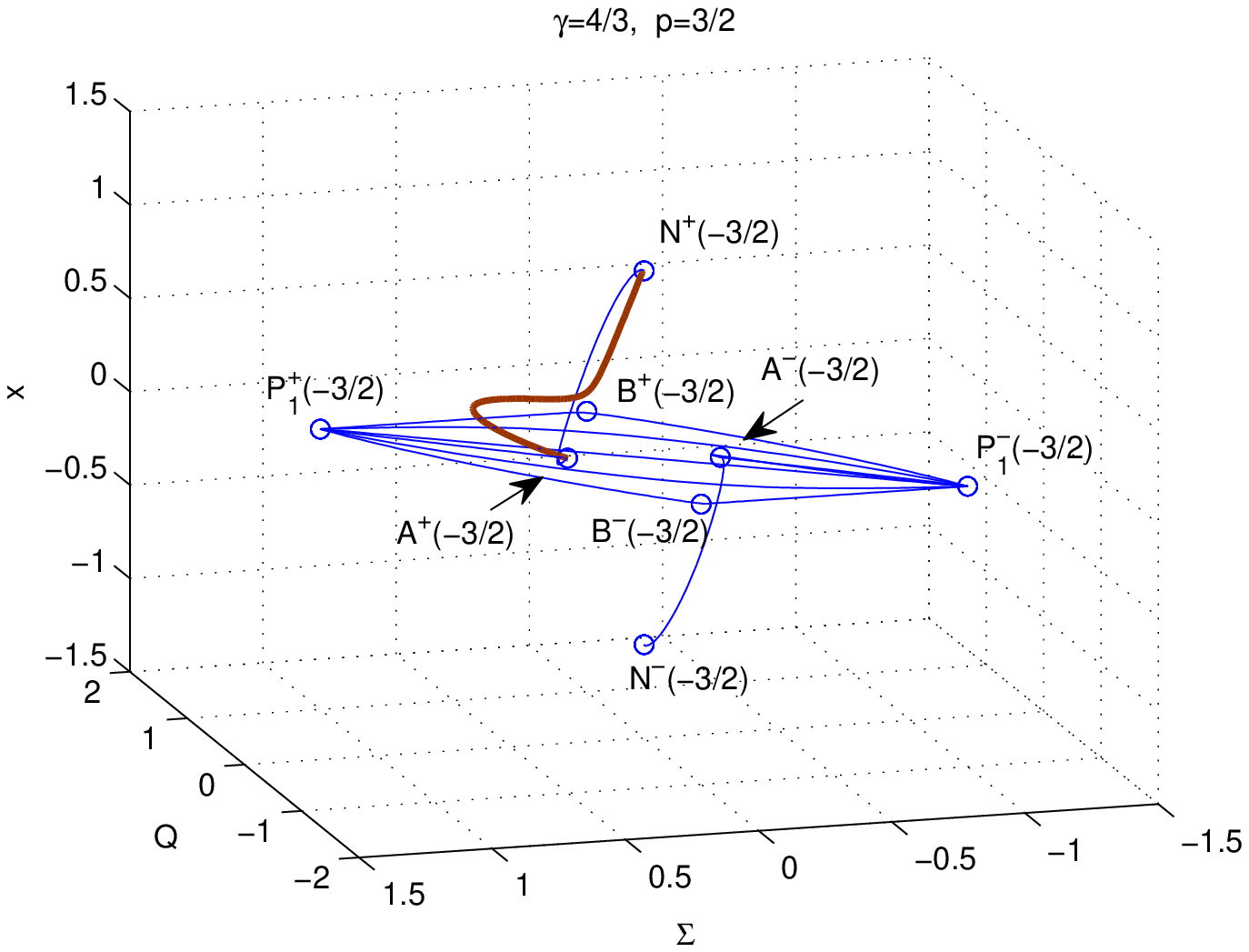}}
\subfigure[]{
\includegraphics[height=5.25cm,width=6cm]{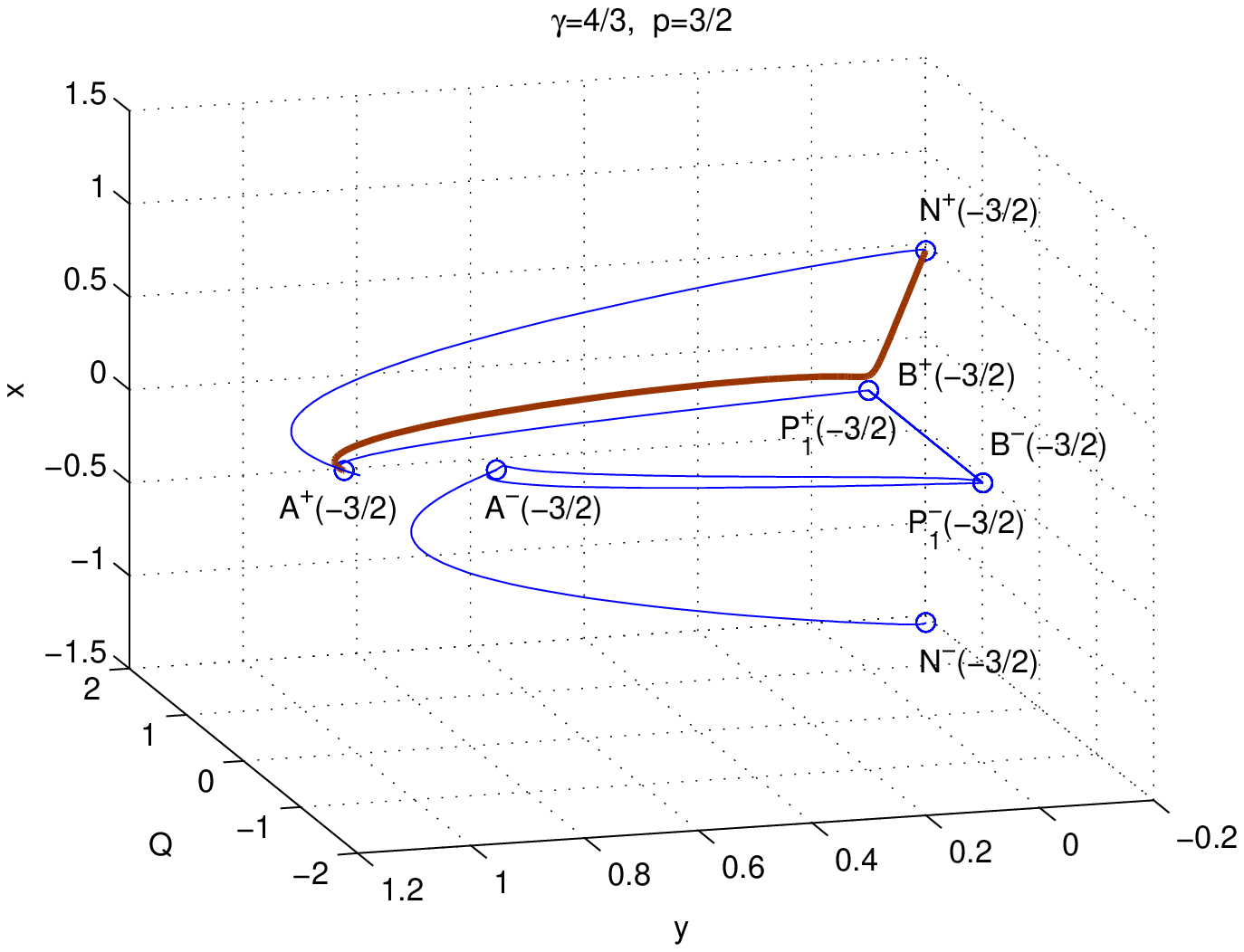}}
\subfigure[]{
\includegraphics[height=5.25cm,width=6cm]{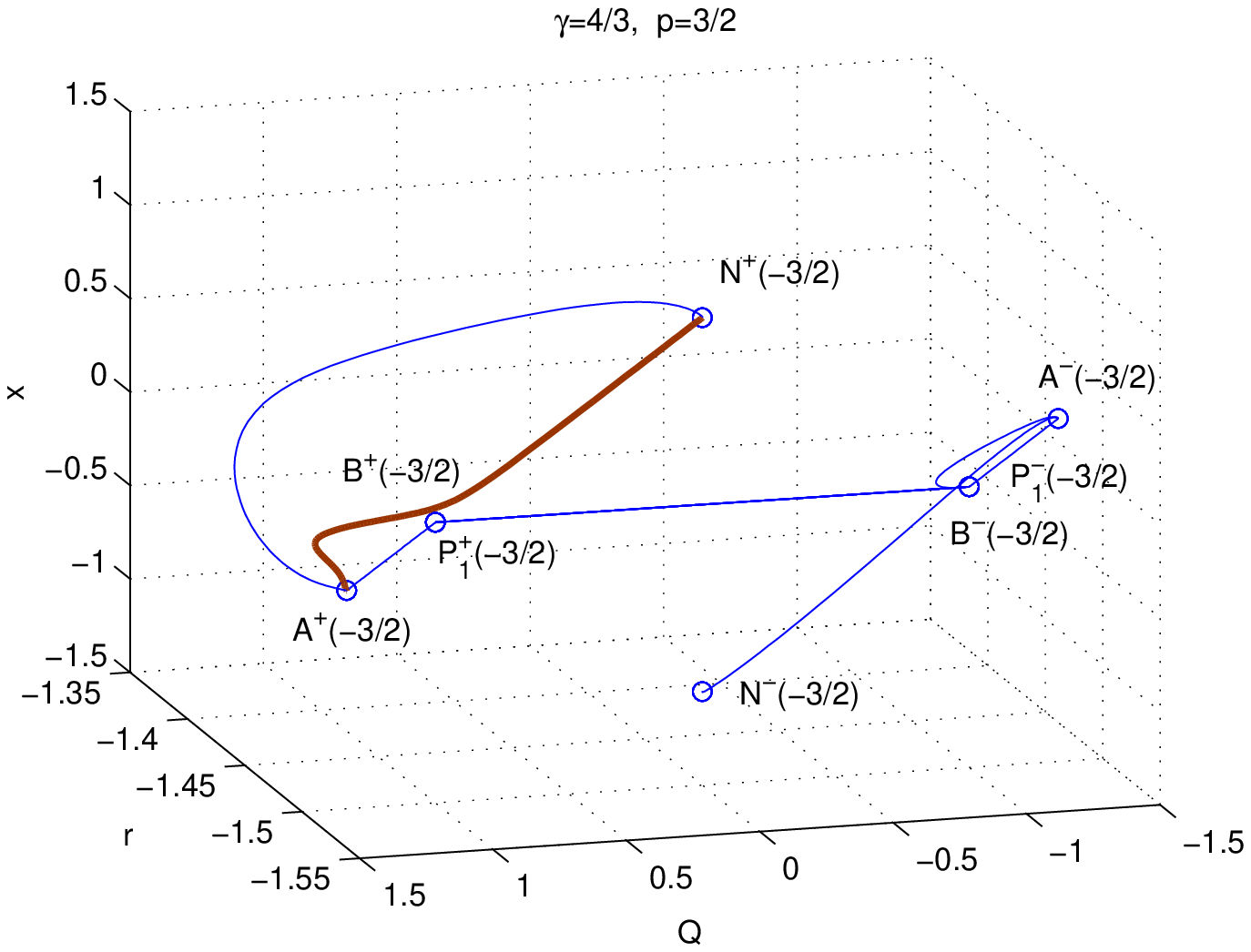}}
\subfigure[]{
\includegraphics[height=5.25cm,width=6cm]{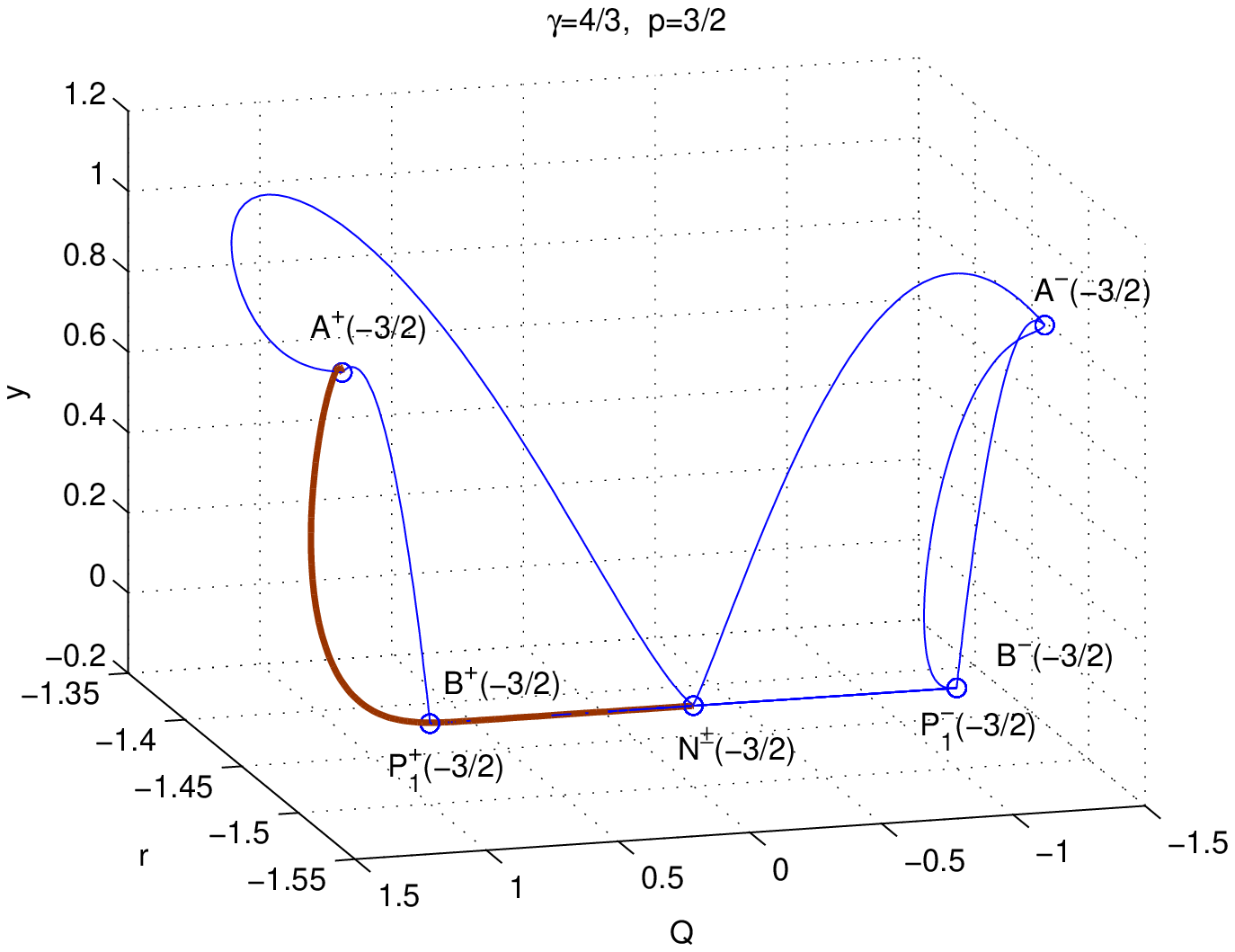}}
\subfigure[]{
\includegraphics[height=5.25cm,width=6cm]{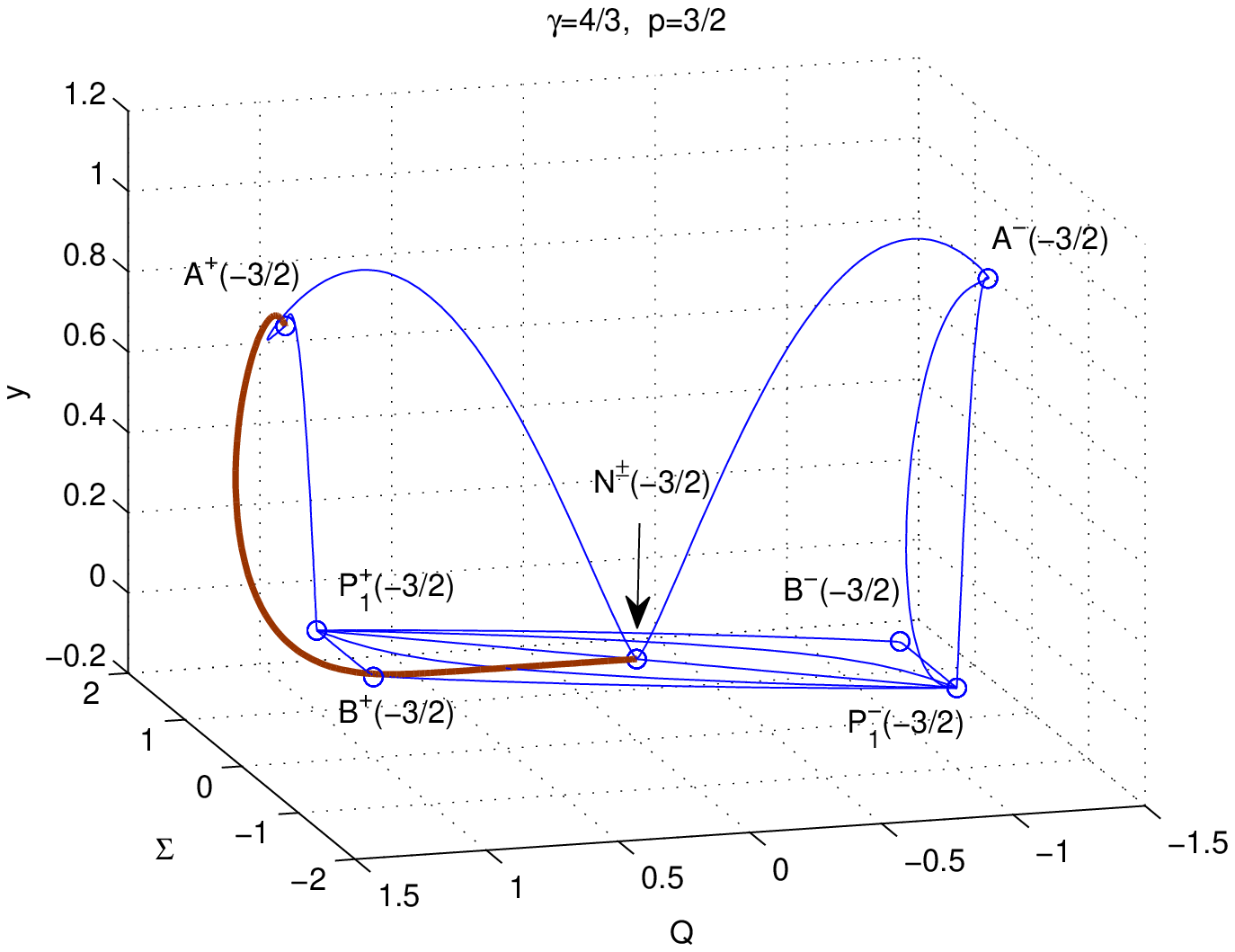}}
\caption{\label{fig3} Some heteroclinic orbits given by
\eqref{heteroclinicsC.2} for the model $f(R)=R^p\exp\left( q R\right)$ with
radiation ($\gamma=4/3$) for $p=\frac{3}{2}$ and $q$ arbitrary. It is
illustrated the transition from an expanding phase to a contracting one. 
The thick (brown) line denotes the orbit that connects the matter-dominated
(radiation like, $\gamma=\frac{4}{3}$) solution $B^+(-3/2)$ (region VII) to
the accelerated de Sitter phase $A^+(-3/2)$ (region III). This orbit,
corresponding to the first line in \eqref{4.11b}, is past asymptotic to the
static solution $N^+(-3/2)$.}
\end{figure}

\begin{figure}[h]
\centering
\subfigure[]{
\includegraphics[height=5.25cm,width=6cm]{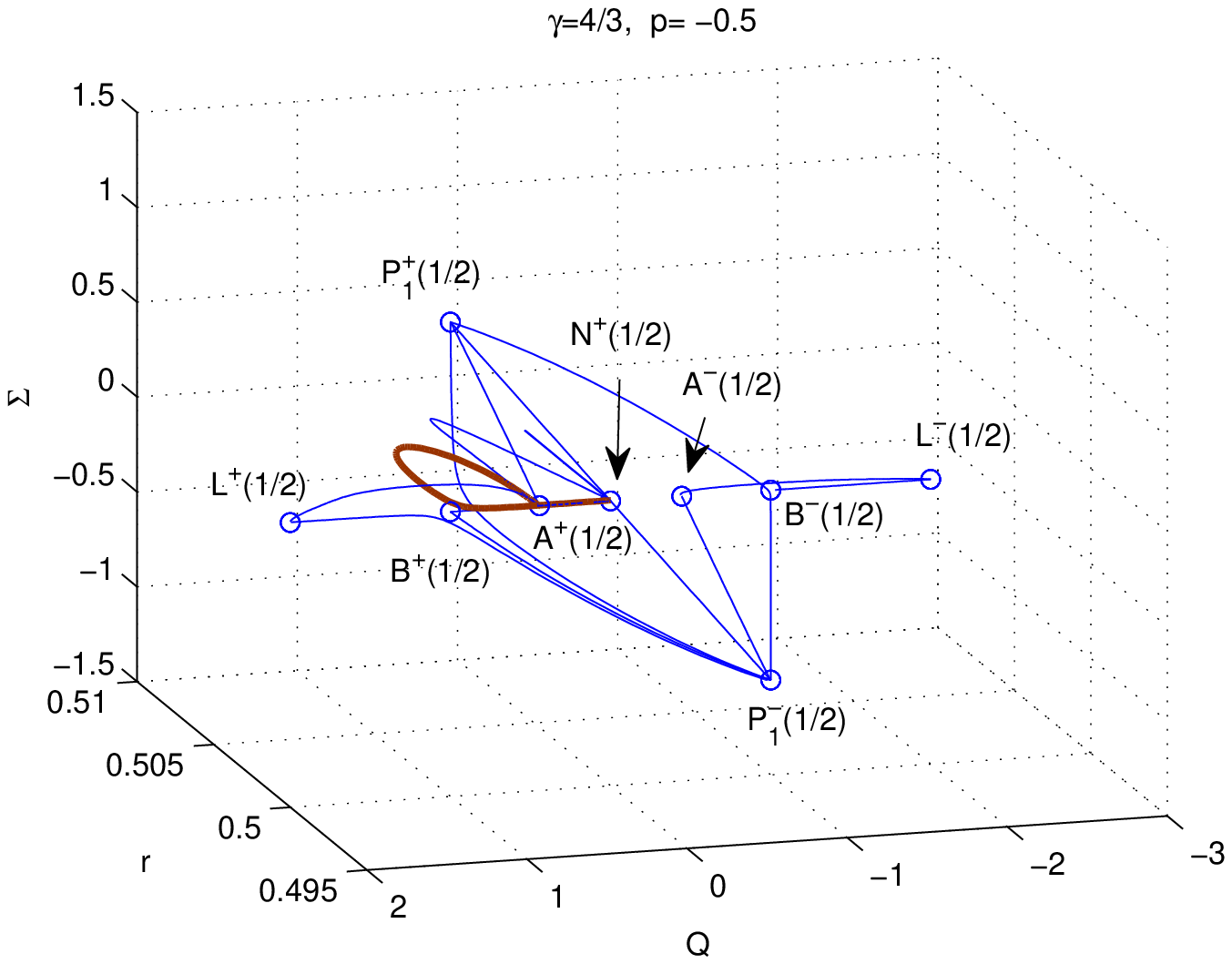}}
\subfigure[]{
\includegraphics[height=5.25cm,width=6cm]{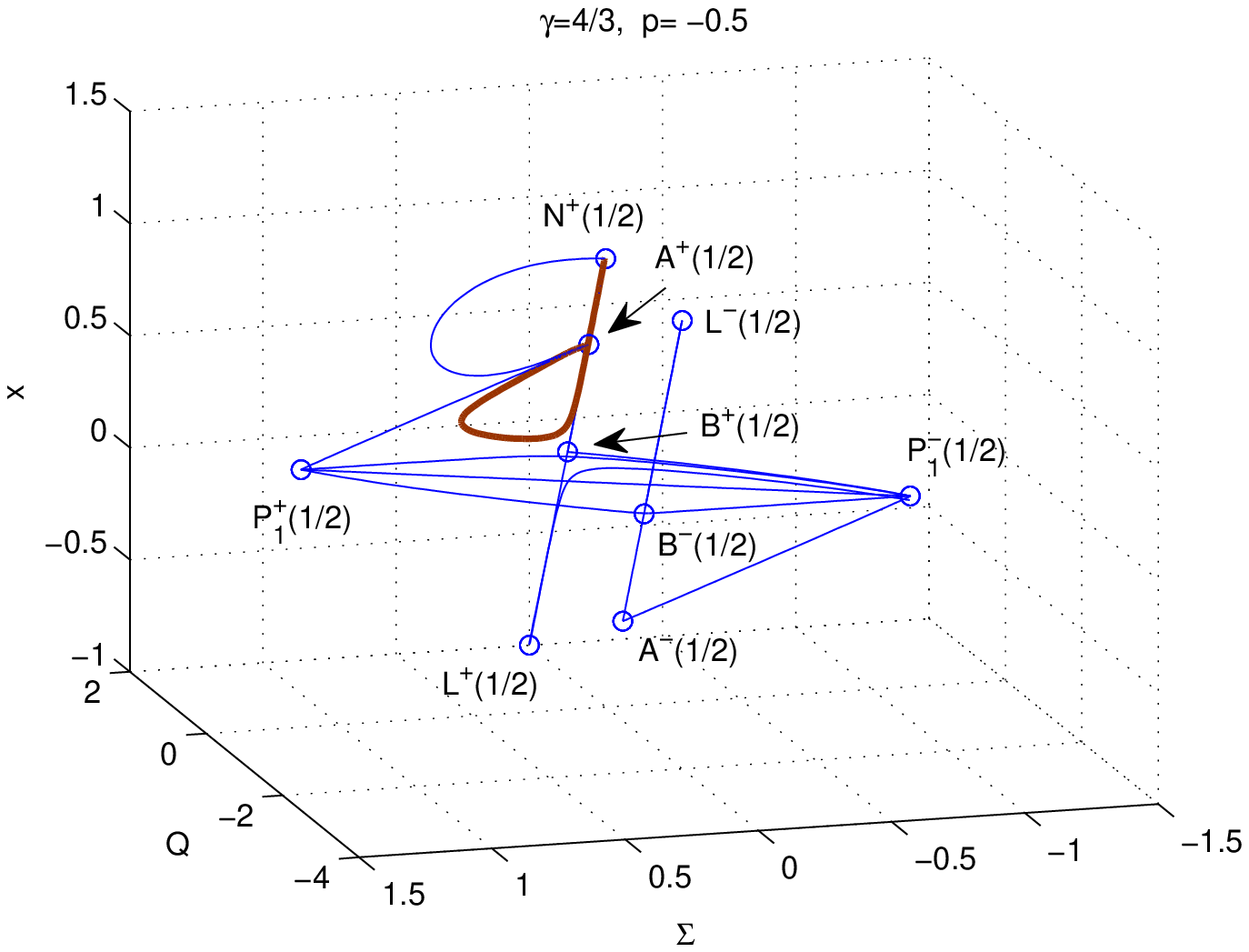}}
\subfigure[]{
\includegraphics[height=5.25cm,width=6cm]{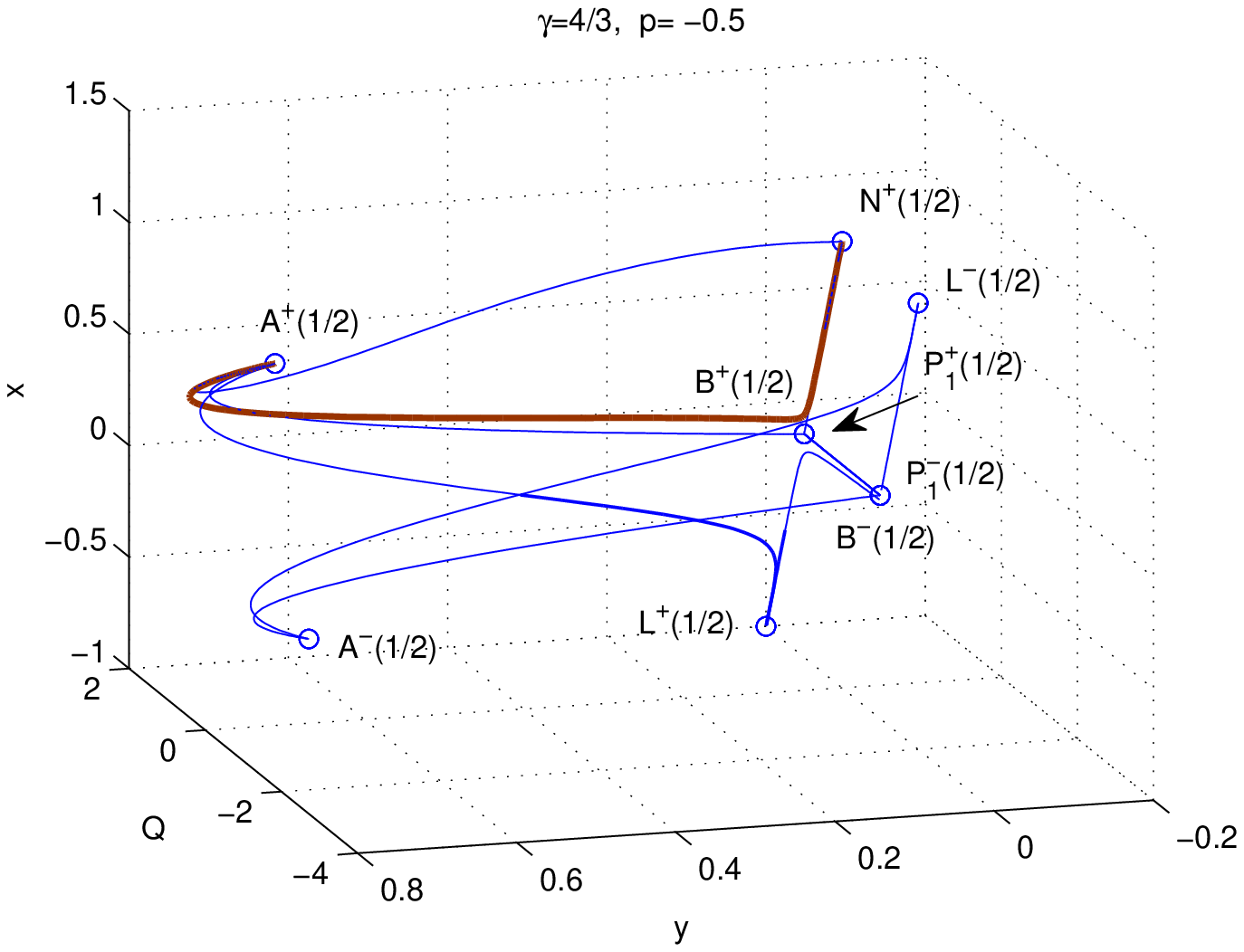}}
\subfigure[]{
\includegraphics[height=5.25cm,width=6cm]{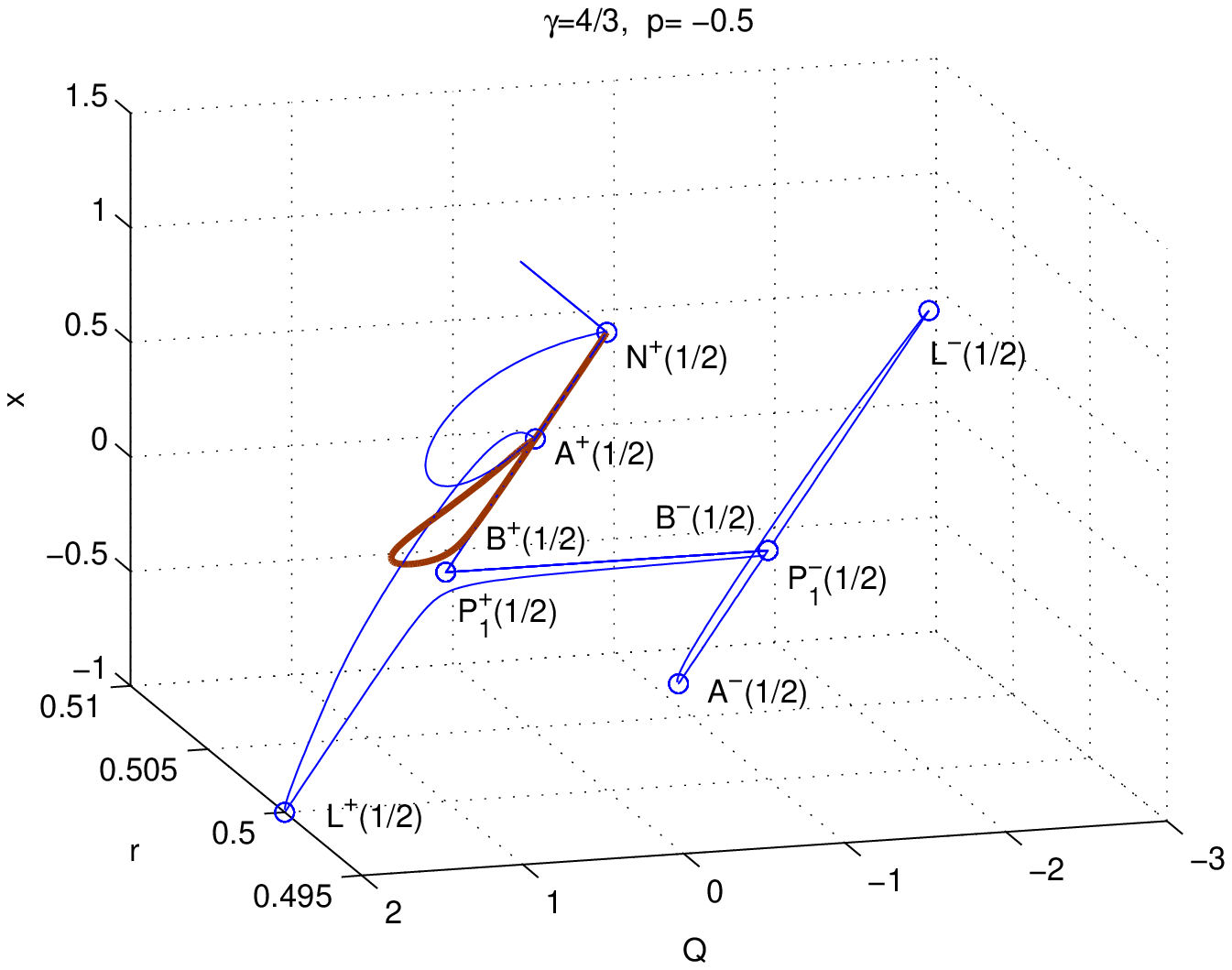}}
\subfigure[]{
\includegraphics[height=5.25cm,width=6cm]{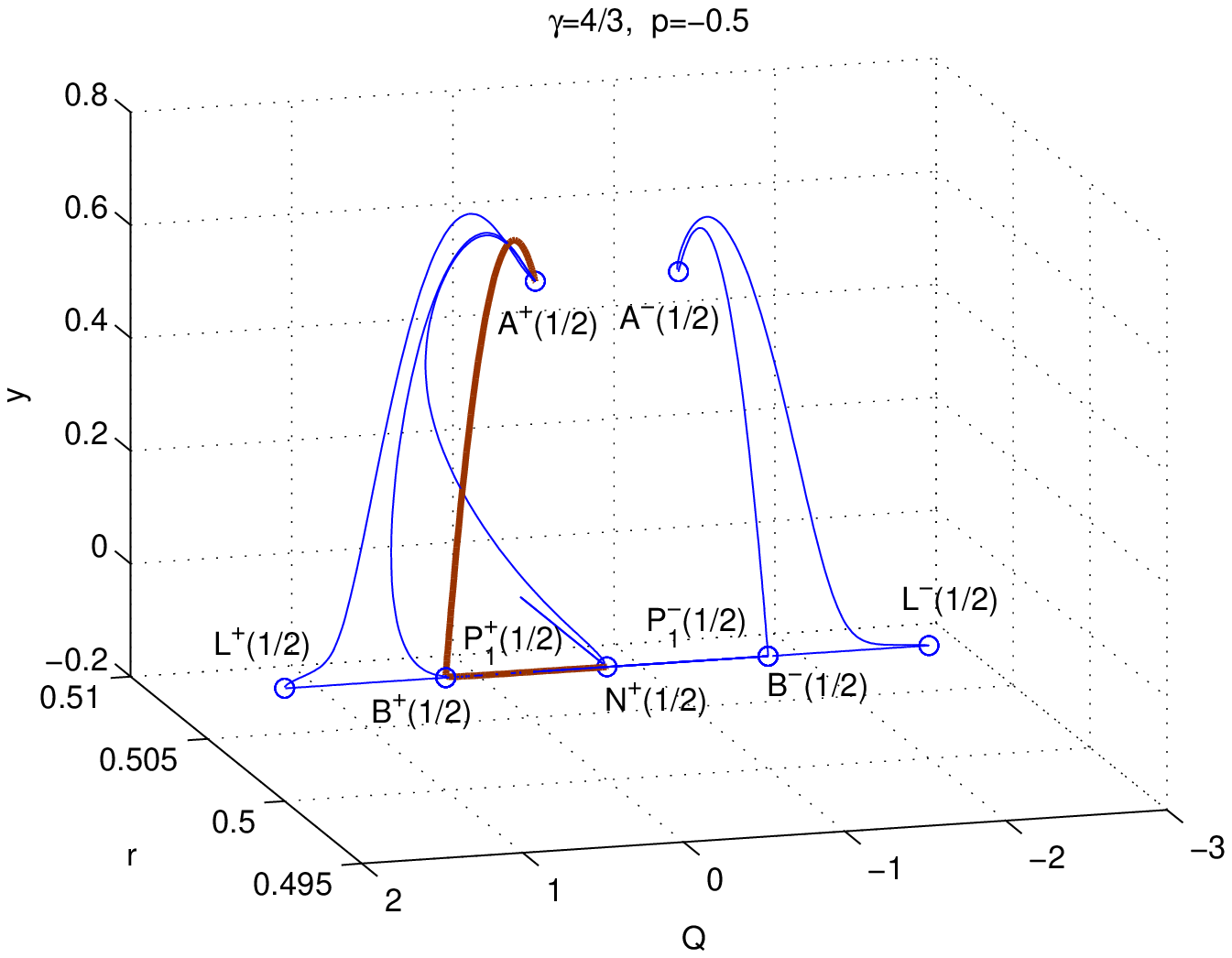}}
\subfigure[]{
\includegraphics[height=5.25cm,width=6cm]{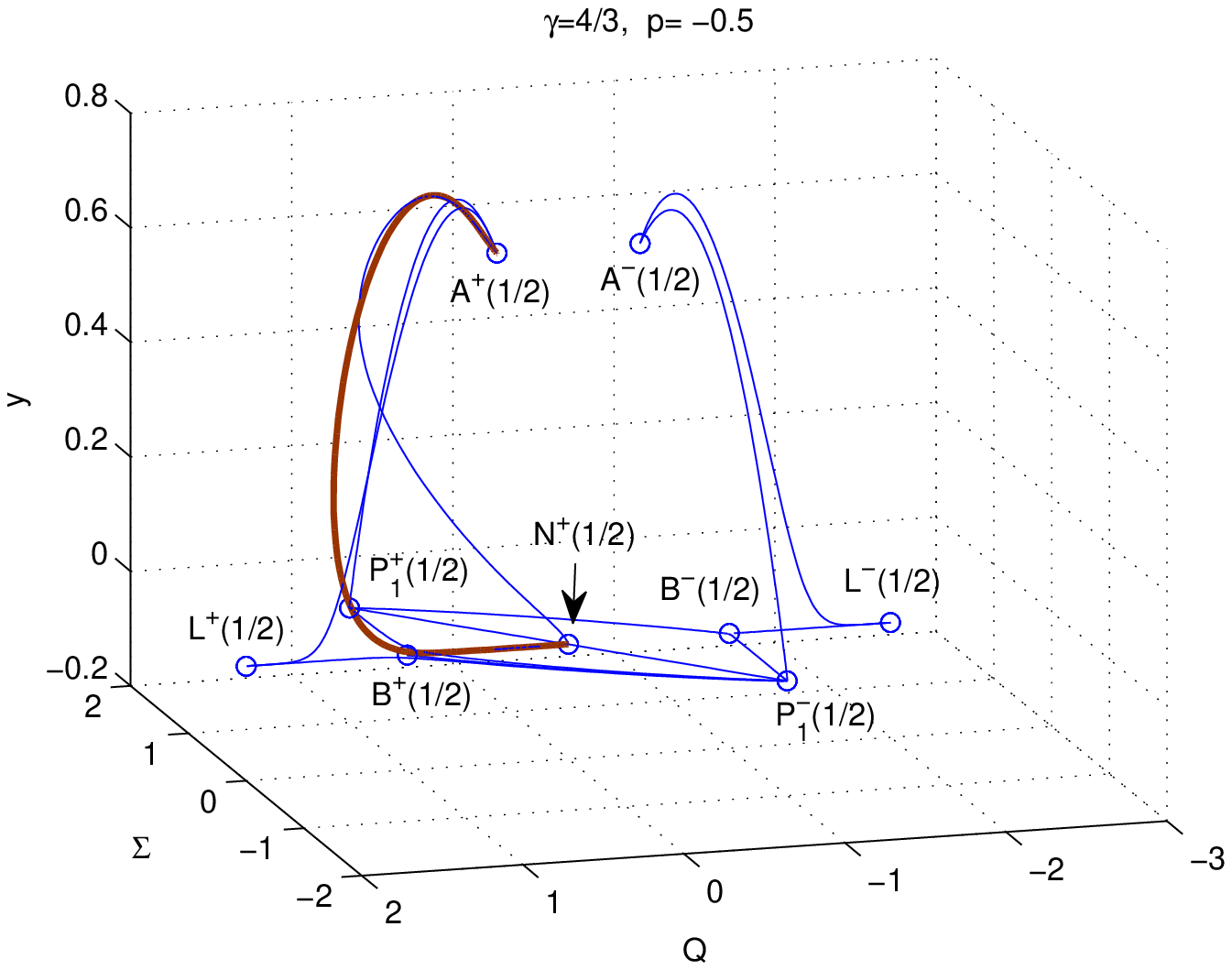}}\caption{\label{fig4} Some heteroclinic
orbits given by \eqref{heteroclinicsC.3} for the model $f(R)=R^p\exp\left( q
R\right)$ with radiation ($\gamma=4/3$) for $p=-\frac{1}{2}$ and $q$
arbitrary. It is illustrated the transition from an expanding phase to a
contracting one.  The thick (brown) line denotes the orbit that connects the
matter-dominated (radiation like, $\gamma=\frac{4}{3}$) solution $B^+(1/2)$
(region VII) to the accelerated phase $A^+(1/2)$ (region IV). This orbit is
past asymptotic to the static solution $N^+(1/2)$ (see the second line in
\eqref{4.12b}).}
\end{figure}

The function $M(r)$ satisfy the following: 
\begin{itemize}
\item it connects the matter-dominated region VII with the region II where
the accelerated solution is a  de Sitter solution provided
$\gamma=\frac{4}{3},\; 2<p<4$ (see figure \ref{fig2}).

 \item it connects the matter-dominated  region VII with the region
III corresponding to an accelerated expansion provided $\gamma=\frac{4}{3},\;
\frac{1+\sqrt{3}}{2}<p<2$ (see figure \ref{fig3}).

 \item it connects the matter-dominated  region VII with the region IV
corresponding to an accelerated expansion provided $\gamma=\frac{4}{3},\;
p<\frac{1-\sqrt{3}}{2}$ (see figure \ref{fig4}). 
 
 \end{itemize}

This model was recently considered in \cite{Ivanov:2011np}.

In order to present the aforementioned results in a transparent way, we
proceed to several numerical
simulations showed in figure \ref{fig2} for the model $f(R)=R^p\exp\left( q
R\right)$ with radiation ($\gamma=4/3$) for $p=3$ and $q$ arbitrary.

In this
figure are presented the heteroclinic orbits: 
\begin{subequations}\label{heteroclinicsC.1}
\begin{align}
& P_{1}^{+}(-3)\longrightarrow \left\{\begin{array}{l}   
   B^{-}(-3)\longrightarrow P_{1}^{-}(-3) \\ 
    B^{+}(-3)\longrightarrow P_{1}^{-}(-3) \\  
                          P_{1}^{-}(-3) \\ 
                          P_{10}^{+} 
                         \end{array}, \right. \label{4.10a} \\
& N^{+}(-3)\longrightarrow \left\{\begin{array}{l}   
    B^{+}(-3)\longrightarrow P_{10}^{+}\\
                         P_{10}^{+}
                         \end{array},    \right. \label{4.10b}\\
& P_{10}^{-}\longrightarrow \left\{\begin{array}{l}
                               P_{10}^{+} \\
                              P_{1}^{-}(-3) 
                             \end{array}.    \right. \label{4.10c}    
\end{align}
\end{subequations}

In the figure \ref{fig2} it is illustrated the transition from an expanding
phase to a contracting one (see the sequences at first, second  and third
lines of \eqref{4.10a}), and viceversa (see sequence at first line of
\eqref{4.10c}). The dotted (red) line corresponds to the orbit joining
directly the contracting de Sitter solution $P_{10}^-$ with the expanding one
 $P_{10}^-$ (first line of \eqref{4.10c}). The thick (brown) line denotes the
orbit that connects the matter-dominated (radiation like,
$\gamma=\frac{4}{3}$) solution $B^+(-3)$ (region VII) to the accelerated de
Sitter phase $P_{10}^+$ (region II). This orbit, corresponding to the first
line in \eqref{4.10b}, is past asymptotic to the static solution $N^+(-3)$.

In the figure \ref{fig3} are presented several numerical simulations for the
model $f(R)=R^p\exp\left( q R\right)$ with radiation ($\gamma=4/3$) for
$p=\frac{3}{2}$ and $q$ arbitrary. 

Some heteroclinic orbits for this case
are:
\begin{subequations}
\label{heteroclinicsC.2}
\begin{align}
& P_{1}^{+}(-3/2)\longrightarrow \left\{\begin{array}{l}  
    B^{-}(-3/2)\longrightarrow P_{1}^{-}(-3/2)
\\    
   B^{+}(-3/2)\longrightarrow P_{1}^{-}(-3/2) \\  
      P_{1}^{-}(-3/2)\\ 
      A^{+}(-3/2)
                         \end{array}, \right. \label{4.11a} \\
& N^{+}(-3/2)\longrightarrow \left\{\begin{array}{l}   
     B^{+}(-3/2)\longrightarrow
A^{+}(-3/2)\\      
   A^{+}(-3/2)
                         \end{array},    \right. \label{4.11b}\\
& A^{-}(-3/2)\longrightarrow \left\{\begin{array}{l}   
       N^{-}(-3/2)\\
                              P_{1}^{-}(-3/2) 
                             \end{array}.    \right. \label{4.11c}
\end{align}
\end{subequations}

In the figure \ref{fig3} it is illustrated the transition from an expanding
phase to a contracting one (see the sequences at first, second  and third
lines of \eqref{4.11a}).  The thick (brown) line denotes the orbit that
connects the matter-dominated (radiation like, $\gamma=\frac{4}{3}$) solution
$B^+(-3/2)$ (region VII) to the accelerated de Sitter phase $A^+(-3/2)$
(region III). This orbit is past asymptotic to the static solution
$N^+(-3/2)$. It corresponds to the first line in \eqref{4.11b}.

In the figure \ref{fig4} are presented several numerical simulations for the
model $f(R)=R^p\exp\left( q R\right)$ with radiation ($\gamma=4/3$) for
$p=-\frac{1}{2}$ and $q$ arbitrary. 

Some heteroclinic orbits in this case
are: 
\begin{subequations}
\label{heteroclinicsC.3}
\begin{align}
& P_{1}^{+}(1/2)\longrightarrow \left\{\begin{array}{l}   
    B^{-}(1/2)\longrightarrow P_{1}^{-}(1/2) \\
B^{-}(1/2)\longrightarrow L^{-}(1/2) \\    
    B^{+}(1/2)\longrightarrow P_{1}^{-}(1/2)\\
       P_{1}^{-}(1/2) \\ 
       A^{+}(1/2) 
                         \end{array}, \right. \label{4.12a} \\
& N^{+}(1/2)\longrightarrow \left\{\begin{array}{l}   
    A^{+}(1/2) \\
                          B^{+}(1/2)\longrightarrow A^{+}(1/2)\\ 
      B^{+}(1/2)\longrightarrow
P_{1}^{-}(1/2)
                         \end{array},    \right. \label{4.12b}\\
& A^{-}(1/2)\longrightarrow \left\{\begin{array}{l}
                              P_{1}^{-}(1/2) \\    
     L^{-}(1/2) 
                             \end{array},    \right. \label{4.12c}\\
& L^{+}(1/2)\longrightarrow \left\{\begin{array}{l}
                              B^{+}(1/2)\longrightarrow P_{1}^{-}(1/2) \\
       A^{+}(1/2) 
                             \end{array}.    \right. \label{4.12d}
\end{align}
\end{subequations}

In figure \ref{fig4} it is illustrated the transition from an expanding phase
to a contracting one (see the sequences at first, second, third  and fourth
lines of \eqref{4.12a}; third line of \eqref{4.12b}; first line of
\eqref{4.12d}).  The thick (brown) line denotes the orbit that connects the
matter-dominated (radiation like, $\gamma=\frac{4}{3}$) solution $B^+(1/2)$
(region VII) to the accelerated phase $A^+(1/2)$ (region IV). This orbit,
corresponding to the second line in \eqref{4.12b}, is past asymptotic to the
static solution $N^+(1/2)$.

%%%%%%%%%%%%%%%%%%%%%%%%%%%%%%%%%%%%%
%%%%%%%%%%%%%%%%%%%%%%%%%%%%%%%%%%%%%%
%%%%%%%%%%%%%%%%%%%%%%%%%%%%%%%%%%%%%
\subsection{Model $f(R)=R^p\exp\left( \frac{q}{R}\right)$.}

In this case $M(r)=\frac{r(p+r)}{r^{2}+2r+p}$, $r^*\in\left\{0,-p\right\}$, 
$M'(0)=1$ and $M'(-p)=\frac{1}{1-p}$.
\begin{itemize}
\item The sufficient conditions for the existence of past-attractors
(future-attractors) are:
   \begin{itemize}
     \item $L^{+}(-p)$ ($L^{-}(-p)$) is a local past (future)-attractor for
$1\leq \gamma<\frac{5}{3}, p>\frac{5}{4}.$
     \item $N^{+}(0)$ ($N^{-}(0)$) is always a past (future)-attractor 
   \item $N^{+}(-p)$ ($N^{-}(-p)$) is a local past
(future)-attractor for $p<\frac{1}{2}$
      \item $A^{-}(-p)$ ($A^{+}(-p)$) is a past (future)-attractor for $p>2$.
      \item The point $P^{-}_{10}$ ($P^{+}_{10}$) is a past
(future)-attractor for $0<p<2.$ Particularly, for $0<p<1.28$, the critical
point $P^{+}_{10}$ is a stable focus.
   \end{itemize}
\item Some saddle points with physical interest are:
   \begin{itemize}
     \item $A^{+}(-p)$ is  a saddle with a 4D stable manifold  for
$\frac{1+\sqrt{3}}{2}<p<2$ or   $p<\frac{1-\sqrt{3}}{2}$  or 
$1\leq\gamma\leq\frac{5}{3},\;
\frac{1-\sqrt{3}}{2}<p<\frac{1}{4}\left(\frac{4+9\gamma}{1+3\gamma}-\sqrt{
\frac{16+48\gamma+9\gamma^{2}}{(1+3\gamma)^{2}}}\right)$.
       \item $B^{+}(-p)$ is a saddle point with a 4D stable manifold for
$p>1,\frac{4p}{3}<\gamma<\frac{5}{3}$.
     \item $P_{4}^{+}(-p)$ is a saddle point with a 4D stable for $p>2$. 
   \end{itemize}
\end{itemize}

\begin{figure}[h]
\centering
\subfigure[]{
\includegraphics[height=5.25cm,width=6cm]{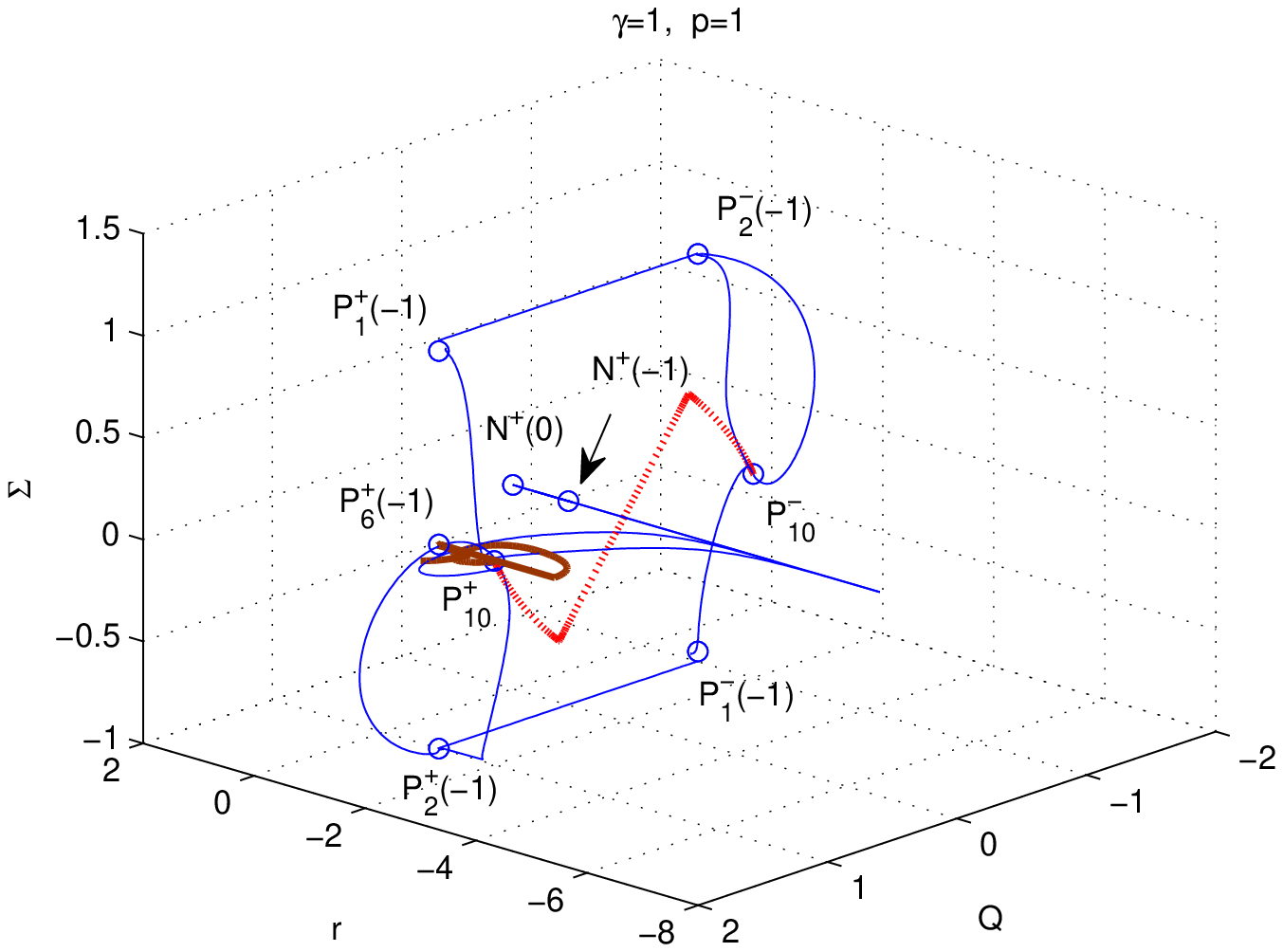}}
\subfigure[]{
\includegraphics[height=5.25cm,width=6cm]{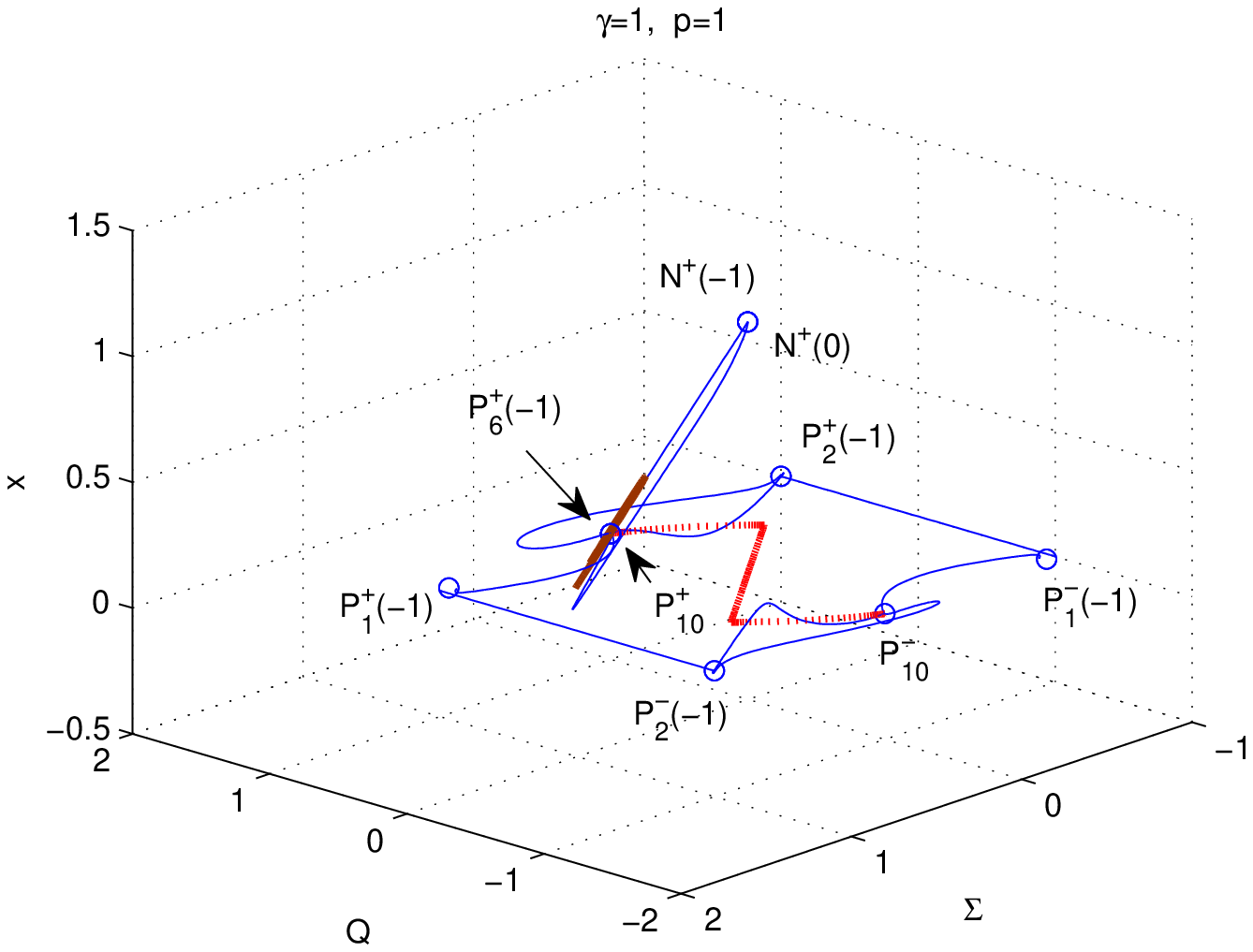}}
\subfigure[]{
\includegraphics[height=5.25cm,width=6cm]{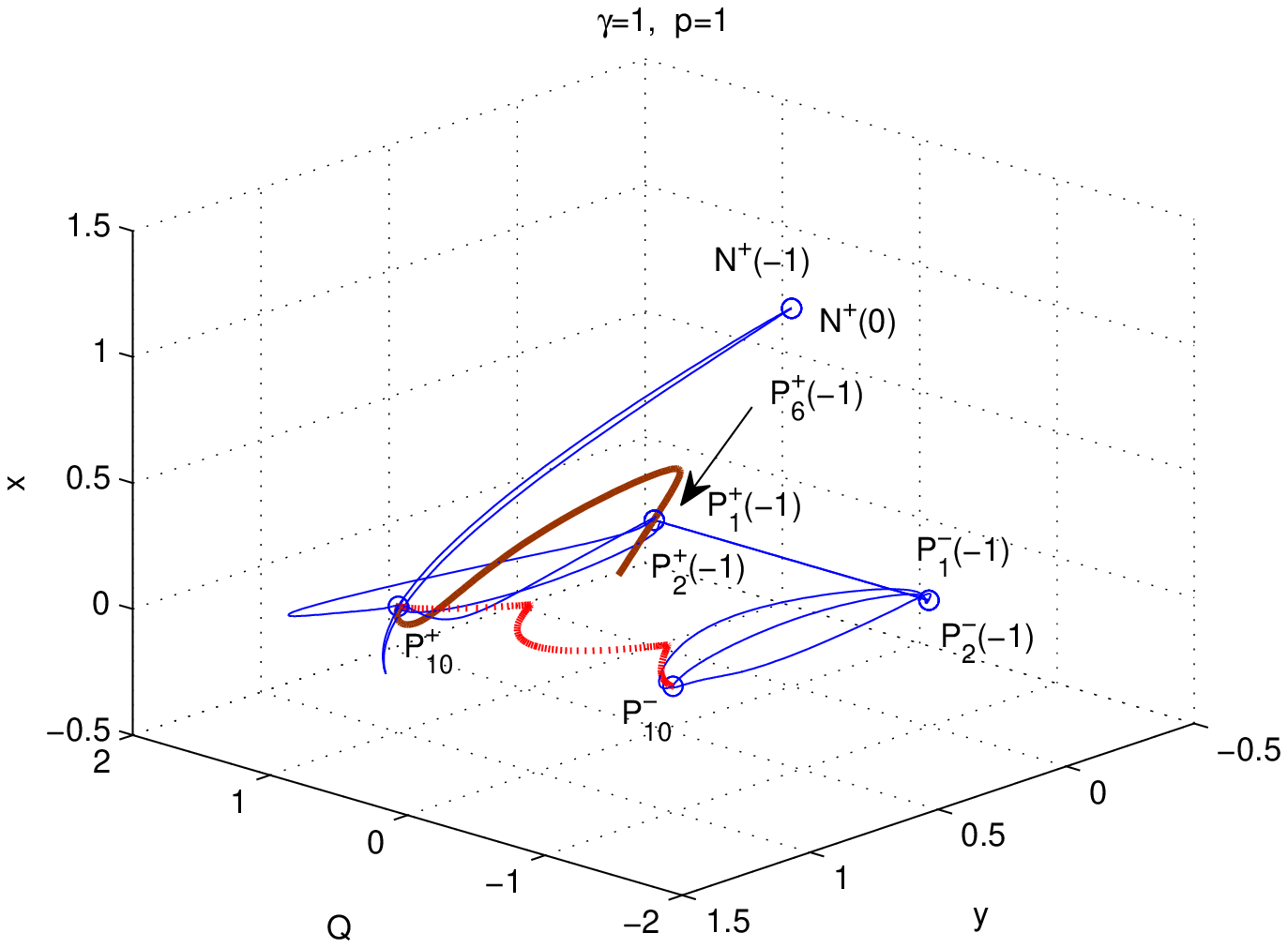}}
\subfigure[]{
\includegraphics[height=5.25cm,width=6cm]{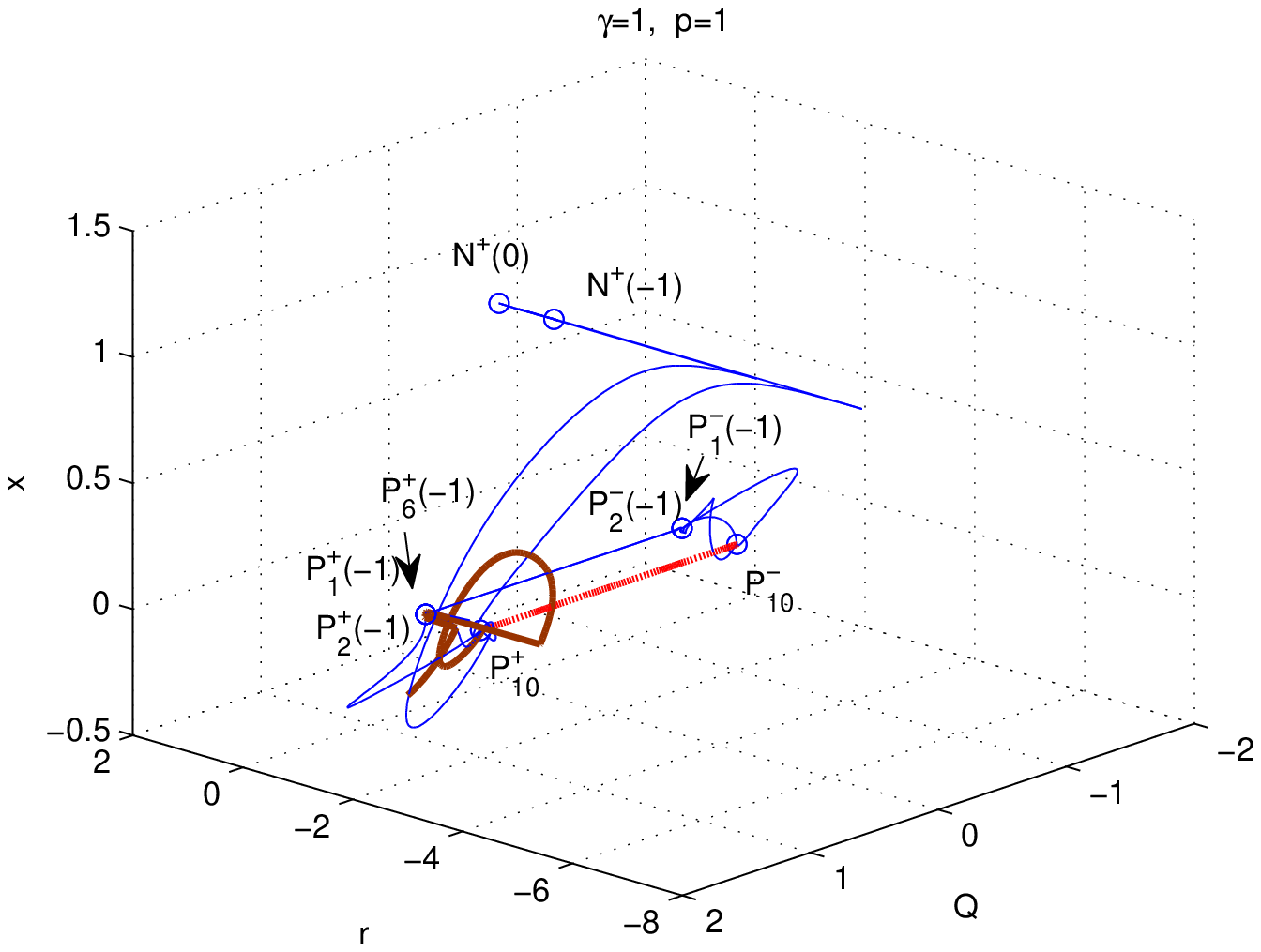}}
\subfigure[]{
\includegraphics[height=5.25cm,width=6cm]{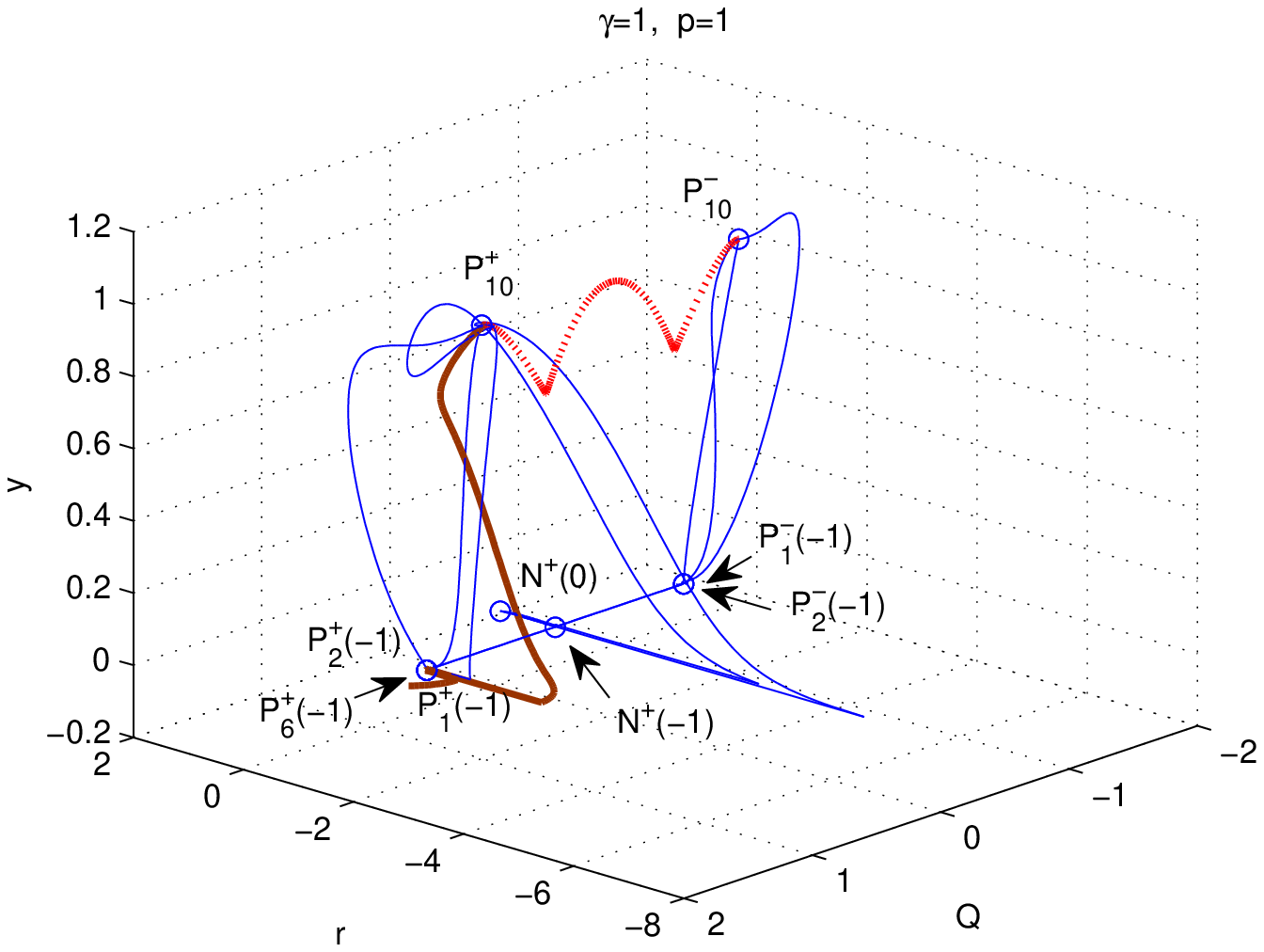}}
\subfigure[]{
\includegraphics[height=5.25cm,width=6cm]{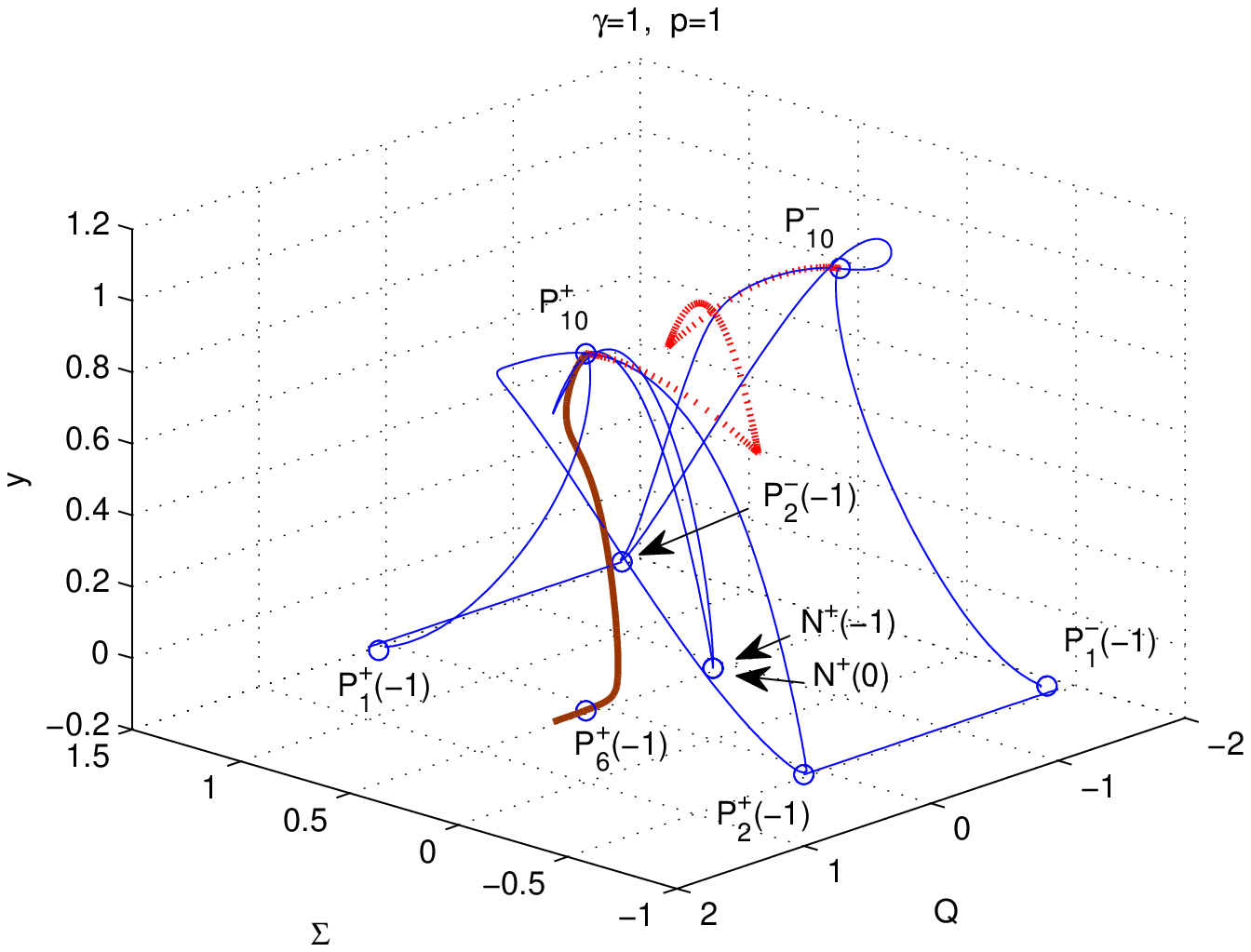}}
\caption{\label{fig5}    Some heteroclinic orbits given by
\eqref{heteroclinicsD.1} for the model $f(R)=R^p\exp\left(
\frac{q}{R}\right)$ for dust ($\gamma=1$) for $p=1$ and $q$ arbitrary. It is
illustrated the transition from an expanding phase to a contracting one, and
from contraction to expansion. The dotted (red) line corresponds to the
orbit joining directly the contracting de Sitter solution $P_{10}^-$ with
the expanding one  $P_{10}^+$ (first line of \eqref{5.1c}).  The thick
(brown) line denotes the orbit that connects the matter-dominated dust 
($\gamma=1$) solution given by $P_6^+(-1)$ (that belongs to region V) to an
accelerated de Sitter phase given by $P_{10}^+$ (that belongs to region
II).}
\end{figure}

\begin{figure}[h]
\centering
\subfigure[]{
\includegraphics[height=5.25cm,width=6cm]{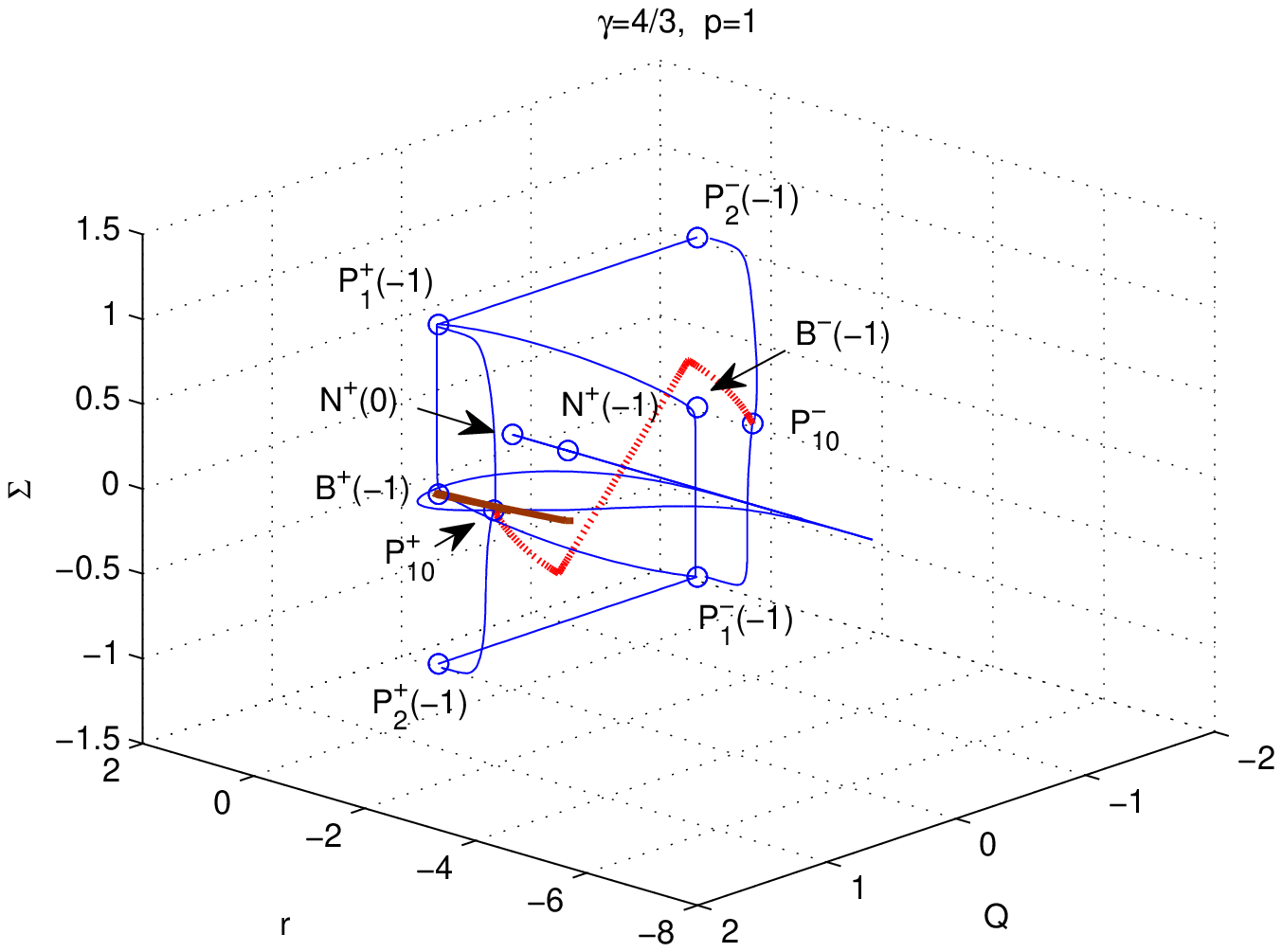}}
\subfigure[]{
\includegraphics[height=5.25cm,width=6cm]{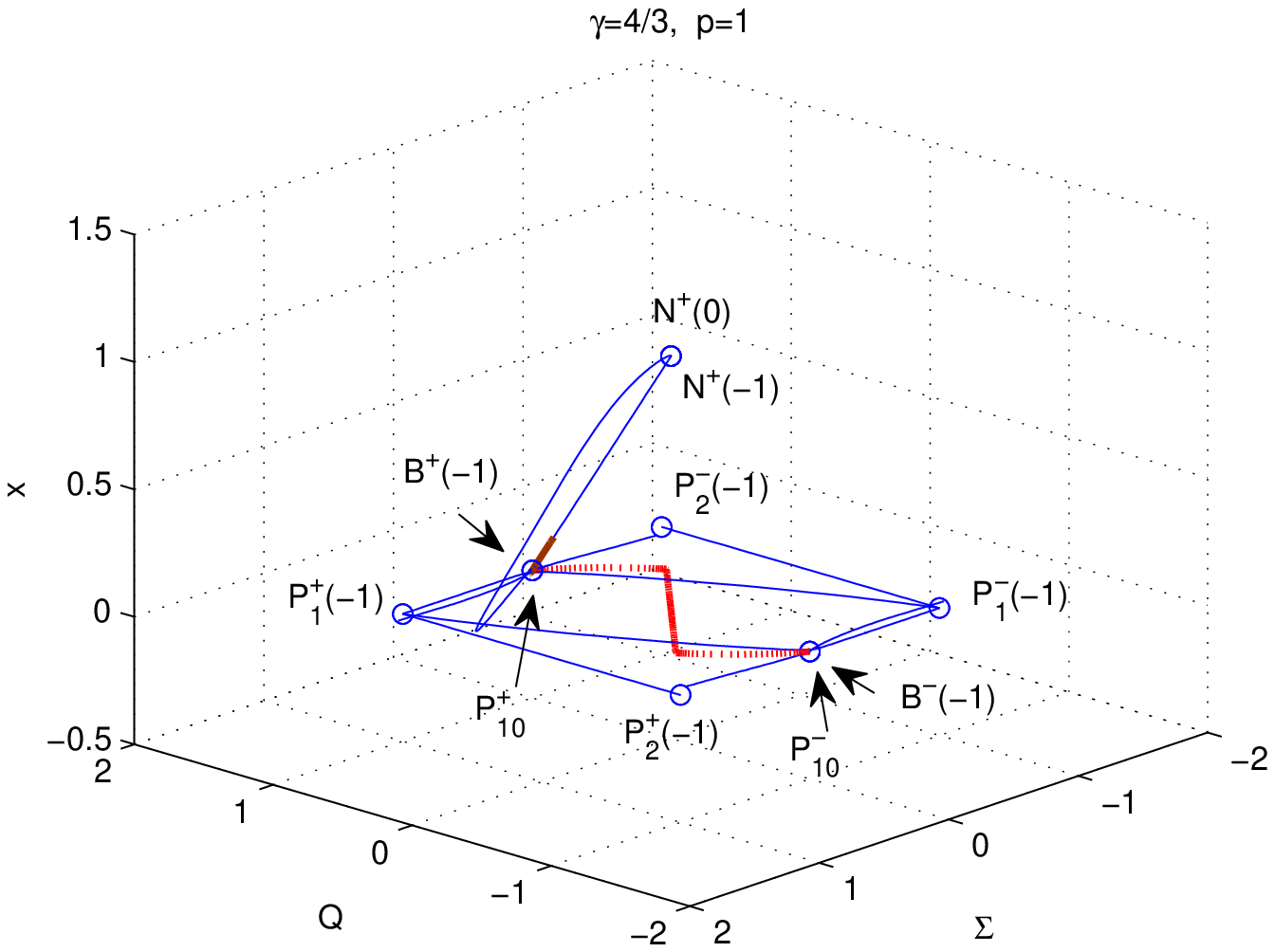}}
\subfigure[]{
\includegraphics[height=5.25cm,width=6cm]{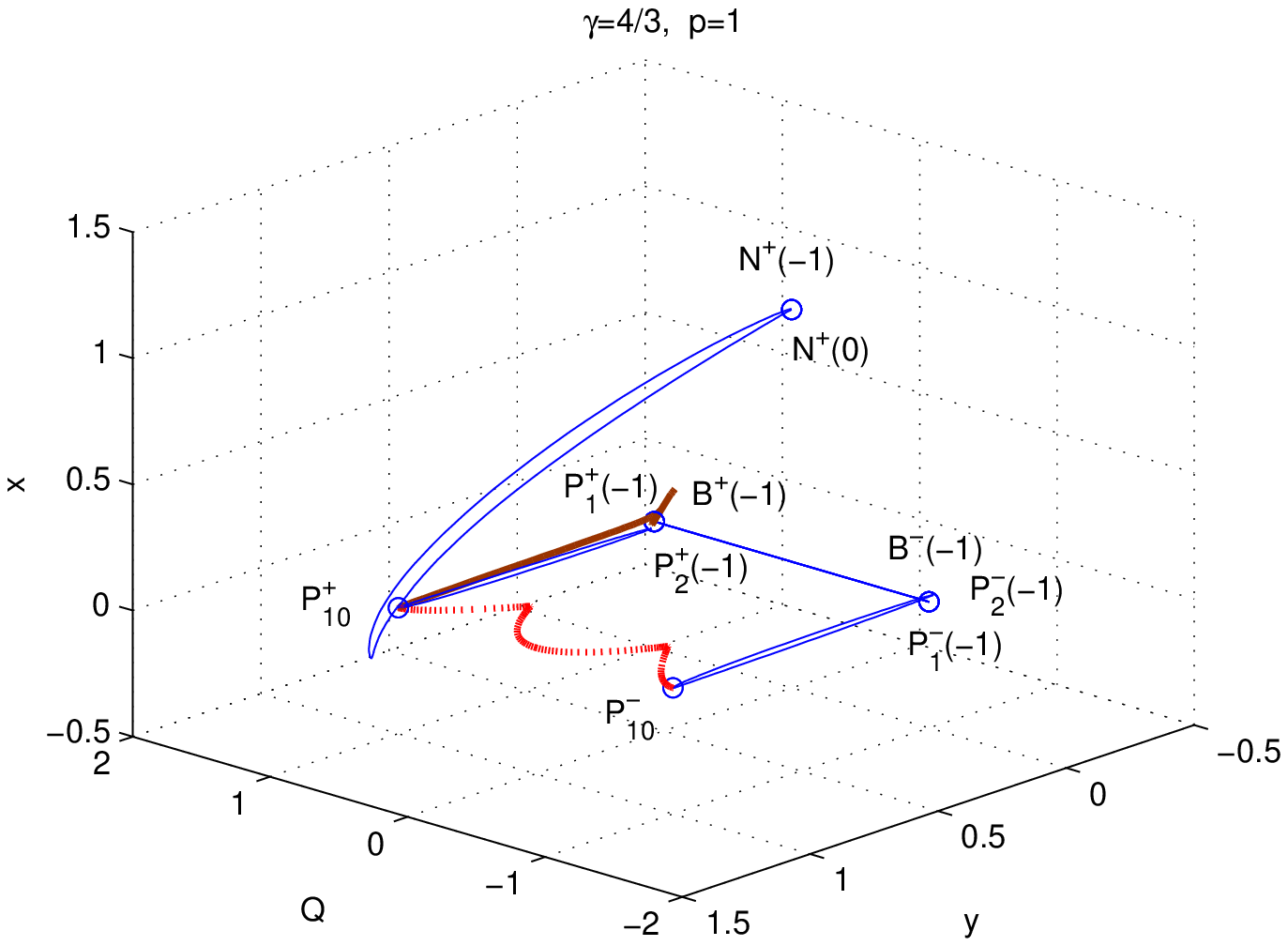}}
\subfigure[]{
\includegraphics[height=5.25cm,width=6cm]{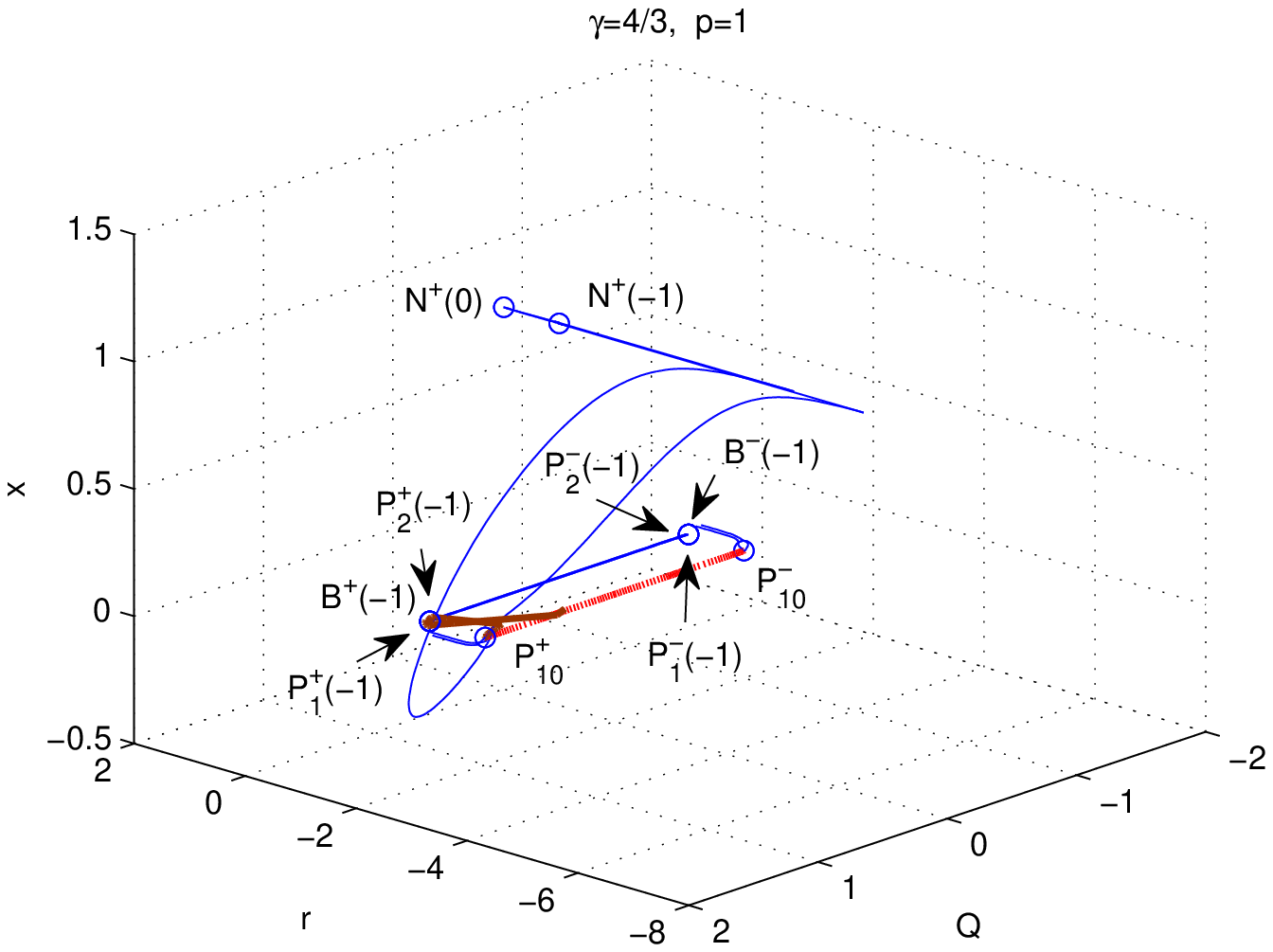}}
\subfigure[]{
\includegraphics[height=5.25cm,width=6cm]{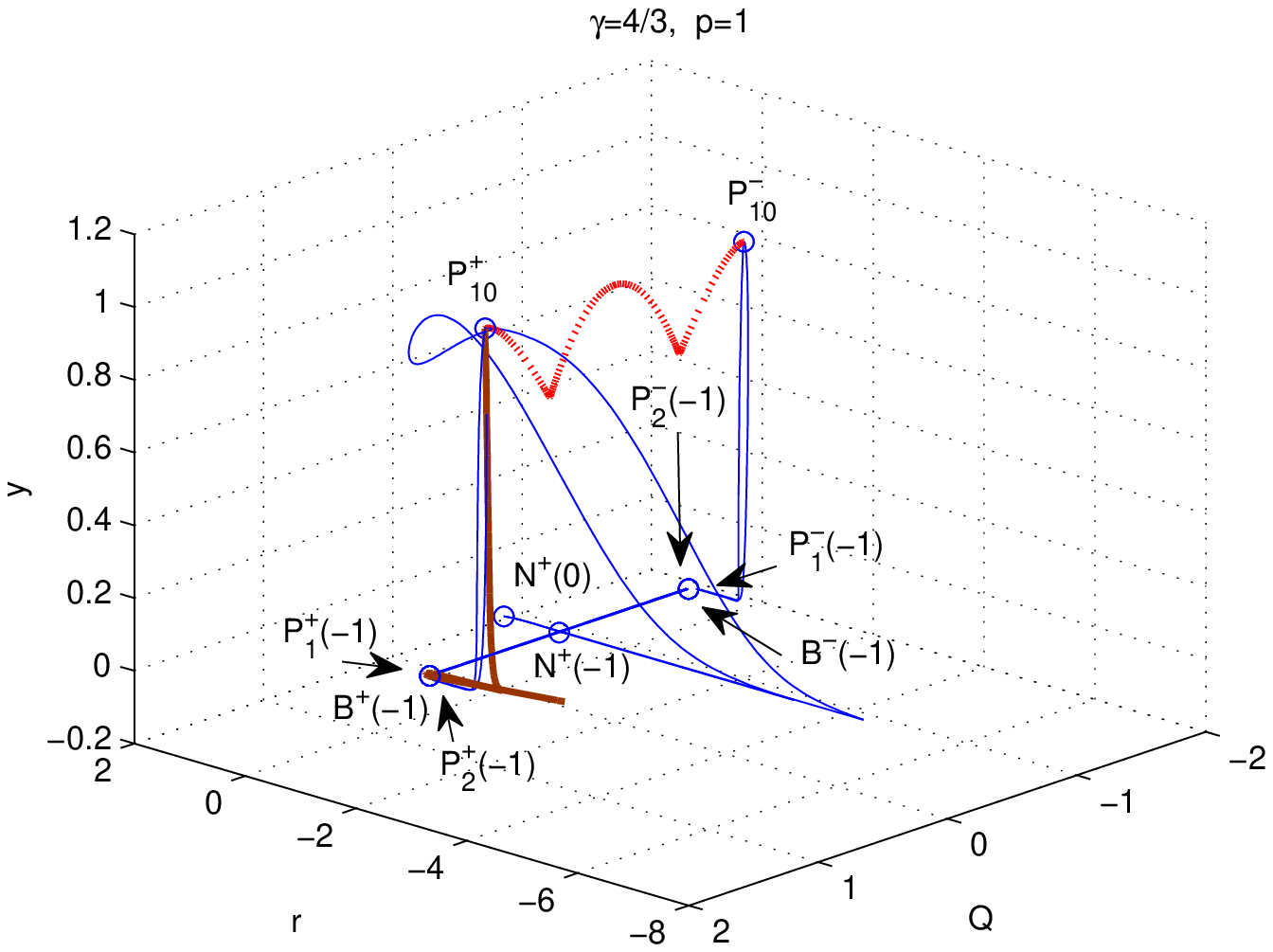}}
\subfigure[]{
\includegraphics[height=5.25cm,width=6cm]{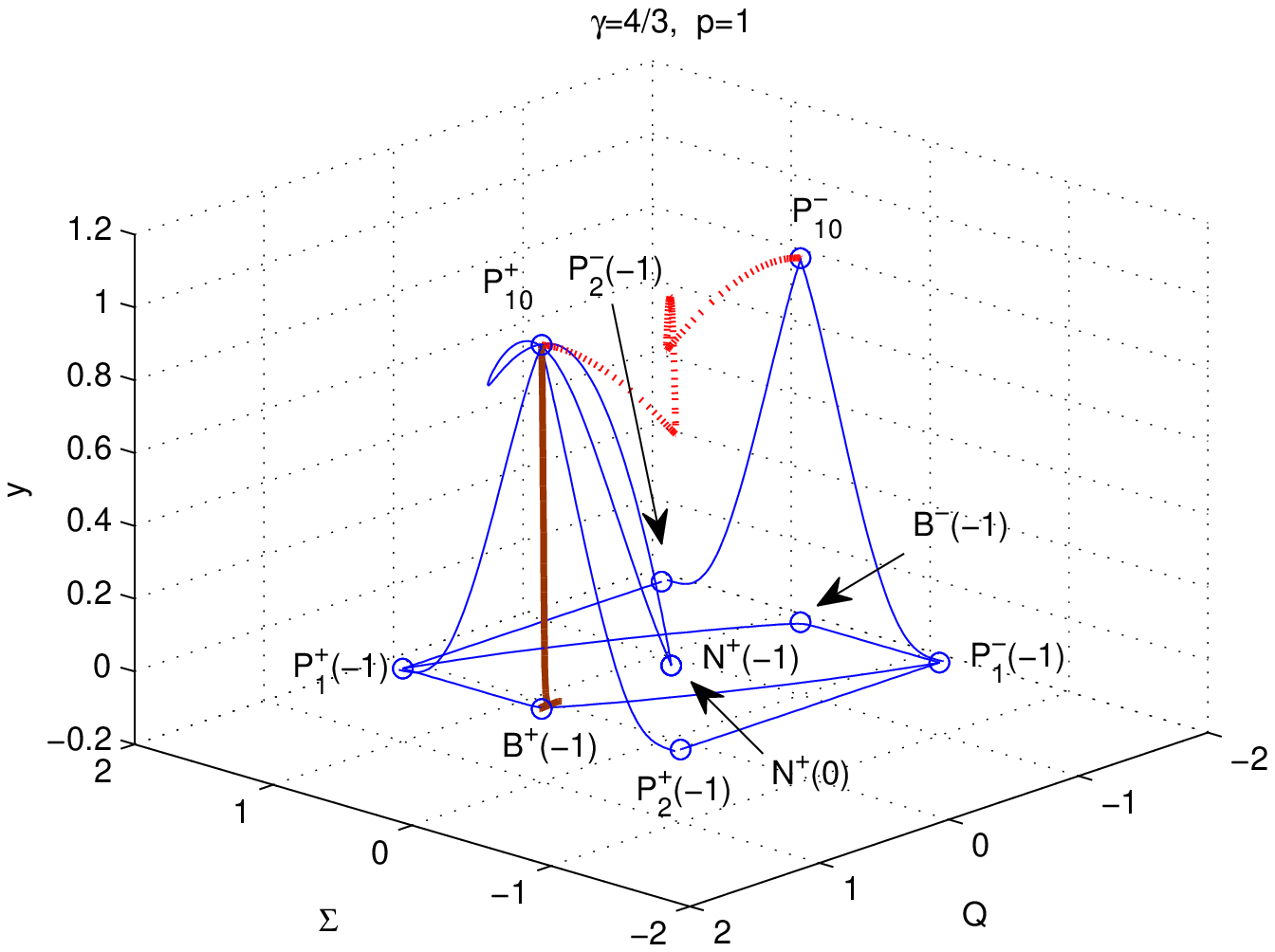}}
\caption{\label{fig6} Some heteroclinic orbits given by
\eqref{heteroclinicsD.2} for the model $f(R)=R^p\exp\left(
\frac{q}{R}\right)$ for radiation ($\gamma=4/3$) for $p=1$ and $q$
arbitrary. It is illustrated the transition from an expanding phase to a
contracting one, and from contraction to expansion. The dotted (red) line
corresponds to the orbit joining directly the contracting de Sitter solution
$P_{10}^-$ with the expanding one  $P_{10}^+$ (first line of \eqref{5.1c}). 
The thick (brown) line denotes the orbit that connects the matter-dominated
radiation ($\gamma=4/3$) solution given by $B^+(-1)$ (that belongs to region
VII) to an accelerated de Sitter phase given by $P_{10}^+$ (that belongs to
region II).}
\end{figure}

\begin{figure}[h]
\centering
\subfigure[]{
\includegraphics[height=5.25cm,width=6cm]{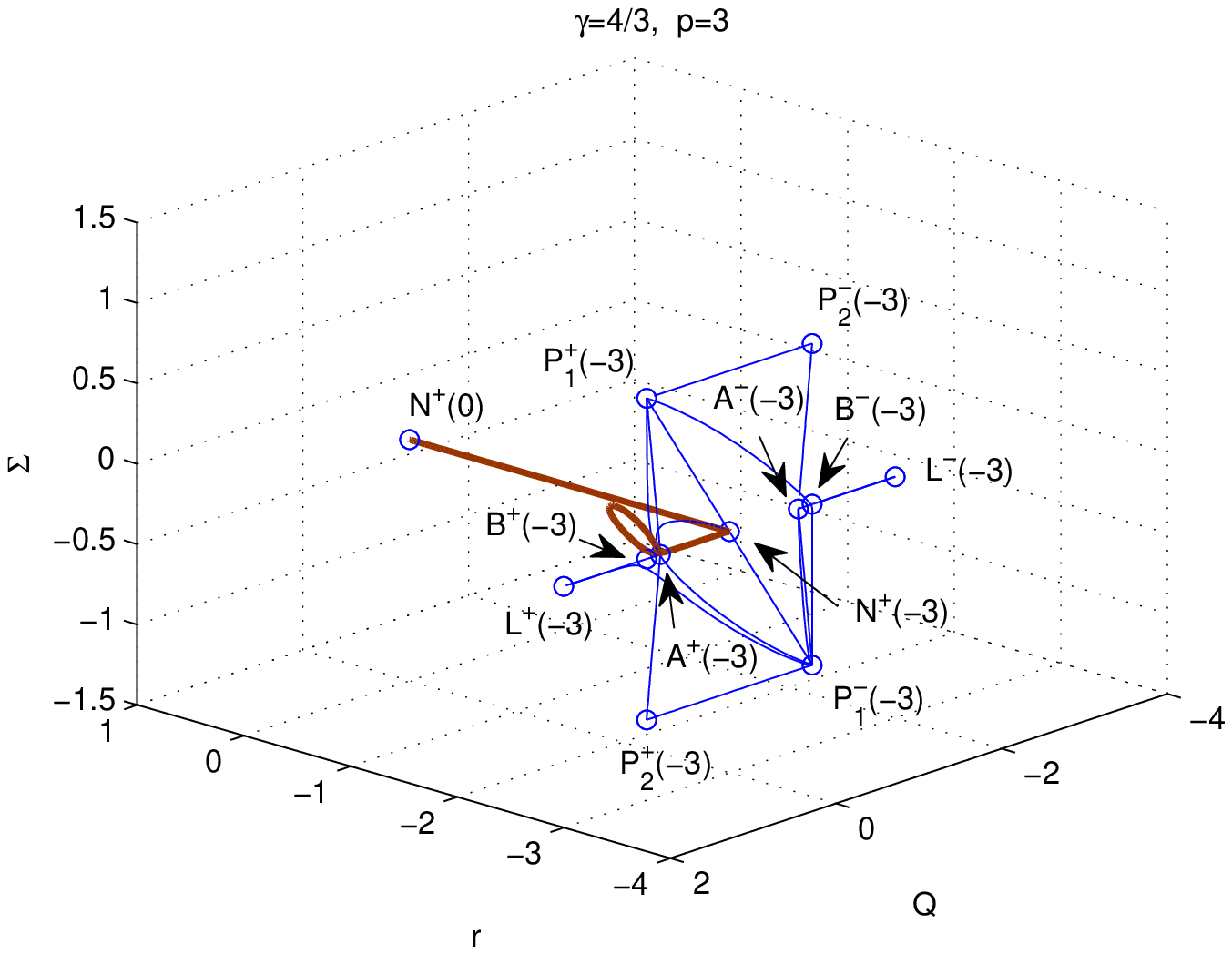}}
\subfigure[]{
\includegraphics[height=5.25cm,width=6cm]{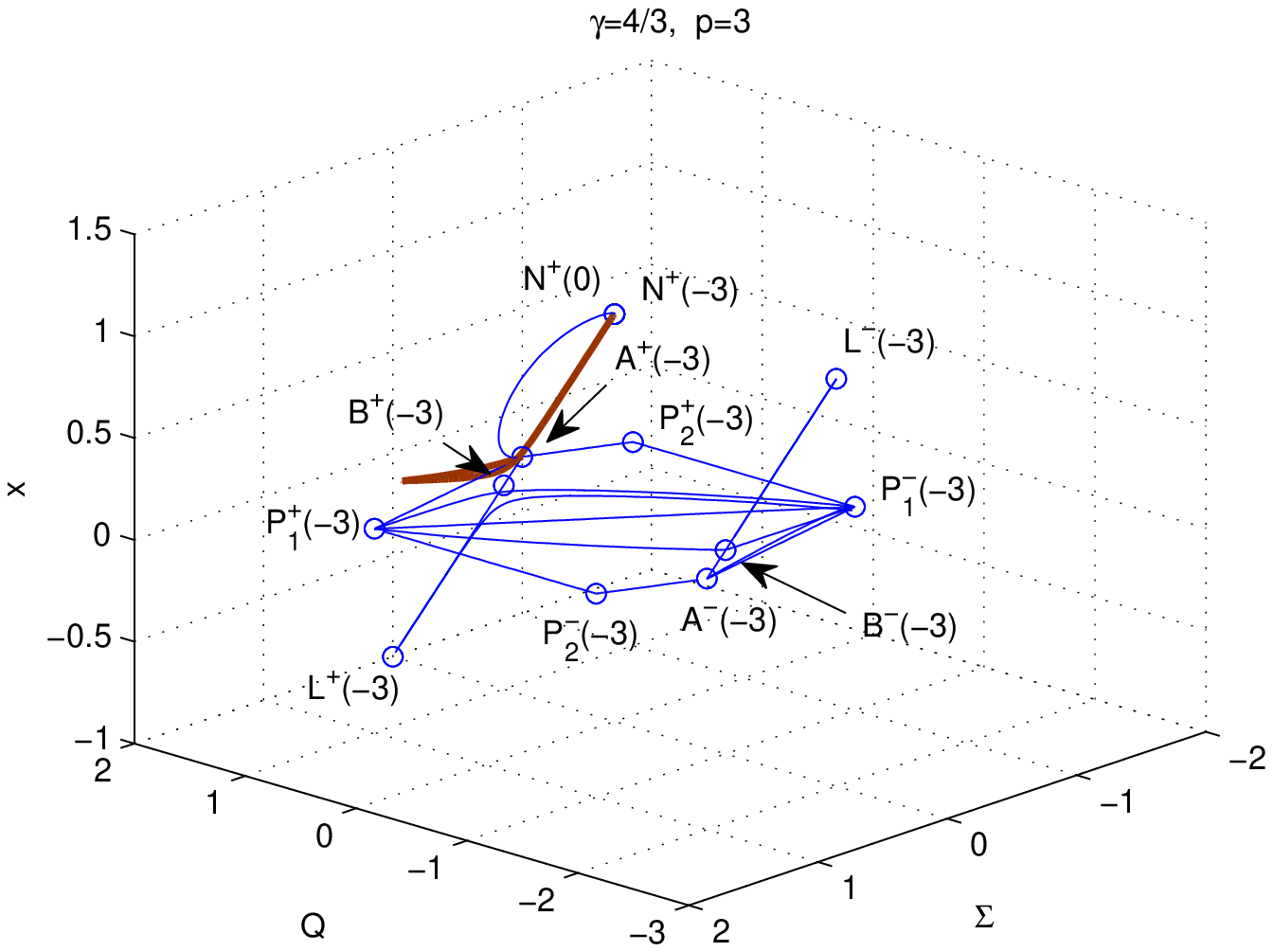}}
\subfigure[]{
\includegraphics[height=5.25cm,width=6cm]{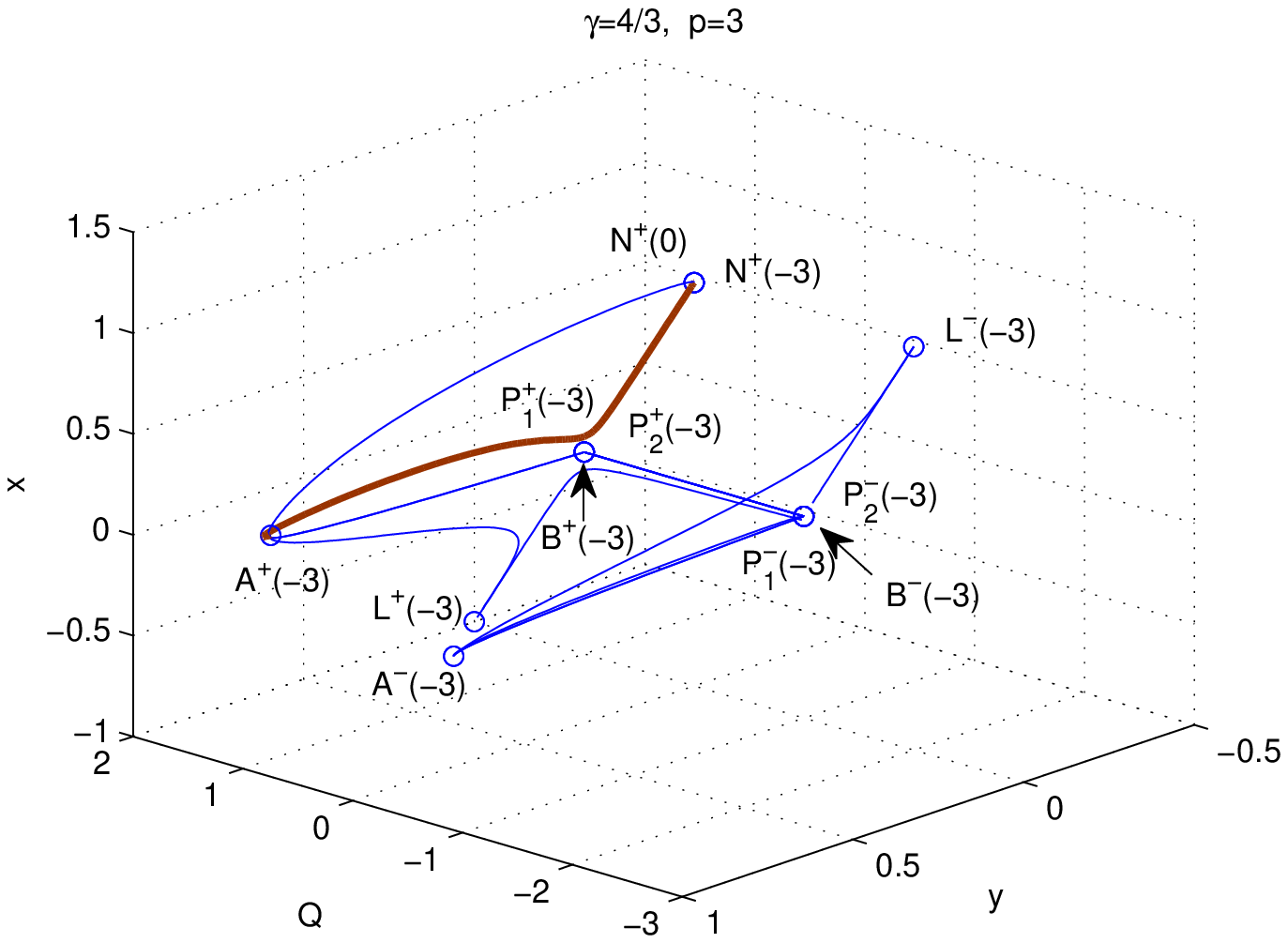}}
\subfigure[]{
\includegraphics[height=5.25cm,width=6cm]{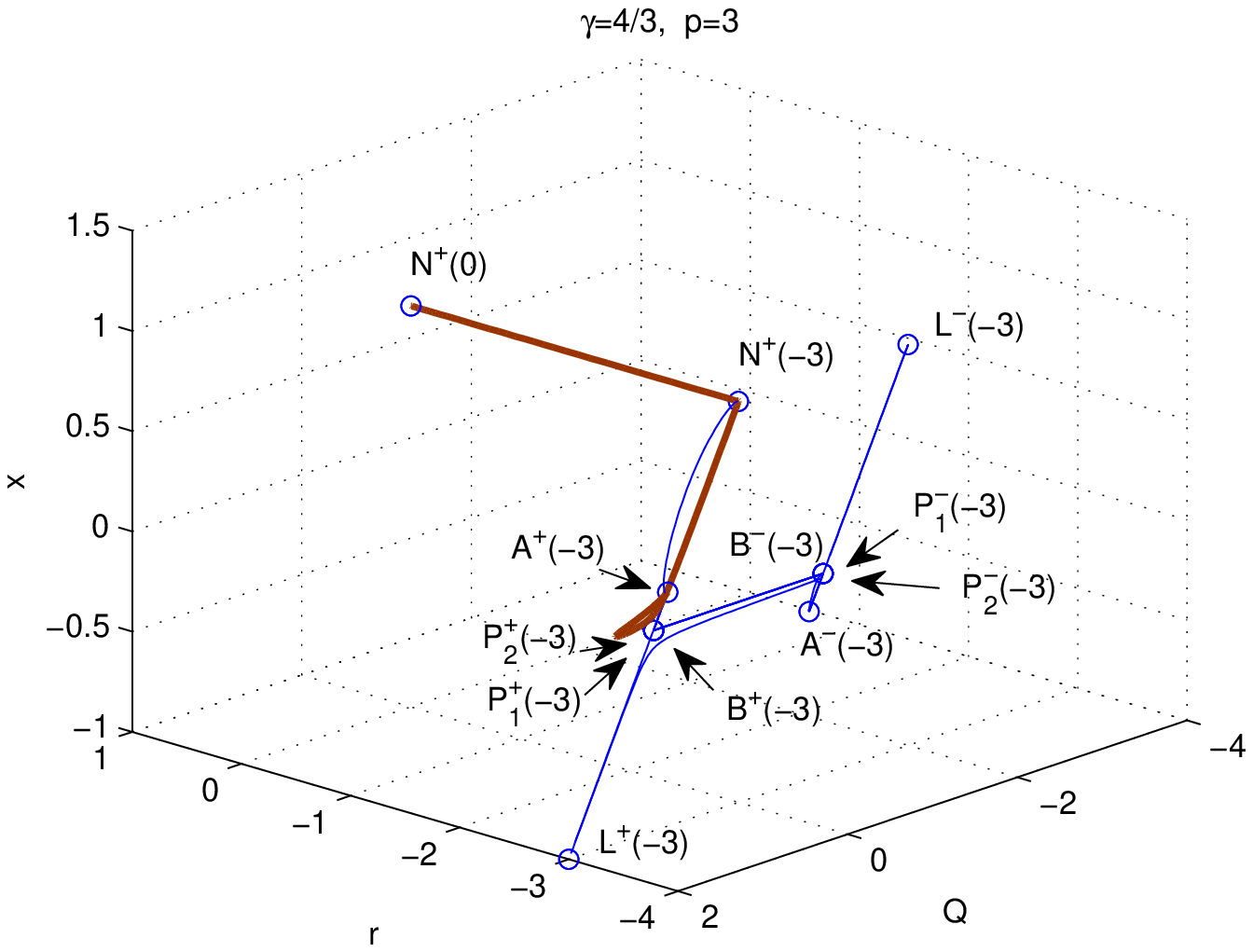}}
\subfigure[]{
\includegraphics[height=5.25cm,width=6cm]{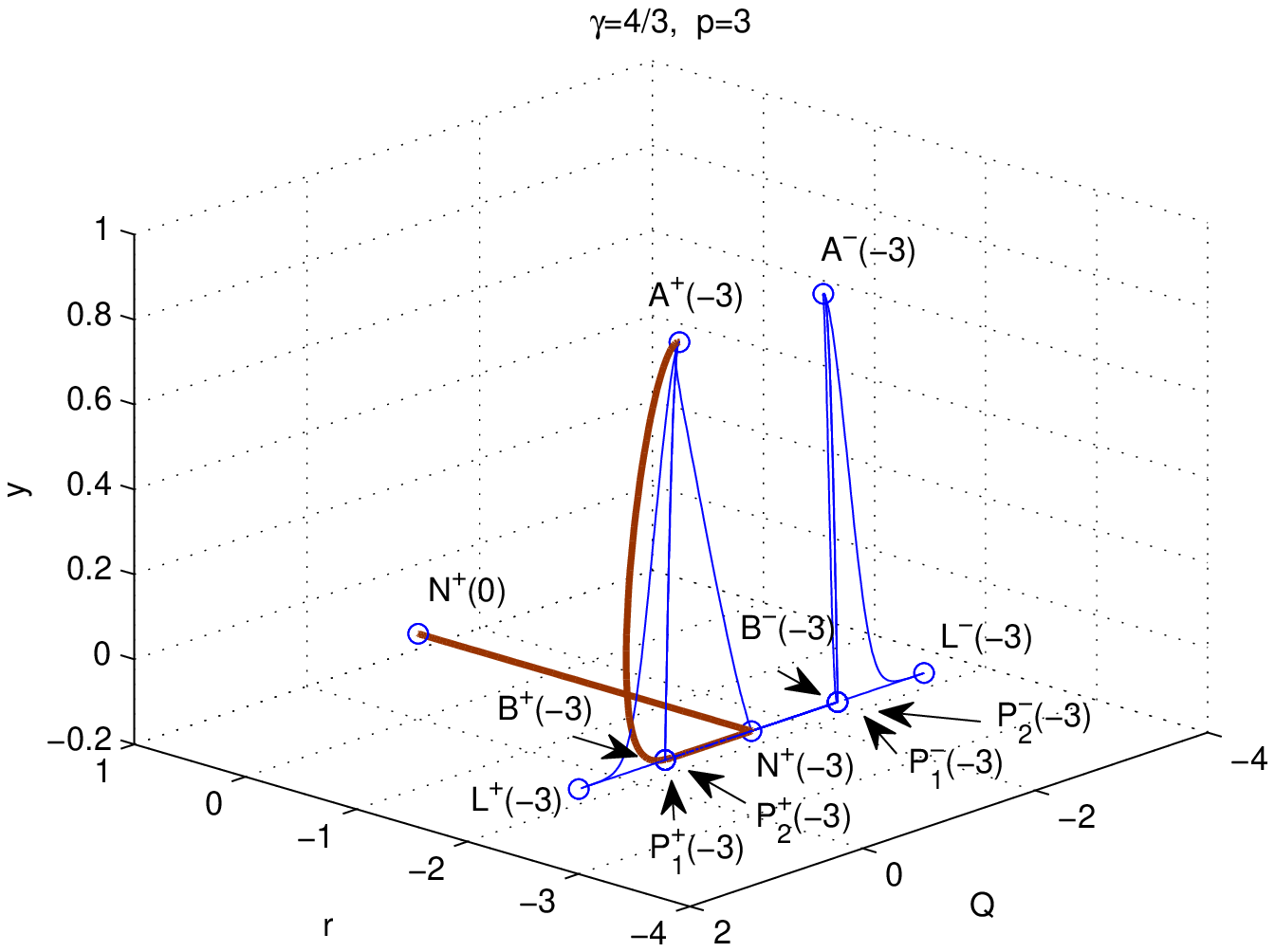}}
\subfigure[]{
\includegraphics[height=5.25cm,width=6cm]{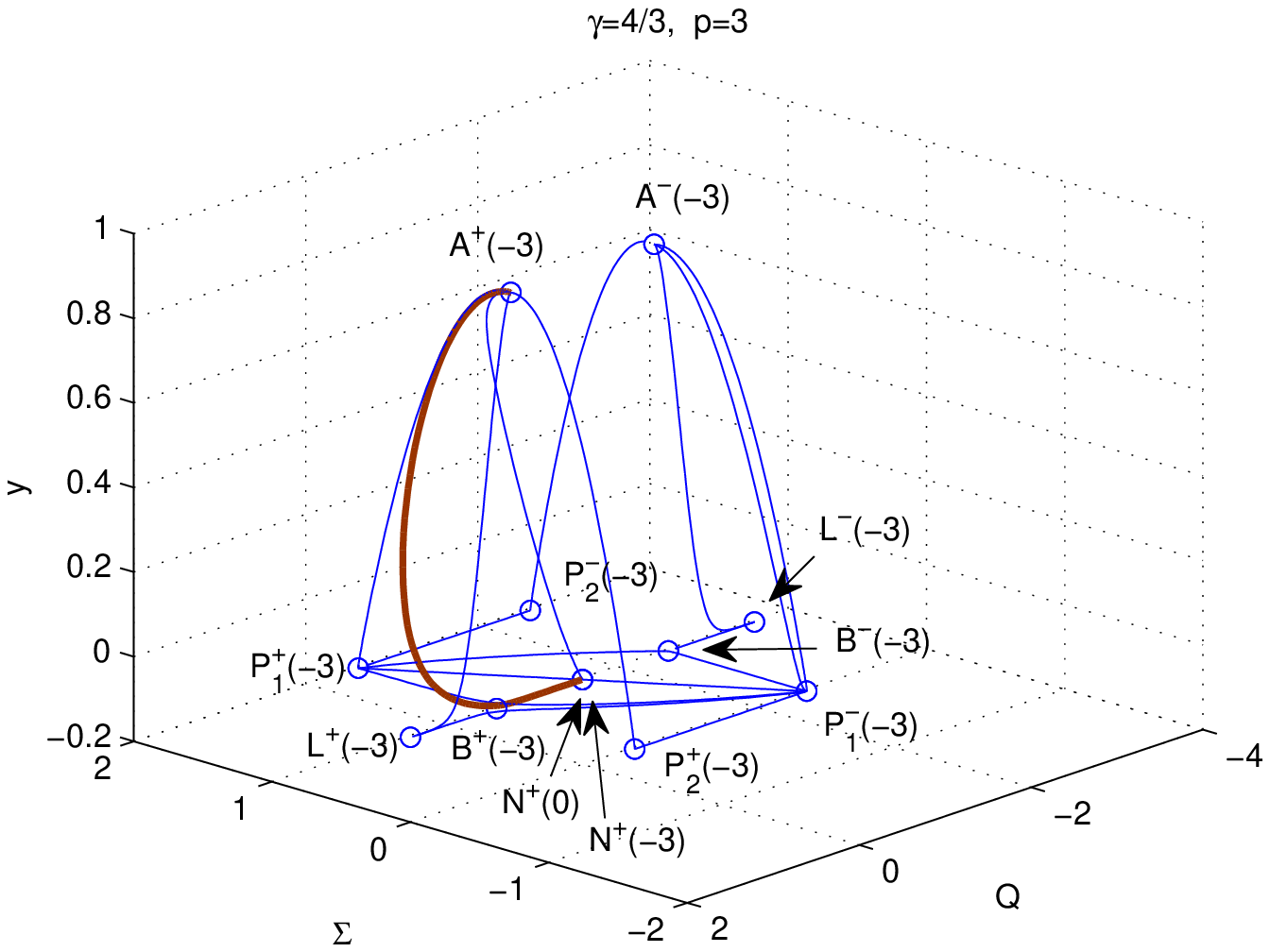}}
\caption{\label{fig7} Some heteroclinic orbits given by
\eqref{heteroclinicsD.3} for the model $f(R)=R^p\exp\left(
\frac{q}{R}\right)$ for radiation ($\gamma=4/3$) for $p=3$ and $q$
arbitrary. It is illustrated the transition from an expanding phase to a
contracting one.  The thick (brown) line denotes the orbit that connects the
matter-dominated radiation  ($\gamma=4/3$) solution given by $B^+(-3)$ (that
belongs to region VII) with  an accelerated phantom phase given by $A^+(-3)$
(that belongs to region I). This orbit is past asymptotic to the static
solution $N^+(0).$}
\end{figure}

The function $M(r)$ satisfy the following:
\begin{itemize}
  \item it connects the matter-dominated region V with the region II
corresponding to a de Sitter accelerated solution for $p\rightarrow\,1$ (see
figure \ref{fig5}). 
 
    \item it connects the matter-dominated region
VII with the region II corresponding to a de Sitter accelerated solution for
$\gamma=\frac{4}{3},\; 0<p<2$ (see figure \ref{fig6}). 
 \item it connects the matter-dominated region VII with the region I
corresponding to an accelerated phantom phase for $\gamma=\frac{4}{3},\; p>2$
(see figure \ref{fig7}). 

\end{itemize}

In order to present the aforementioned results in a transparent way, we
proceed to several numerical
simulations as follows. 

In the figure \ref{fig5} are presented several numerical simulations for the
model $f(R)=R^p\exp\left( \frac{q}{R}\right)$ for dust ($\gamma=1$) for $p=1$
and $q$ arbitrary. 

In this figure are displayed the heteroclinic sequences: 
 \begin{subequations}
 \label{heteroclinicsD.1}
\begin{align}
& P_{1}^{+}(-1)\longrightarrow \left\{\begin{array}{l}
                          P^{-}_{2}(-1) \\ 
                          P^{+}_{10}
                         \end{array} \right., \label{5.1a}\\
& P_{2}^{+}(-1)\longrightarrow \left\{\begin{array}{l}
                          P^{+}_{10}\\     
   P^{-}_{1}(-1) 
                         \end{array} \right., \label{5.1b}\\
& P_{10}^{-}\longrightarrow \left\{\begin{array}{l}
                         P^{-}_{2}(-1)\\    
    P^{-}_{1}(-1)\\   
    P^{+}_{10} 
                         \end{array} \right., \label{5.1c}\\
& N^{+}(0)\longrightarrow N^{+}(-1)\longrightarrow P^{+}_{10},\label{5.1d}\\
& N^{+}(-1)\longrightarrow  P^{+}_{10}, \label{5.1e}\\
& P^{+}_{6}(-1)\longrightarrow P^{+}_{10}.  \label{5.1f}
\end{align}
\end{subequations}

In the figure \ref{fig5} it is illustrated the transition from an expanding
phase to a contracting one (see the sequences at first line of \eqref{5.1a}
and first line of \eqref{5.1b}) and from contraction to expansion (see first
line in sequence \eqref{5.1c}). The dotted (red) line corresponds to the
orbit joining directly the contracting de Sitter solution $P_{10}^-$ with the
expanding one  $P_{10}^+$ (first line of \eqref{5.1c}).  The thick (brown)
line denotes the orbit that connects the matter-dominated dust  ($\gamma=1$)
solution given by $P_6^+(-1)$ (that belongs to region V) to an accelerated de
Sitter phase given by $P_{10}^+$ (that belongs to region II).

In the figure \ref{fig6} are presented several numerical solutions for the
model $f(R)=R^p\exp\left( \frac{q}{R}\right)$ for radiation ($\gamma=4/3$)
for $p=1$ and $q$ arbitrary. 

In this figure are presented the  heteroclinic
orbits:
\begin{subequations}
\label{heteroclinicsD.2}
\begin{align}
& P_{1}^{+}(-1)\longrightarrow \left\{\begin{array}{l}
                          P^{-}_{2}(-1) \\ 
                          B^{-}(-1) \longrightarrow P^{-}_{1}(-1) \\ 
     B^{+}(-1) \longrightarrow
P^{-}_{1}(-1) \\     
   P^{+}_{10}
                         \end{array} \right., \label{5.2a}\\
& P_{2}^{+}(-1)\longrightarrow \left\{\begin{array}{l}
                           P^{-}_{1}(-1)\\   
     P^{+}_{10} 
                         \end{array} \right. \label{5.2b}\\
& P_{10}^{-}\longrightarrow \left\{\begin{array}{l}
                          P^{+}_{10} \\ 
                          P^{-}_{2}(-1)\\   
    P^{-}_{1}(-1)
                         \end{array} \right., \label{5.2c}\\
& N^{+}(0)\longrightarrow N^{+}(-1)\longrightarrow P^{+}_{10}\label{5.2d},\\
& N^{+}(-1)\longrightarrow  P^{+}_{10} \label{5.2e},\\
& B^{+}(-1)\longrightarrow P^{+}_{10}.  \label{5.2f}
\end{align}
\end{subequations}

These sequences show the transition from an expanding phase to a contracting
one (see the sequences at first, second and third lines of \eqref{5.2a} and
first line of \eqref{5.1b}) and from contraction to expansion (see first line
in sequence \eqref{5.1c}). The dotted (red) line corresponds to the orbit
joining directly the contracting de Sitter solution $P_{10}^-$ with the
expanding one  $P_{10}^+$ (first line of \eqref{5.1c}).  The thick (brown)
line denotes the orbit that connects the matter-dominated radiation 
($\gamma=4/3$) solution given by $B^+(-1)$ (that belongs to region VII) to an
accelerated de Sitter phase given by $P_{10}^+$ (that belongs to region II).

Finally, in the figure \ref{fig7} are presented some numerical solutions for 
the model $f(R)=R^p\exp\left( \frac{q}{R}\right)$ for radiation
($\gamma=4/3$) for $p=3$ and $q$ arbitrary. 

Specially, are represented the 
heteroclinic orbits:
\begin{subequations}
\label{heteroclinicsD.3}
\begin{align}
& N^{+}(0)\longrightarrow N^{+}(-3) \longrightarrow \left\{\begin{array}{l}
       A^{+}(-3) \\
                          B^{+}(-3)\longrightarrow A^{+}(-3)\\  
    \end{array}    \right., \label{5.3a}\\
& N^{+}(-3) \longrightarrow \left\{\begin{array}{l}   
    A^{+}(-3) \\
                          B^{+}(-3)\longrightarrow A^{+}(-3)\\  
    \end{array}    \right., \label{5.3b}\\
& P_{1}^{+}(-3)\longrightarrow \left\{\begin{array}{l}   
    P_{2}^{-}(-3) \\   
    P_{1}^{-}(-3) \\   
    B^{-}(-3)\longrightarrow P_{1}^{-}(-3) \\
   B^{-}(-3)\longrightarrow L^{-}(-3) \\
   B^{+}(-3)\longrightarrow P_{1}^{-}(-3) \\
   A^{+}(-3) 
                         \end{array} \right., \label{5.3c}\\
& P_{2}^{+}(-3)\longrightarrow \left\{\begin{array}{l}
   P_{1}^{-}(-3) \\
   A^{+}(-3)
   \end{array} \right., \label{5.3d} \\
& A^{-}(-3)\longrightarrow \left\{\begin{array}{l}
                              P_{1}^{-}(-3) \\
     L^{-}(-3) \\
     P_{2}^{-}(-3)
                             \end{array}    \right., \label{5.3e}\\
& L^{+}(-3)\longrightarrow \left\{\begin{array}{l}
                              B^{+}(-3)\longrightarrow P_{1}^{-}(-3) \\      
       A^{+}(-3) 
                             \end{array}    \right..\label{5.3f}    
\end{align}
\end{subequations}

These sequences show  the transition from an expanding phase to a contracting
one (see the first five lines of \eqref{5.2a}; the first line of \eqref{5.3d}
and the first line of \eqref{5.3f}).  The thick (brown) line denotes the
orbit that connects the matter-dominated radiation  ($\gamma=4/3$) solution
given by $B^+(-3)$ (that belongs to region VII) with  an accelerated phantom
phase given by $A^+(-3)$ (that belongs to region I). This orbit is past
asymptotic to the static solution $N^+(0).$

%%%%%%%%%%%%%%%%%%%%%%%%%%%%%%%%%%%%%
%%%%%%%%%%%%%%%%%%%%%%%%%%%%%%%%%%%%%%
%%%%%%%%%%%%%%%%%%%%%%%%%%%%%%%%%%%%%
\subsection{Model $f(R)=R^p\left(\ln \alpha R\right)^q$.}

In this case $M(r)=\frac{r(p+r)^{2}}{(p+r)^{2}-r(r+1)q}$, 
$r^*\in\left\{0,-p\right\}$,  $M'(0)=1$ and $M'(-p)=0$.

\begin{itemize}
\item The sufficient conditions for the existence of past-attractors
(future-attractors) are:
   \begin{itemize}
    \item [-] $N^{+}(0)$ ($N^{-}(0)$) is always a local past
(future)-attractor.
   \end{itemize}
\end{itemize}

\begin{figure}[h]
\centering
\subfigure[]{
\includegraphics[height=5.25cm,width=6cm]{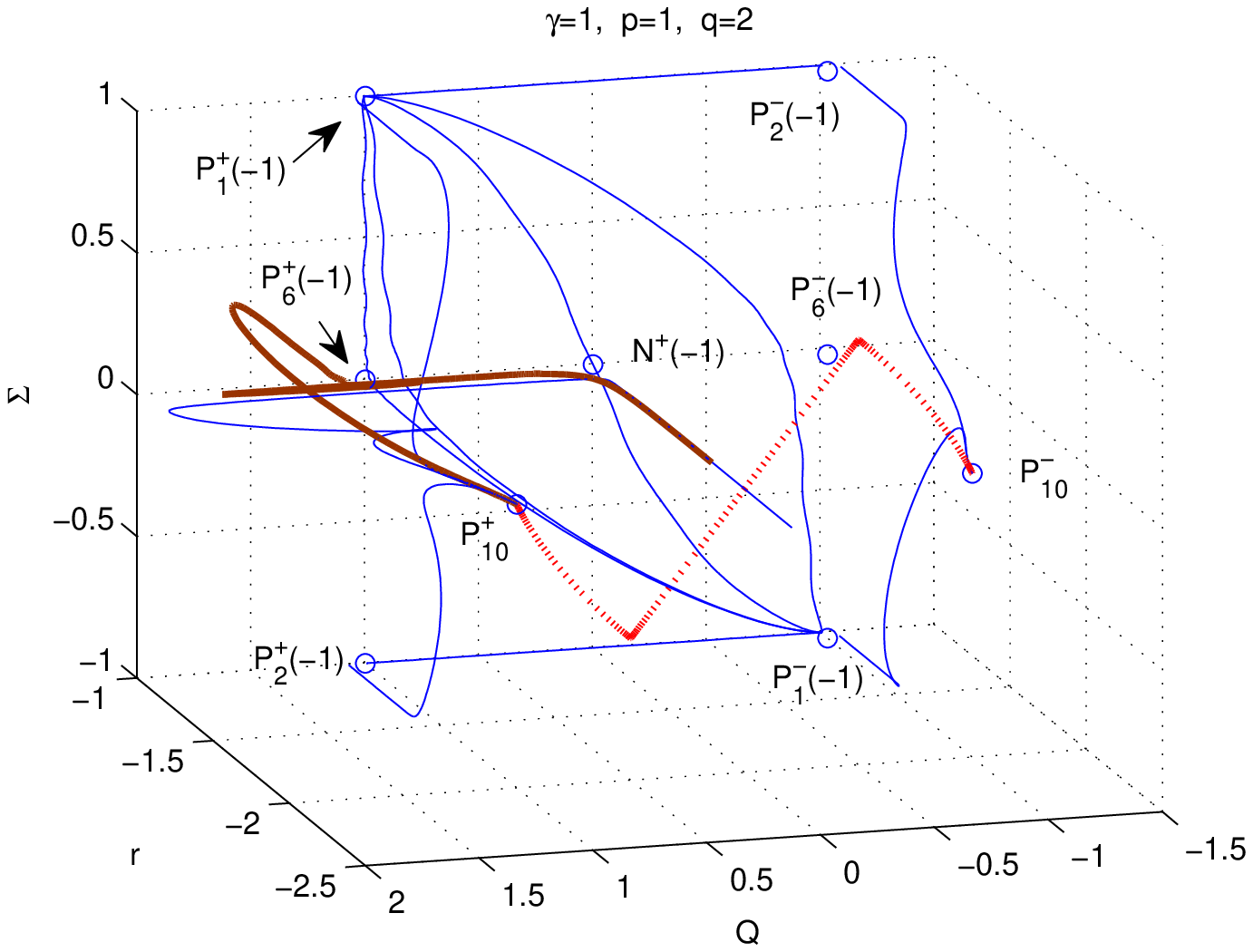}}
\subfigure[]{
\includegraphics[height=5.25cm,width=6cm]{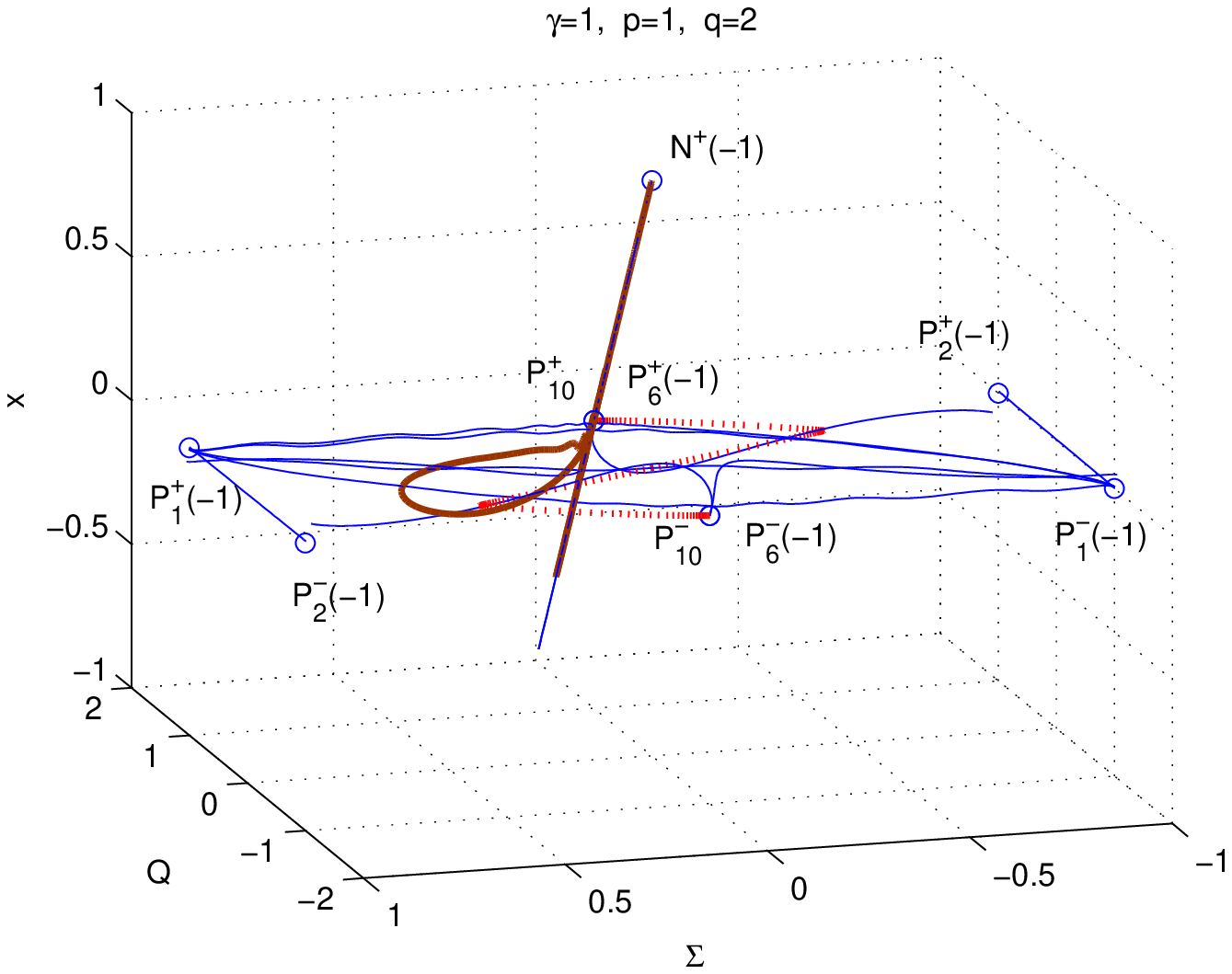}}
\subfigure[]{
\includegraphics[height=5.25cm,width=6cm]{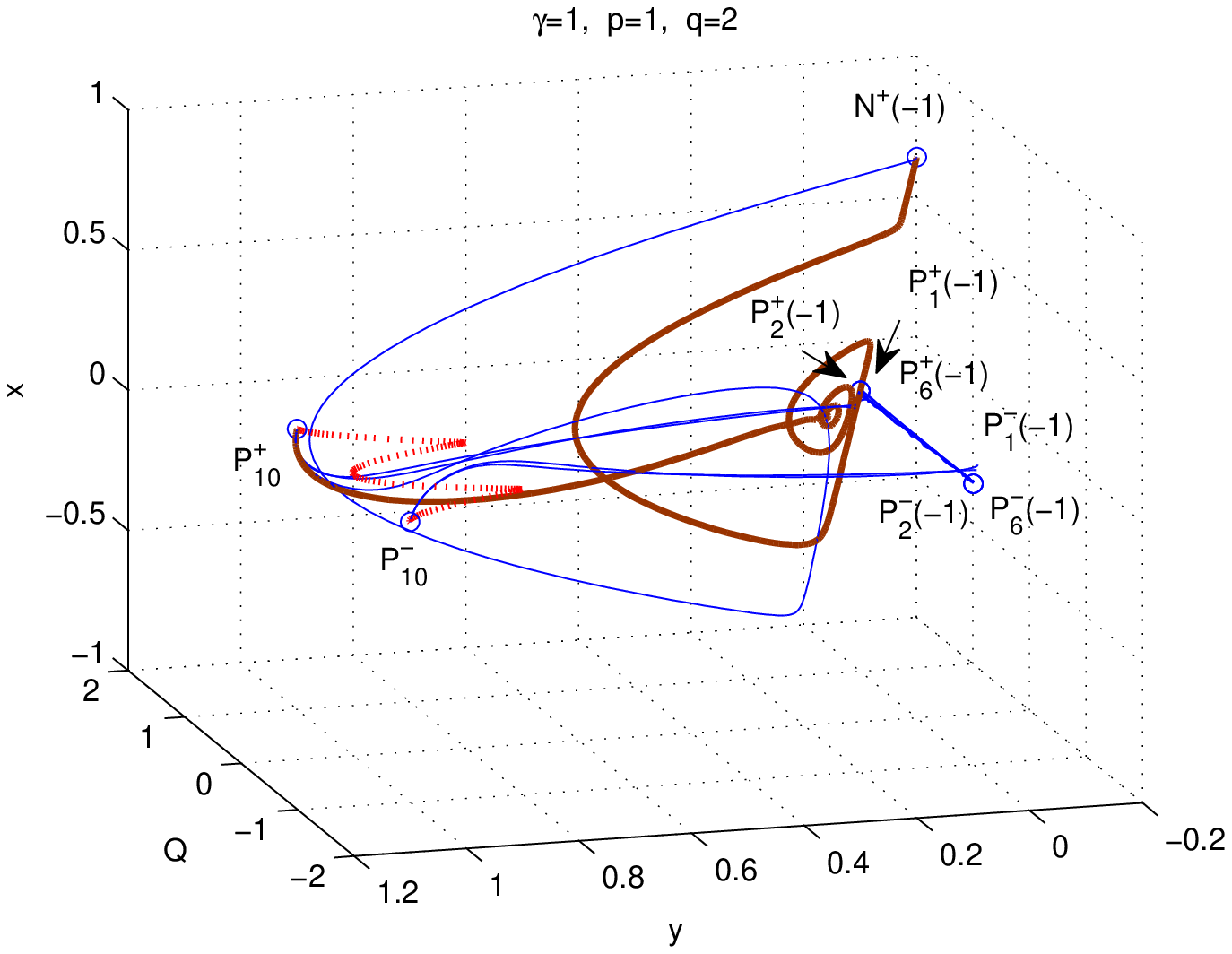}}
\subfigure[]{
\includegraphics[height=5.25cm,width=6cm]{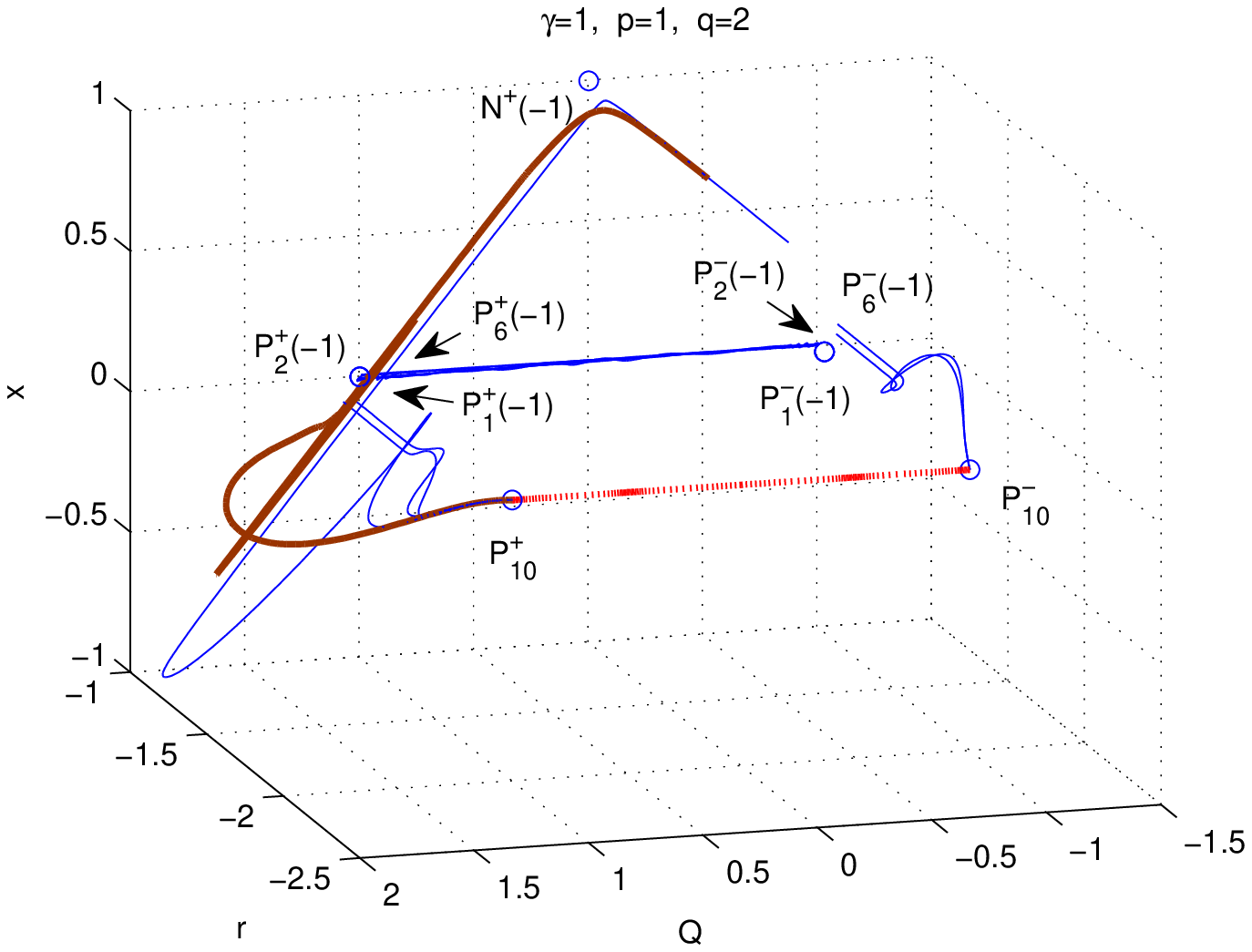}}
\subfigure[]{
\includegraphics[height=5.25cm,width=6cm]{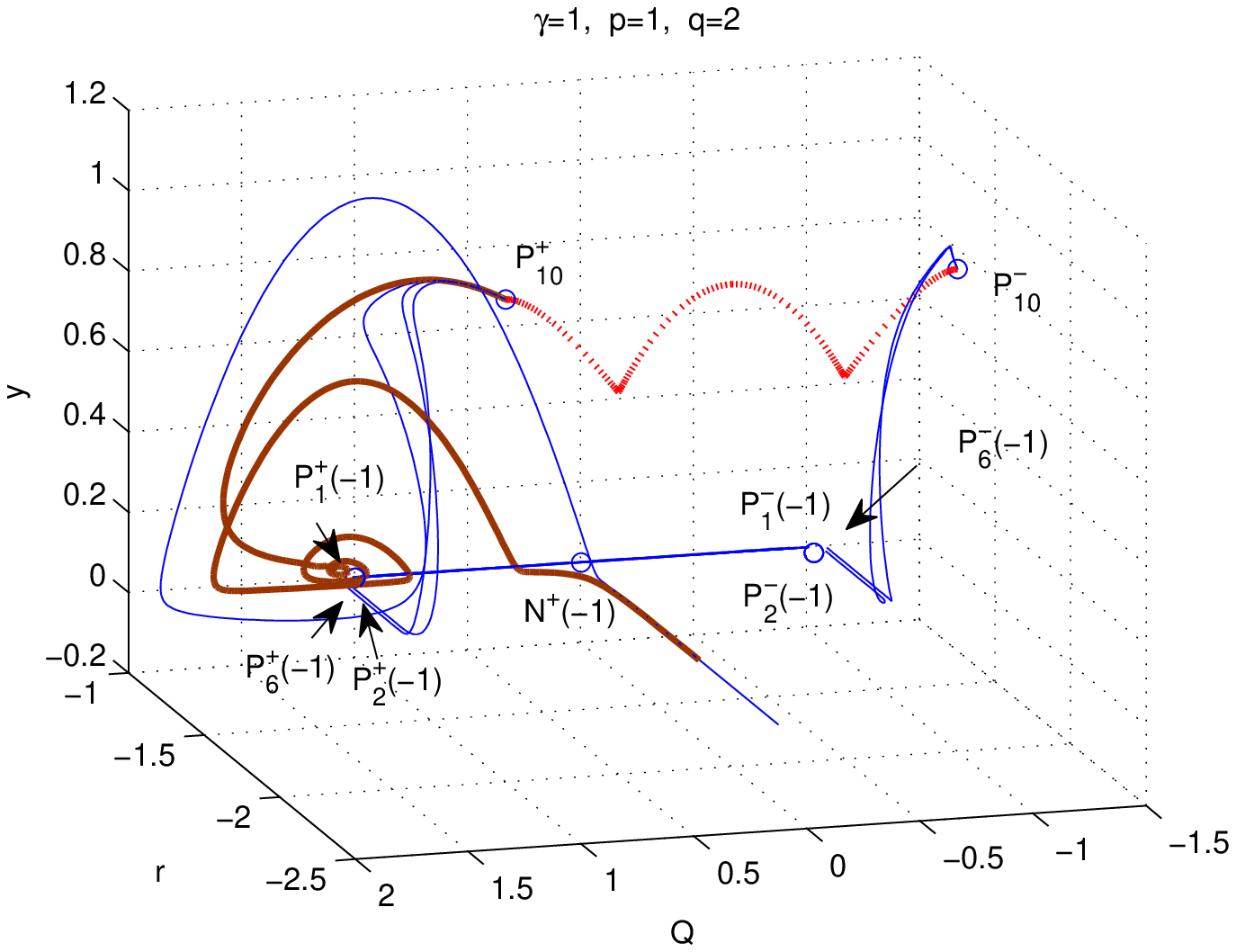}}
\subfigure[]{
\includegraphics[height=5.25cm,width=6cm]{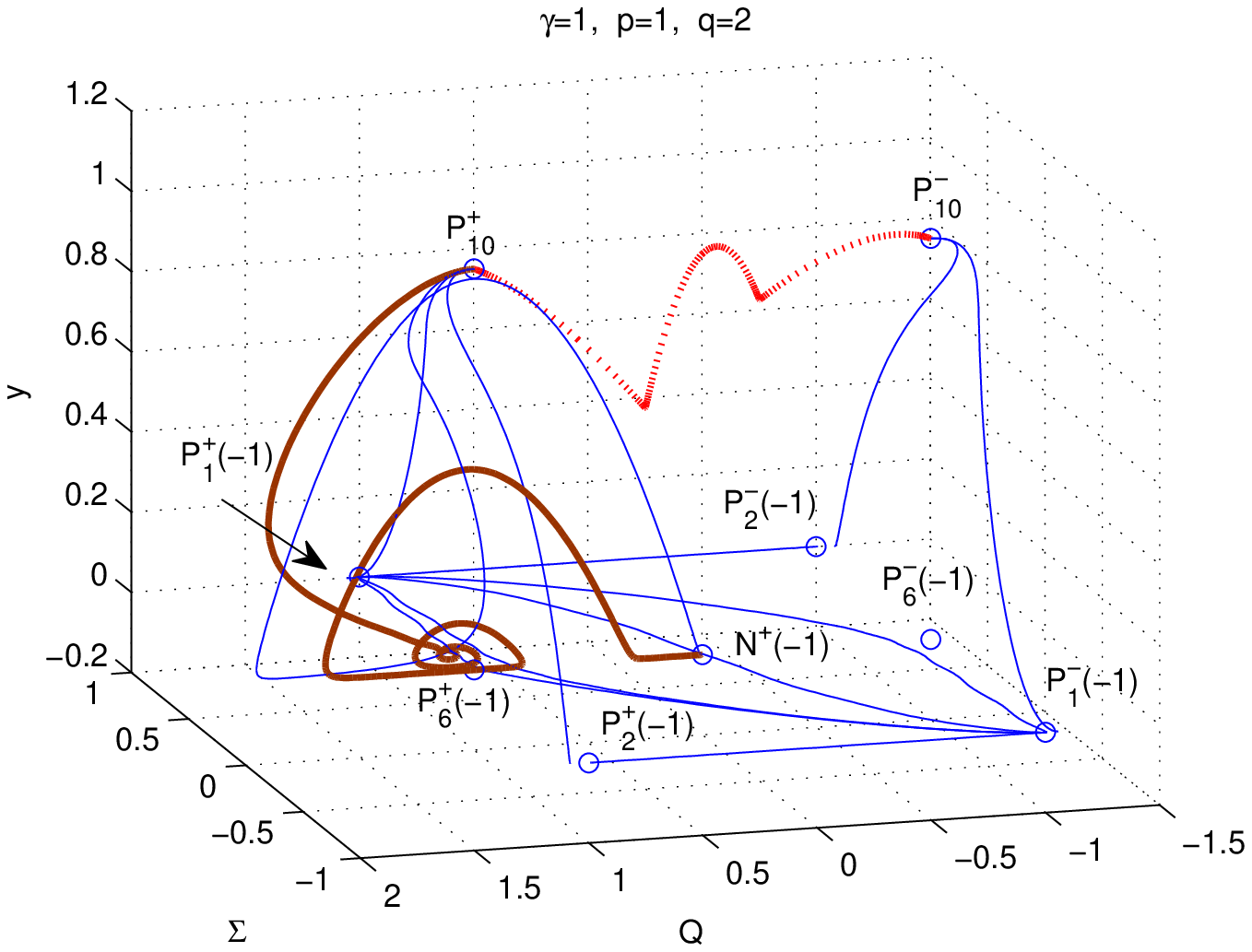}}
\caption{\label{fig8} Some heteroclinic orbits  for the model
$f(R)=R^p\left(\ln \alpha R\right)^q$ for $p=1, q=2$, $\alpha$ arbitrary,
and for dust ($\gamma=1$). The dotted (red) line corresponds to the orbit
joining directly the contracting de Sitter solution $P_{10}^-$ with the
expanding one $P_{10}^+$. The thick (brown) line denotes the orbit that
connects the matter (dust) dominated solution $P_6^+(-1)$ (that belongs to
the region V) with the de Sitter accelerated solution $P_{10}^+$ (that
belongs to region II).}
\end{figure}

\begin{figure}[h]
\centering
\subfigure[]{
\includegraphics[height=5.25cm,width=6cm]{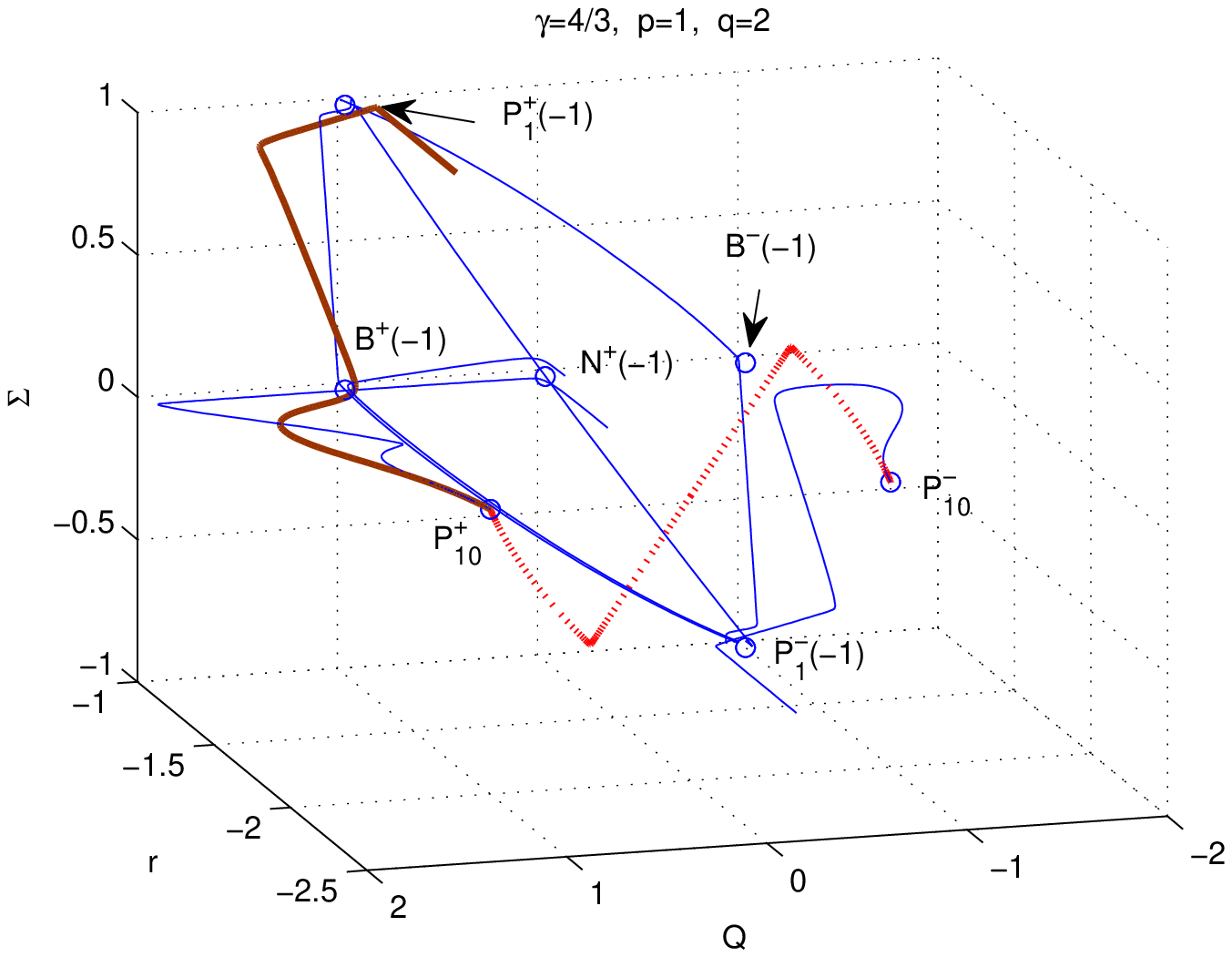}}
\subfigure[]{
\includegraphics[height=5.25cm,width=6cm]{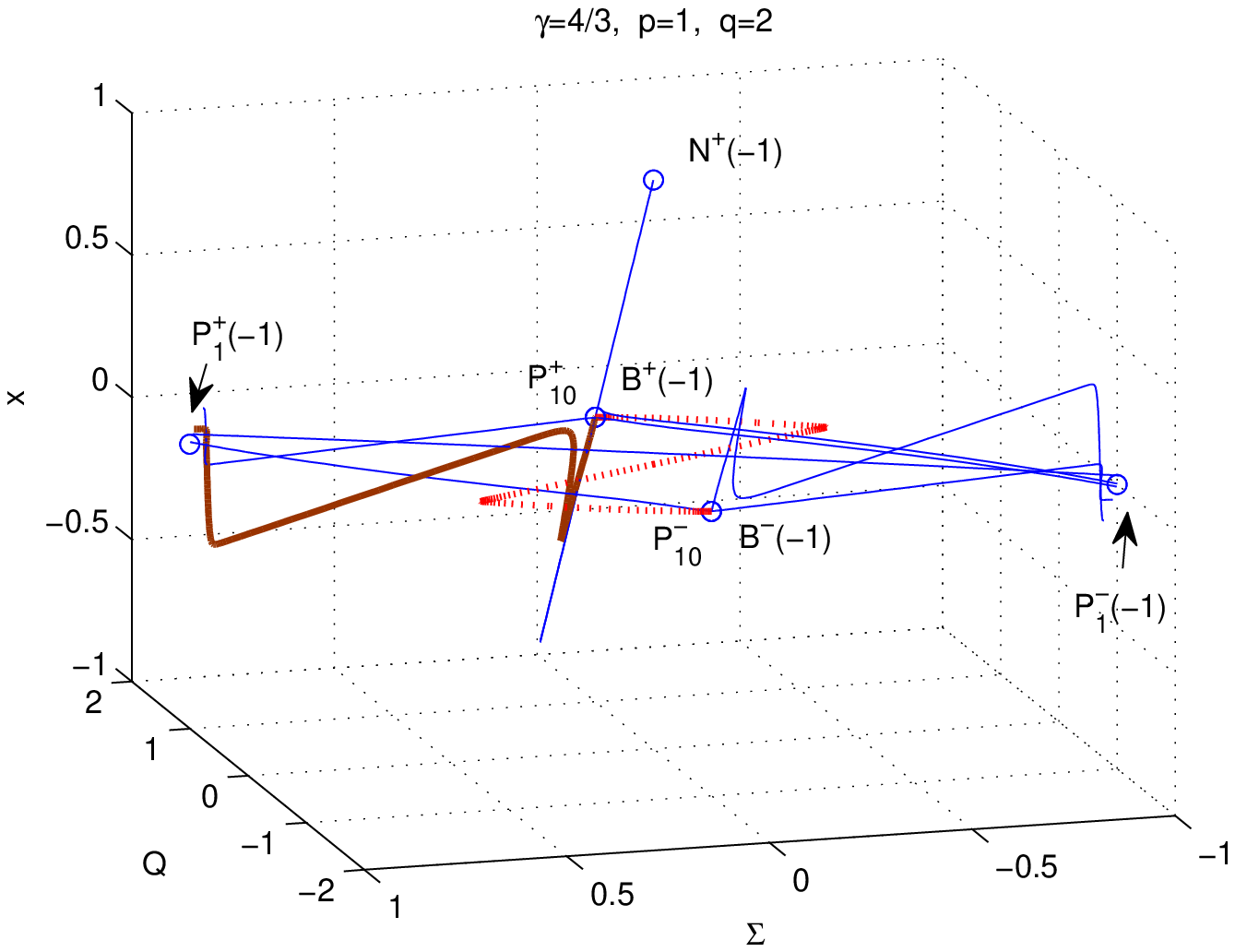}}
\subfigure[]{
\includegraphics[height=5.25cm,width=6cm]{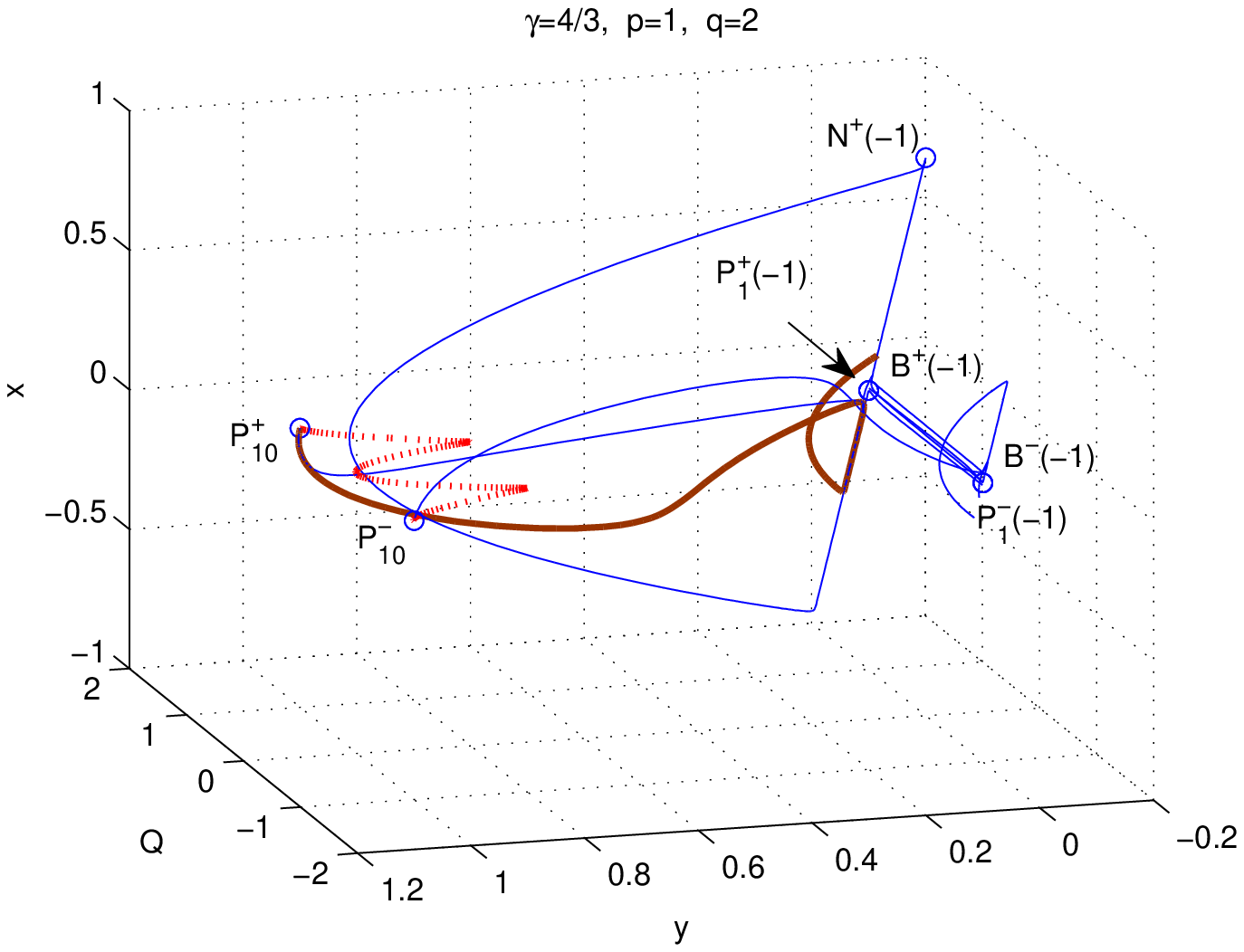}}
\subfigure[]{
\includegraphics[height=5.25cm,width=6cm]{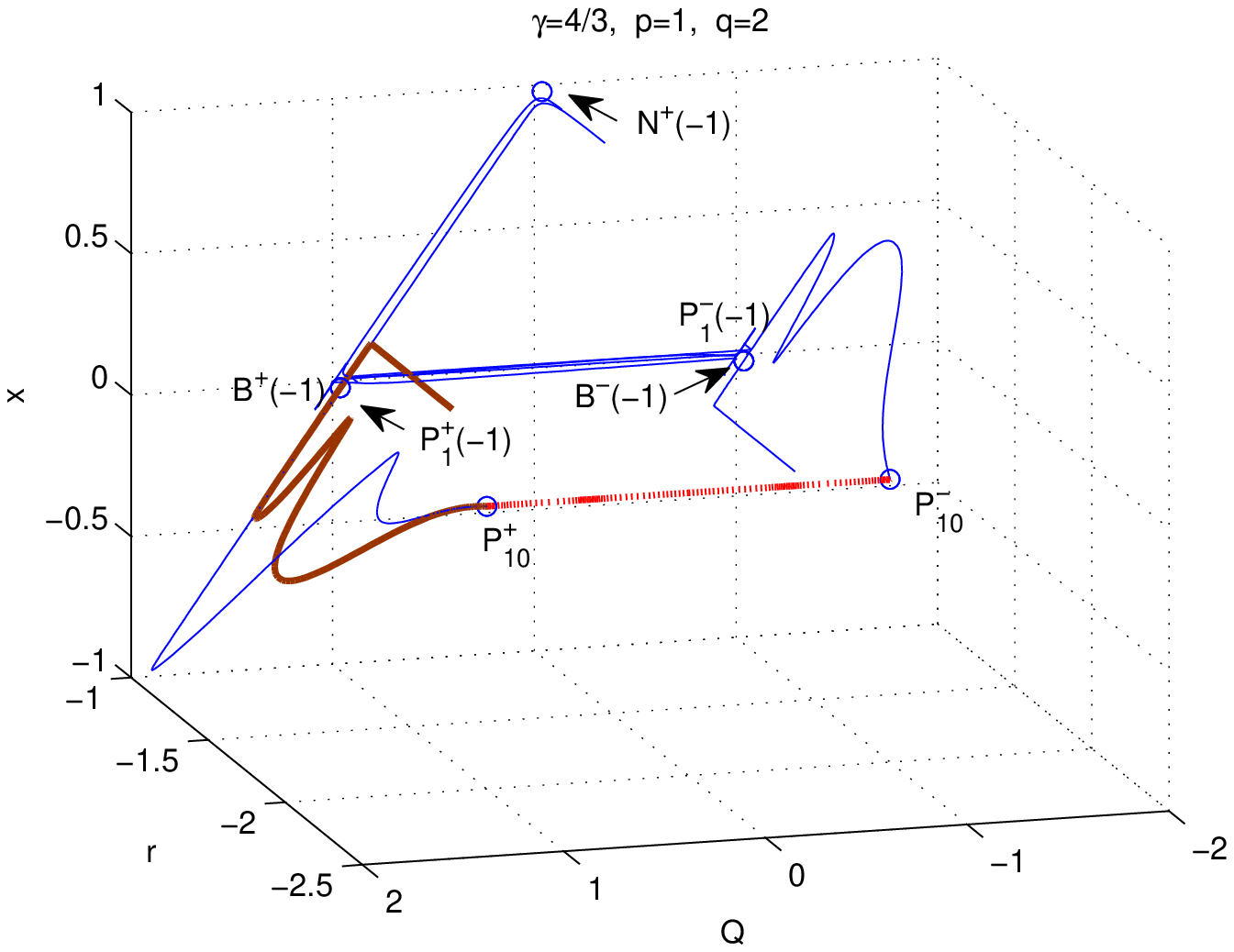}}
\subfigure[]{
\includegraphics[height=5.25cm,width=6cm]{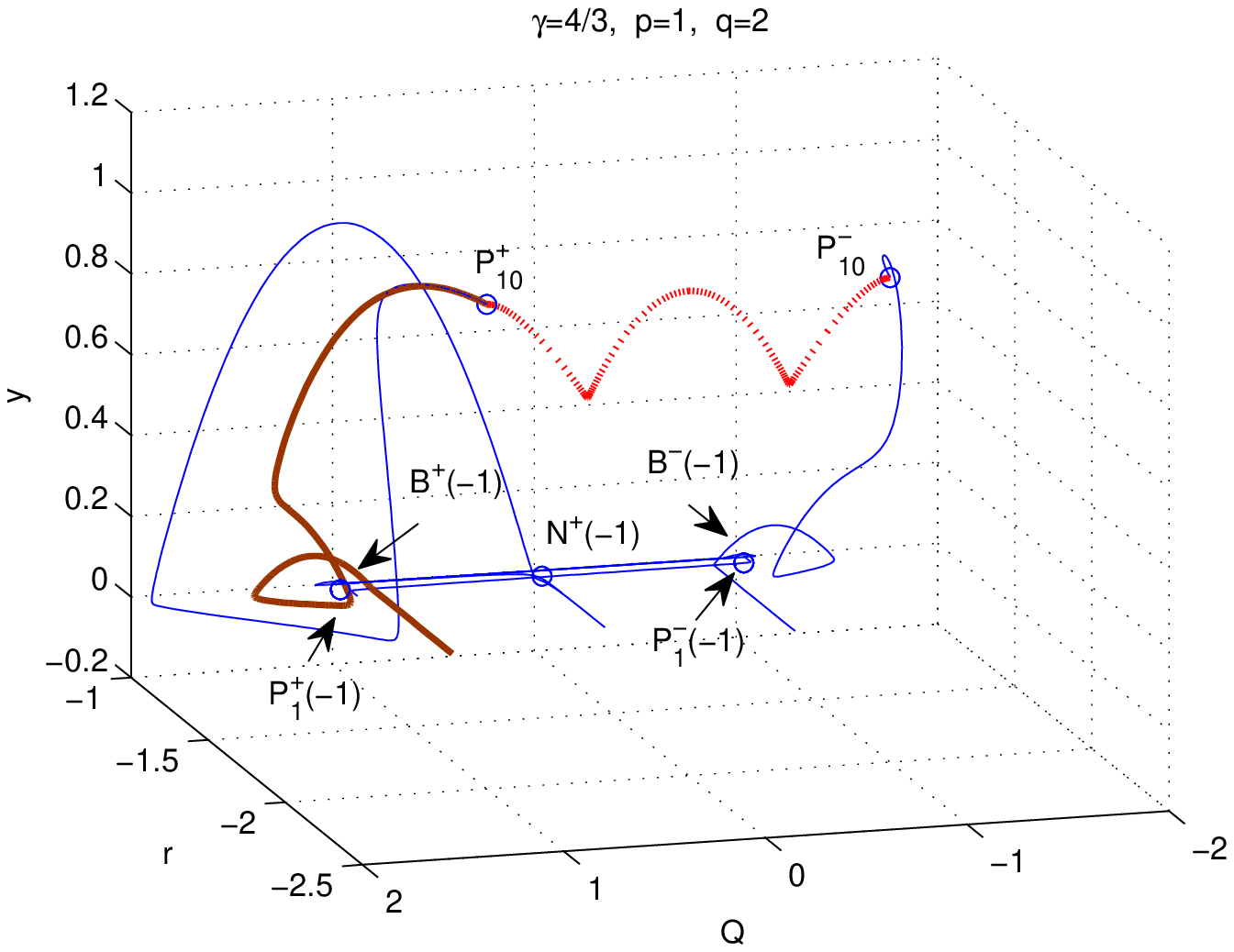}}
\subfigure[]{
\includegraphics[height=5.25cm,width=6cm]{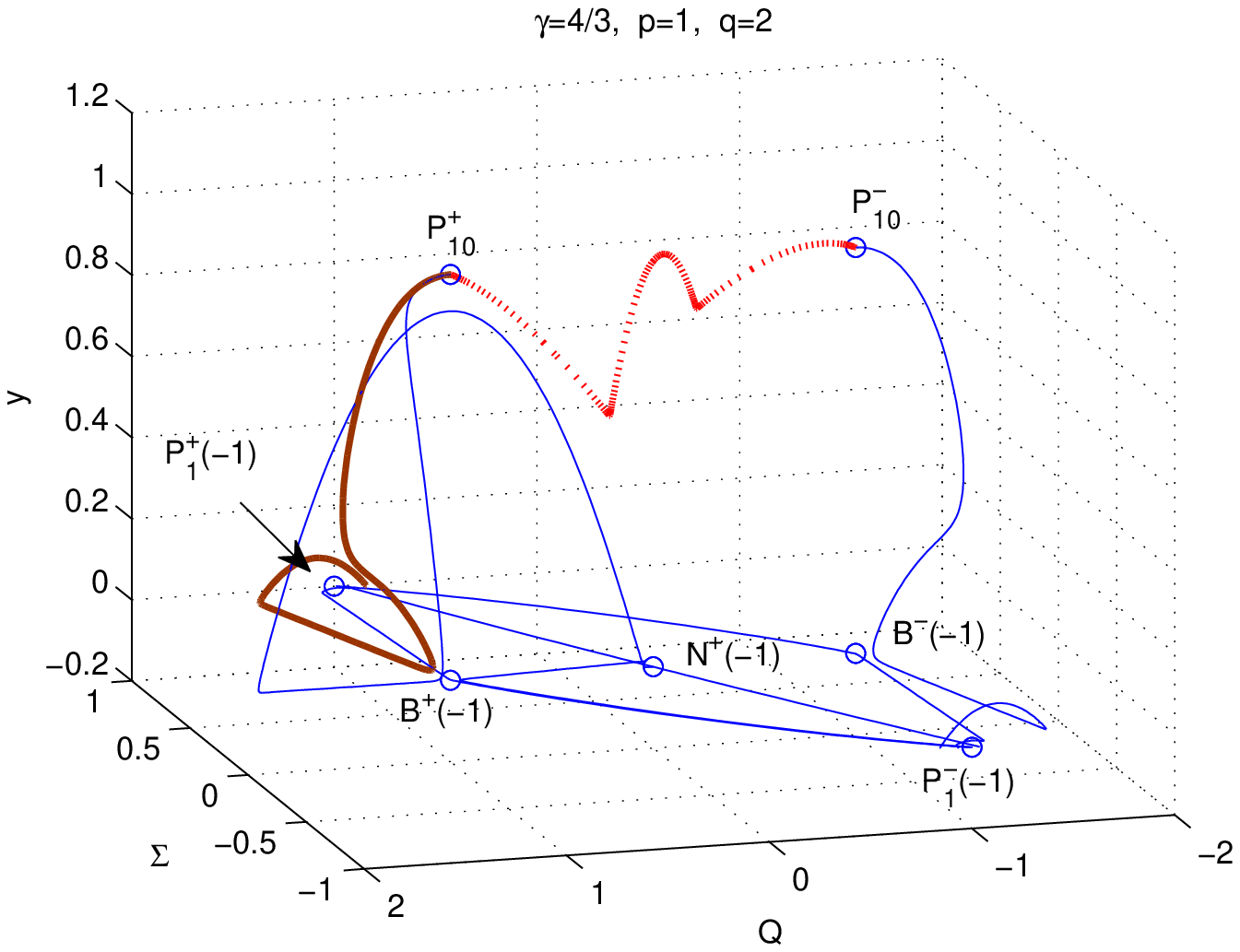}}
\caption{\label{fig9} Some heteroclinic orbits  for the model
$f(R)=R^p\left(\ln \alpha R\right)^q$ for $p=1, q=2$, $\alpha$ arbitrary,
and radiation ($\gamma=4/3$). The dotted (red) line corresponds to the orbit
joining directly the contracting de Sitter solution $P_{10}^-$ with the
expanding one  $P_{10}^+$. The thick (brown) line denotes the orbit that
connects the matter (radiation) dominated solution $B^+(-1)$ (that belongs
to the region VII) with the de Sitter accelerated solution $P_{10}^+$ (that
belongs to region II).}
\end{figure}

The function $M(r)$ satisfy the following:
\begin{itemize}
  \item it connects the matter-dominated region V with the region II
associated to accelerated  de Sitter expansion provided $p\rightarrow\,1$ and
$q>\frac{1}{2}$ (see figure \ref{fig8}). 

  \item it connects the matter-dominated region VII with the region II
associated to accelerated  de Sitter expansion provided
$\gamma=\frac{4}{3},\;p\neq2,\; q>\frac{(-2+p)^{2}}{2}$ (see figure
\ref{fig9}).
\end{itemize}

 In this case, for some values of the parameters, the function $M(r)$
is cosmologically viable. 

In order to present the above results in a transparent way, we perform
several numerical investigations.

%%%%%%%%%%%%%%%%%%%%%%%%%

In the figure \ref{fig8} are presented some orbits for the model
$f(R)=R^p\left(\ln \alpha R\right)^q$ for $p=1, q=2$, $\alpha$ arbitrary, and
for dust ($\gamma=1$).

 Specially are presented the heteroclinic sequences: 
\begin{subequations}
\label{heteroclinicsB.1}
\begin{align}
& P_{1}^{+}(-1)\longrightarrow \left\{\begin{array}{l}
   P_{1}^{-}(-1) \\
   P_{2}^{-}(-1) \\
   P_{6}^{-}(-1)\longrightarrow P_{1}^{-}(-1) \\
   P_{6}^{+}(-1)\longrightarrow P_{1}^{-}(-1) \\  
                          P_{10}^{+} 
                         \end{array} \right. \label{6.8a} \\
& N^{+}(-1)\longrightarrow \left\{\begin{array}{l}
   P_{10}^{+} \\
                          P_{6}^{+}(-1)\longrightarrow P_{10}^{+} 
                         \end{array}    \right. \label{6.8b}\\
& P_{10}^{-}\longrightarrow \left\{\begin{array}{l}
                              P_{10}^{+} \\
                              P_{1}^{-}(-1) \\
     P_{2}^{-}(-1) 
                             \end{array}    \right. \label{6.8c}\\
& P_{2}^{+}(-1)\longrightarrow \left\{\begin{array}{l}
   P_{1}^{-}(-1) \\ 
                          P_{10}^{+} \\
                         \end{array} \right. \label{6.8d}    
\end{align}
\end{subequations}

In figure \ref{fig8} the dotted (red) line corresponds to the orbit joining
directly the contracting de Sitter solution $P_{10}^-$ with the expanding one
 $P_{10}^-$. The thick (brown) line denotes the orbit that connects the
matter (dust) dominated solution $P_6^+(-1)$ (that belongs to the region V)
with the de Sitter accelerated solution $P_{10}^+$ (that belongs to region
II). From the sequences  \eqref{heteroclinicsB.1} follow that the transition
from expansion to contraction is possible (see the first four sequences in
\eqref{6.8a} and the first line of \eqref{6.8d}). Additionally, the
transition from contraction to expansion is also possible as shown in the
first line of the sequence \eqref{6.8c}.

%%%%%%%%%%%%%%%%%%%%%%%%%%%%%%%%
%%%%%%%%%%%%%%%%%%%%%%%%%%%%%%%%%

Finally,  in the figure \ref{fig9} are represented some orbits for the model
$f(R)=R^p\left(\ln \alpha R\right)^q$ for $p=1, q=2$, $\alpha$ arbitrary, and
radiation ($\gamma=4/3$).

 Specially are drawn the heteroclinic sequences:
   \begin{subequations}
 \label{heteroclinicsB.3}
\begin{align}
& P_{1}^{+}(-1)\longrightarrow \left\{\begin{array}{l}
   P_{1}^{-}(-1) \\
   B^{-}(-1)\longrightarrow P_{1}^{-}(-1) \\
   B^{+}(-1)\longrightarrow P_{1}^{-}(-1) \\  
                          B^{+}(-1)\longrightarrow P_{10}^{+} 
                         \end{array} \right. \label{6.9a} \\
& N^{+}(-1)\longrightarrow \left\{\begin{array}{l}
   P_{10}^{+} \\
                          B^{+}(-1)\longrightarrow P_{1}^{-}(-1)
                         \end{array}    \right. \label{6.9b}\\
& P_{10}^{-}\longrightarrow \left\{\begin{array}{l}
                              P_{10}^{+} \\
                              B^{-}(-1)\longrightarrow P_{1}^{-}(-1)
                             \end{array}    \right. \label{6.9c}    
\end{align}
\end{subequations}

In the figure \ref{fig9} the dotted (red) line corresponds to the orbit
joining directly the contracting de Sitter solution $P_{10}^-$ with the
expanding one  $P_{10}^+$. The thick (brown) line denotes the orbit that
connects the matter (radiation) dominated solution $B^+(-1)$ (that belongs to
the region VII) with the de Sitter accelerated solution $P_{10}^+$ (that
belongs to region II). 
From the sequences  \eqref{heteroclinicsB.3} follow that the transition from
expansion to contraction is possible (see the first three sequences in
\eqref{6.9a} and the second line of \eqref{6.9b}). Additionally, the
transition from contraction to expansion is also possible as shown in the
first line of the sequence \eqref{6.9c}.

\section{Concluding Remarks}\label{Sect:7}

In this paper we have investigated from the dynamical systems perspective the
viability of cosmological models based on Kantowski-Sachs metrics for a
generic class of $f(R)$, allowing for a cosmic evolution with an acceptable
matter era, in correspondence to the modern cosmological paradigm. 

Introducing an input function $M(r)$ constructed in terms of the auxiliary
quantities  $m=\frac{R f''(R)}{f'(R)}=\frac{d \ln{f'(R)}}{d \ln{R}}$ and
$r=-\frac{R f'(R)}{f(R)}=-\frac{d \ln{f(R)}}{d \ln(R)},$ and considering very
general mathematical properties such as differentiability, existence of
minima, monotony intervals, etc, we have obtained cosmological solutions
compatible with the modern cosmological paradigm. With the introduction of
the variables $M$ and $r$, one adds an extra direction in the phase-space,
whose neighboring points correspond to ``neighboring'' $f(R)$-functions.
Therefore, after the general analysis has been completed, the substitution of
the specific $M(r)$ for the desired function $f(R)$ gives immediately the
specific results. Is this crucial
aspect of the method the one that make it very powerful, enforcing its
applicability. 

We have discussed which $f(R)$ theories allows for a cosmic evolution with an
acceptable matter era, in correspondence to the modern cosmological paradigm.
We have found a very rich behavior, and amongst others the universe can
result in isotropized solutions with observables in agreement with
observations, such as de Sitter, quintessence-like, or phantom solutions.
Additionally, we find that a cosmological bounce and turnaround are realized
in a part of the parameter-space as a consequence of the metric choice.
Particularly, we found that 
\begin{itemize}
\item $L^+(r^{*})$ is a local repulsor if $1\leq \gamma<\frac{5}{3},
r^{*}<-\frac{5}{4}, M'(r^{*})<0;$ or $1\leq\gamma<\frac{5}{3}, r^{*}>-1,
M'(r^{*})<0.$
\item $N^+(r^{*})$ is a local repulsor if $r^{*}<-1, M'(r^{*})>0;$ or
$r^{*}>-\frac{1}{2}, M'(r^{*})>0.$
\item $A^{-}(r^{*})$ is a repulsor if
$-2<r^{*}<\frac{-1-\sqrt{3}}{2},\;M'(r^{*})>0;$ or $r^{*}<-2,\;M'(r^{*})<0;$
or $r^{*}>\frac{1}{2}(-1+\sqrt{3}),\;M'(r^{*})<0.$  
 \item The point $P_{10}^{-}$ is a repulsor for $M(-2)>0.$
\end{itemize}
Additionally, 
\begin{itemize}
\item $L^-(r^{*})$ is a local attractor if $1\leq \gamma<\frac{5}{3},
r^{*}<-\frac{5}{4}, M'(r^{*})<0;$ or $1\leq\gamma<\frac{5}{3}, r^{*}>-1,
M'(r^{*})<0.$
\item $N^-(r^{*})$ is a local attractor if $r^{*}<-1, M'(r^{*})>0;$ or
$r^{*}>-\frac{1}{2}, M'(r^{*})>0.$
\item $A^{+}(r^{*})$ is a attractor if
$-2<r^{*}<\frac{-1-\sqrt{3}}{2},\;M'(r^{*})>0;$ or $r^{*}<-2,\;M'(r^{*})<0;$
or $r^{*}>\frac{1}{2}(-1+\sqrt{3}),\;M'(r^{*})<0.$  
 \item The point $P_{10}^{+}$  represent a future attractor for $M(-2)>0$ and
if $M(-2)>\frac{9}{8}$, it is a stable focus. In the  special case 
$M(-2)=0,$ $P_{10}^{+}$ coincides with $A^{+}(-2)$ and becomes
non-hyperbolic.
 \end{itemize}
Now, a crucial observation here is that from the critical points enumerated
just before,  the ones that belong to the contracting $\epsilon=-1$ branch,
say $A^{-}(r^{*})$, $L^-(r^{*})$, $N^-(r^{*})$ and $P_{10}^{-}$, have the
reversal stability behavior compared with their analogous $A^{+}(r^{*})$,
$L^+(r^{*})$, $N^+(r^{*})$ and $P_{10}^{+}$, respectively, in the
accelerating $\epsilon=+1$ branch, for the same conditions in the parameter
space. This means that there is a large probability that an orbit initially
at contraction connects an expanding region. 
Of course the real possibility of this kind of solutions depends  on how the
initial points are attracted/repelled by the saddle points with higher
dimensional stable/unstable manifold. 
Some saddle points of physical interest, that may represent intermediate
solutions in the cosmic evolution are:
\begin{itemize}
 \item $A^{+}(r^{*})$ is saddle with stable manifold 4D if
%\begin{footnotesize}
   \begin{eqnarray}
   -2<r^{*}&<&\frac{-1-\sqrt{3}}{2},M'(r^{*})<0\;\;\text{or}
\nonumber\\
   r^{*}&<&-2,M'(r^{*})>0 \;\;\text{or}\nonumber\\
   r^{*}&>&\frac{-1+\sqrt{3}}{2},M'(r^{*})>0
\;\;\text{or}\nonumber\\
   \frac{5}{3}<\gamma\leq2,
\frac{1}{2}(-1-\sqrt{3})<r^{*}&<&-\frac{5}{4},M'(r^{*})>0\;\;\text{or}
\nonumber\\
   1\leq\gamma\leq\frac{5}{3},
\frac{1}{2}(-1-\sqrt{3})<r^{*}&<&\frac{-4-9\gamma}{4(1+3\gamma)}-\frac{1}{4}
\sqrt{\frac{16+48\gamma+9\gamma^{2}}{(1+3\gamma)^{2}}},M'(r^{*})>0\;\;\text{
or}\nonumber\\
\frac{-4-9\gamma}{4(1+3\gamma)}+\frac{1}{4}\sqrt{\frac{16+48\gamma+9\gamma^{2
}}{(1+3\gamma)^{2}}}<r^{*}&<&\frac{1}{2}(-1+\sqrt{3}),M'(r^{*})<0.\nonumber
     \end{eqnarray}
%\end{footnotesize}
\item $B^{+}(r^{*})$ is saddle with stable manifold 4D if
%\begin{footnotesize}
 \begin{eqnarray}
 1\leq\gamma&<&\frac{4}{3},
-1<r^{*}<-\frac{3\gamma}{4},M'(r^{*})<0\;\;\text{or} \nonumber\\
   \frac{4}{3}<\gamma&<&\frac{5}{3},
-\frac{3\gamma}{4}<r^{*}<-1,M'(r^{*})<0.\nonumber
 \end{eqnarray}
%\end{footnotesize}
\item $P_{4}^{+}(r^{*})$ is saddle with stable manifold 4D if
%\begin{footnotesize}
   \begin{eqnarray}
   -2<r^{*}&<&\frac{-1-\sqrt{3}}{2},M'(r^{*})>0\;\;\text{or}
\nonumber\\
   r^{*}&<&-2,M'(r^{*})<0 \;\;\text{or}\nonumber\\
   r^{*}&>&\frac{-1+\sqrt{3}}{2},M'(r^{*})<0.\nonumber
   \end{eqnarray}
%\end{footnotesize}
\end{itemize}

Furthermore, we have presented a reconstruction method for the $f(R)$-theory
given the $M(r)$ function. This procedure was introduced first in the
isotropic context in the reference \cite{Amendola:2006we}. However, in this
paper  we have formalized and extended the geometric procedure discussed in
\cite{Amendola:2006we} in such way that the problems cited in
\cite{Carloni:2007br} do not arise,  and we have applied  the procedure to
``generic'' $f(R)$ models for the case of a Kantowski-Sachs metric.  

Summarizing, in this paper we have extended the results in
\cite{Amendola:2006we,Leon:2010ai,Leon:2010pu} to the Kantowski-Sachs metric.
Additionally, we extent the results obtained in the reference
\cite{Leon:2010ai} related to the stability analysis for the de Sitter
solution (with unbounded scalar field) for the homogeneous but anisotropic
Kantowski-Sachs metric and we have extended to generic  $f(R)$ models the
results in \cite{Leon:2010pu} obtained for $R^n$-cosmologies. Our results are
also in agreement with the related ones in \cite{Shabani:2013djy} for the
choice $f(R,T)=g(R)+c_1 \sqrt{-T}+c_2$ in the isotropic regime. However, our
results are more general since we consider also anisotropy. Finally, we have
presented several heteroclinic sequences for four classes of $f(R)$ models
showing the transition from an expanding phase to a contracting one, and
viceversa. So the realization of a bounce or a turnaround indeed occurs. In
fact, if the universe start from an expanding initial conditions and 
result in a contracting phase then we have the realization of a cosmological
turnaround, while if it start from contracting initial conditions and result
in an expanding phase then we have the realization of a cosmological bounce.
These behaviors were known to be possible in Kantowski-Sachs geometry
\cite{Leon:2010pu,Fadragas:2013ina}. We argue that this behaviors are not
just mathematical elaborations that are possible for some specific classes of
$f(R)$ models, but they are generic features of a Kantowski-Sachs scenario.

\begin{acknowledgments}
The authors wish to thank Emmanuel N. Saridakis and  Carlos R. Fadragas
for useful comments.
G. L. wishes to thank his colleagues Joel Saavedra, and Yoelsy Leyva because
enlightening discussions, and he also thanks his colleagues at Instituto de
F\'{\i}sica, Pontificia Universidad  Cat\'{o}lica de Valpara\'{\i}so for
providing a warm working environment. A. A. Roque wishes to thank his
colleagues at Grupo de Estudios Avanzados (GEA) at Universidad de Cienfuegos
for their support and advice. G.L. was partially supported by PUCV through
Proyecto DI Postdoctorado 2013, by COMISI\'ON NACIONAL DE CIENCIAS Y
TECNOLOG\'IA through Proyecto FONDECYT DE POSTDOCTORADO 2014  grant  3140244
and by DI-PUCV grant 123.730/2013. A. A. Roque acknowledges the Ministerio
de Educaci\'on Superior (MES) from Cuba for partial financial support.

\end{acknowledgments}

\begin{appendix}

\section*{Appendix}

\section{The curves of fixed points $C^{\pm}(r^{*}).$}

In this appendix we present the stability results for the curves of fixed
points $C^{\pm}(r^{*}).$
In order to be more transparent in the stability analysis, let us consider
the parametrization:
\begin{displaymath}
 C^{\pm}(r^{*}) = \left\{ \begin{array}{lll}
Q= \pm\,1 + \sin u \\
\Sigma=\cos u & \;\;\; u\in[0,2\pi]\\
x= - \sin u 
\end{array} \right.,
\end{displaymath}
resulting that the eigenvalues of the linearization of  $C^{\pm}(r^{*})$
are:
\begin{eqnarray}
 \{0&,&(\pm\,4-2\cos u),\;\pm\,(6-3\gamma)+(4-3\gamma)\sin u,
\nonumber\\
 &&  2\left(\pm\,3+\left[1+\frac{1}{1+r^{*}}\right]\,\sin
u\right),\;-2M'(r^{*})\sin u \}.
\end{eqnarray}

Now, let us assume that  $M'(r^{*})\neq 0$ and that no other eigenvalue
vanishes. Then, since the eigenvector associated to the only zero eigenvalue
is tangent to the equilibrium curve, the curves are actually normally
hyperbolic \cite{BAulbach1981} and therefore the stability can be determined 
by the signs of the real parts of the nonzero eigenvalues.
Thus, the curve $C^{+}(r^{*})$ have a 4D unstable manifold (and it is
unstable) for either
\begin{itemize}
\item $\pi<u<2\pi,\,r^{*}<-1,\,M'(r^{*})>0$ or
\item $\pi<u<2\pi,\,r^{*}>\frac{3}{3+\sin u}- 2,\,M'(r^{*})>0$ or 
\item $0<u<\pi,\,\gamma \neq 2,r^{*}>-1,\,u\neq\frac{\pi}{2},\,M'(r^{*})<0$
or
\item $0<u<\pi,\,\gamma \neq
2,\,\gamma\neq\frac{5}{3},\,r^{*}>-1,\,M'(r^{*})<0$ or
\item $0<u<\pi,\,\gamma>\frac{2}{3}(2+\frac{1}{1+\sin u}),\,\gamma \neq
2,\,r^{*}>-1,\,M'(r^{*})<0$ or 
\item $0<u<\pi,\,\gamma\neq 2,r^{*}<\frac{3}{3+\sin u}-
2,\,u\neq\frac{\pi}{2},\,M'(r^{*})<0$ or
\item $0<u<\pi,\,\gamma \neq
2,\,\gamma\neq\frac{5}{3},\,r^{*}<\frac{3}{3+\sin u}- 2,\,M'(r^{*})<0$ or
\item $0<u<\pi,\,\gamma>\frac{2}{3}(2+\frac{1}{1+\sin u}),\,\gamma \neq
2,\,r^{*}<\frac{3}{3+\sin u}- 2,\,M'(r^{*})<0.$
\end{itemize}
On the other hand, $C^{-}(r^{*})$ have a 4D stable manifold (so, it is
stable) for either 
\begin{itemize}
\item $0<u<\pi,\,r^{*}>\frac{-3}{-3+\sin u}- 2,\,M'(r^{*})>0$ or 
\item $0<u<\pi,\,r^{*}<-1,\,M'(r^{*})>0$ or 
\item $\pi<u<2\pi,\,\gamma \neq
2,\,r^{*}>-1,\,u\neq\frac{3\pi}{2},\,M'(r^{*})<0$ or 
\item $\pi<u<2\pi,\,\gamma \neq
2,\,\gamma\neq\frac{5}{3},\,r^{*}>-1,\,M'(r^{*})<0$ or 
\item $\pi<u<2\pi,\,\gamma>\frac{4}{3}-\frac{2}{3-3\sin u},\,\gamma \neq
2,\,r^{*}>-1,\,M'(r^{*})<0$ or 
\item $\pi<u<2\pi,\,\gamma \neq 2,\,r^{*}<\frac{-3}{-3+\sin u}-
2,\,u\neq\frac{3\pi}{2},\,M'(r^{*})<0$ or 
\item $\pi<u<2\pi,\,\gamma \neq
2,\,\gamma\neq\frac{5}{3},\,r^{*}<\frac{-3}{-3+\sin u}- 2,\,M'(r^{*})<0$ or 
\item $\pi<u<2\pi,\,\gamma>\frac{4}{3}-\frac{2}{3-3\sin u},\,\gamma \neq
2,\,r^{*}<\frac{-3}{-3+\sin u}- 2,\,M'(r^{*})<0.$ 
\end{itemize}

Among the curves of $C^{\pm}(r^{*})$ we can find the special critical points
$L^{+}(r^{*})=C^{+}(r^{*})|_{u=\frac{\pi}{2}}$,
\mbox{$N^{-}(r^{*})=C^{-}(r^{*})|_{u=\frac{\pi}{2}}$,}
$L^{-}(r^{*})=C^{-}(r^{*})|_{u=\frac{3\pi}{2}}$,
$N^{+}(r^{*})=C^{+}(r^{*})|_{u=\frac{3\pi}{2}}$,
$P_{1}^{+}(r^{*})=C^{+}(r^{*})|_{u=0}$,
$P_{1}^{-}(r^{*})=C^{-}(r^{*})|_{u=\pi}$ y
$P_{2}^{+}(r^{*})=C^{+}(r^{*})|_{u=\pi}$,
$P_{2}^{-}(r^{*})=C^{-}(r^{*})|_{u=0}.$ Their stability conditions read from
the cases discussed above. Now, for $R^n$ cosmology are recovered the
stability properties discussed in \cite{Leon:2010pu} for the special $n$-$w$
relation discussed there. As commented before, for $R^n$-gravity the function
$r$ is a constant and it is not required to examine the stability along the
$r$-direction. 

\section{Stability analysis of the \emph{de Sitter} solution in Quadratic
Gravity}\label{stabilityAplus}

In order to analyze the stability of the \emph{de Sitter} solution in
Quadratic Gravity ($A^+(-2)$) we use the Center Manifold to prove that this
solution is locally asymptotically unstable (saddle type) irrespectively the
value of $\gamma$. We proceed as follows.

{\bf Case $\gamma\neq 1$}. Let's introduce the new variable  $(u,v_1,v_2,v_3,
v_4)\equiv\mathbf{x}$ defined by 
\begin{subequations}\label{lin1}
\begin{align} & u=r+2, \\& v_1=\frac{r}{3}+x+\frac{y (4-3 \gamma )}{6-6
\gamma }+\frac{\gamma }{6 (\gamma -1)}, \\&v_2= -2 Q-2
   x+\Sigma +2, \\&v_3=y-1, \\&v_4=2 Q+2 x-2.\end{align}
\end{subequations} 

Taylor-expanding  the system \eqref{Sistema00} with $M(r)=\frac{1}{2}r(r+2),$
around $A^+(-2)$ up to fourth order in the vector norm and applying the
linear transformation \eqref{lin1} we obtain
\begin{equation}
\left(\begin{array}{c}u'\\v_1'\\v_2'\\v_3'\\v_4'
\end{array}\right)=\left(\begin{array}{ccccc}0& 0 &0 &0&0\\0& -3 &0 &0&0\\0&
0 &-3&0&0\\0& 0 &0 &-3\gamma&0\\0& 0 &0 &0&-2
\end{array}\right)\left(\begin{array}{c}u\\v_1\\v_2\\v_3\\v_4
\end{array}\right)+\left(\begin{array}{c}f(u,v_1,v_2,v_3,v_4)\\g_1(u,v_1,v_2,
v_3,v_4)\\g_2(u,v_1,v_2,v_3,v_4)\\g_3(u,v_1,v_2,v_3,v_4)\\g_4(u,v_1,v_2,v_3,
v_4)\end{array}\right)\label{center2}
\end{equation}
where the second term in the right hand side of \eqref{center2} are
homogeneous polynomials of order $r\geq 2$ truncated up to fourth order. 
Then, the system \eqref{center2} can be written in diagonal form as 
\begin{subequations}\label{center3}
\begin{align}
u'  &  =Cu+f\left(u,\mathbf{v}\right),\\
\mathbf{v}'  &  =P\mathbf{v}+\mathbf{g}\left(u,\mathbf{v}\right),
\end{align}
\end{subequations}
where $\left( u,\mathbf{v}\right)  \in\mathbb{R}\times\mathbb{R}^{4},$ $C$ is
 $1\times1$ zero matrix, $P$ is a $4\times4$ matrix with negative
eigenvalues, $f,\mathbf{g}$ is zero at $\mathbf{0}$ and have zero derivative
(Jacobian) matrix at $\mathbf{0.}$ The Center Manifold theorem asserts that
there exists a local invariant 1-dimensional center manifold  
$W^{c}\left(\mathbf{0}\right) $ of \eqref{center3} tangent at the origin to
the center subspace. $W^{c}\left(
\mathbf{0}\right)  $ can be represented by the graph 
\[
W^{c}\left(  \mathbf{0}\right)  =\left\{  \left(u,\mathbf{v}\right)
\in\mathbb{R}\times\mathbb{R}^{4}:\mathbf{v}=\mathbf{h}\left(u\right)
,\;\left\vert u\right\vert <\delta\right\}  ;\;\;\;\mathbf{h}\left(  0\right)
=\mathbf{0},\;D\mathbf{h}\left(  0\right)  =\mathbf{0},
\]
for $\delta$ small enough. The restriction of the flow of 
(\ref{center3}) to the center manifold is given by 
\begin{equation}
x'=f\left( x,\mathbf{h}\left(  x\right)  \right)  . \label{rest}
\end{equation}
According tho the Center Manifold theorem, if the origin $u=0$ of
\eqref{rest}
is stable (asymptotically stable) (unstable) then the origin of
\eqref{center3} is also stable (asymptotically stable) (unstable). 
Hence, to find the local center manifold is equivalent to obtain
$\mathbf{h}\left(u\right).$

Substituting  $\mathbf{v}=\mathbf{h}\left(u\right)$ in the second component
of \eqref{center3} and using the chain rule, $\mathbf{v
}'=D\mathbf{h}\left( u\right) u'$ it can be proved that the function
$\mathbf{h}\left( u\right)$ defining the local center manifold is the
solution of the quasi-linear partial differential equation 
\begin{equation}
D\mathbf{h}\left( u\right)  \left[  f\left( u,\mathbf{h}\left( u\right)
\right)  \right]  -P\mathbf{h}\left(  u\right)  -\mathbf{g}\left(
u,\mathbf{h}\left(u\right)  \right)  =0. \label{h}
\end{equation}
The equation \eqref{h} can be solved with good accuracy Taylor expanding the
function  $\mathbf{h}\left(u\right)  $
around $u=0.$ Since $\mathbf{h}\left(  0\right)  =\mathbf{0\ }
$ y $D\mathbf{h}\left(  0\right)  =\mathbf{0},$ is obvious that
$\mathbf{h}\left(u\right)  $ must start with quadratic terms. Substituting
\[
\mathbf{h}\left( u\right)  =:\left[
\begin{array}
[c]{c}%
h_{1}\left(u\right) \\
h_{2}\left(u\right) \\
h_{3}\left(u\right)\\
h_{4}\left(u\right)
\end{array}
\right]  =\left[
\begin{array}
[c]{c}%
a_{1}u^{2}+a_{2}u^{3}+O\left(u^{4}\right) \\
b_{1}u^{2}+b_{2}u^{3}+O\left(u^{4}\right) \\
c_{1}u^{2}+c_{2}u^{3}+O\left(u^{4}\right)
\\
d_{1}u^{2}+d_{2}u^{3}+O\left(u^{4}\right)
\end{array}
\right]
\]
in \eqref{h} and comparing coefficients with equal powers of  $u$ with zero
we obtain $a_1=\frac{1}{54} \left(\frac{1}{\gamma -1}-17\right),
a_2=\frac{1}{243} \left(\frac{7}{\gamma
 -1}-69\right),b_1=0,b_2=0,c_1=-\frac{1}{9},c_2=-\frac{14}{81},d_1=0,d_2=0.$
Henceforth, \eqref{rest} leads to the differential equation 
\begin{equation} u'=\frac{2 u^2}{3}+\frac{5 u^3}{27}+\frac{11
u^4}{81}+O\left(u^5\right).\label{rest1}\end{equation}
From \eqref{rest1} is deduced the the origin $u=0$  is locally asymptotically
unstable (saddle point). Then, the origin $\mathbf{x}=\mathbf{0}$ of the 
5-dimensional system is unstable (saddle type).

{\bf Case $\gamma=1$}. Introducing the new variables  $(u,v_1,v_2,v_3,
v_4)\equiv\mathbf{x}$ given by 
\begin{subequations}\label{lin2}
\begin{align}& u=r+2, \\& v_1=\frac{r}{3}+x, \\&v_2= -2 Q-2
   x+\Sigma +2, \\&v_3=-\frac{y-1}{2},\\&v_4=2 Q+2 x-2,\end{align}
\end{subequations} 
   
Taylor-expanding  the system \eqref{Sistema00} with $M(r)=\frac{1}{2}r(r+2),$
around $A^+(-2)$ up to fourth order in the vector norm and applying the
linear transformation \eqref{lin2} we obtain
\begin{equation}
\left(\begin{array}{c}u'\\v_1'\\v_2'\\v_3'\\v_4'
\end{array}\right)=\left(\begin{array}{ccccc}0& 0 &0 &0&0\\0& -3 &0 &0&0\\0&
0 &-3&0&0\\0& 0 &0 &-3&0\\0& 0 &0 &0&-2
\end{array}\right)\left(\begin{array}{c}u\\v_1\\v_2\\v_3\\v_4
\end{array}\right)+\left(\begin{array}{c}f(u,v_1,v_2,v_3,v_4)\\g_1(u,v_1,v_2,
v_3,v_4)\\g_2(u,v_1,v_2,v_3,v_4)\\g_3(u,v_1,v_2,v_3,v_4)\\g_4(u,v_1,v_2,v_3,
v_4)\end{array}\right)\label{center2a}
\end{equation}
where

$f(u,v_1,v_2,v_3,v_4)=-\frac{u^3}{3}+u^2 v_1+\frac{2 u^2}{3}-2 u v_1+O(4),$

$g_1(u,v_1,v_2,v_3,v_4)=-\frac{7 u^3}{6}+\frac{5 u^2 v_1}{6}+2 u^2
v_3-\frac{5 u^2}{6}-\frac{3 u
   v_1^2}{2}-\frac{u v_1}{3}-\frac{u v_2^2}{2}-\frac{2 u v_2
   v_4}{3}-\frac{u v_3 v_4}{2}+u v_3-\frac{u v_4^2}{6}+\frac{u
   v_4}{2}+\frac{3 v_1^3}{2}-\frac{v_1^2}{2}+\frac{3 v_1 v_2^2}{2}+2
   v_1 v_2 v_4+\frac{3 v_1 v_3 v_4}{2}+3 v_1
   v_3+\frac{v_1 v_4^2}{2}-\frac{3 v_1
   v_4}{2}-\frac{v_2^2}{2}+\frac{3 v_2 v_3 v_4}{2}-v_2
   v_4+v_3-\frac{v_4^2}{2}+O(4),$

$g_2(u,v_1,v_2,v_3,v_4)=\frac{u^2 v_2}{6}-\frac{u^2 v_4}{6}-u v_1 v_2+u v_1
v_4+\frac{3
   v_1^2 v_2}{2}-\frac{3 v_1^2 v_4}{2}+\frac{3
   v_2^3}{2}+\frac{v_2^2 v_4}{2}+3 v_2 v_3-2 v_2
   v_4^2+\frac{v_2 v_4}{2}-3 v_3 v_4-v_4^3+\frac{5
   v_4^2}{4}+O(4),$

$g_3(u,v_1,v_2,v_3,v_4)=\frac{u^3}{3}-u^2 v_1-\frac{u^2 v_3}{3}-\frac{u^2
v_4}{12}+\frac{u^2}{6}+\frac{u
   v_1 v_4}{2}+3 v_1^2 v_3-\frac{3 v_1^2 v_4}{4}-\frac{3
   v_1^2}{2}+3 v_2^2 v_3-\frac{3 v_2^2 v_4}{4}-\frac{3
   v_2^2}{2}-\frac{5 v_2 v_4^2}{4}-2 v_2 v_4+3 v_3^2
   v_4+6 v_3^2+v_3 v_4^2-\frac{3 v_3
   v_4}{2}-\frac{v_4^3}{2}-\frac{v_4^2}{2}+O(4),$  and

$g_4(u,v_1,v_2,v_3,v_4)=\frac{u^2 v_4}{3}-2 u v_1 v_4+3 v_1^2 v_4+3 v_2^2
   v_4+\frac{9 v_2 v_4^2}{2}-2 v_2 v_4+\frac{3 v_3
   v_4^2}{2}+6 v_3 v_4+\frac{3 v_4^3}{2}-\frac{5 v_4^2}{2}+O(4).$
\\
Using the same approach as before we obtain
$a_1=-\frac{7}{27},a_2=-\frac{16}{81},b_1=0,b_2=0,c_1=\frac{1}{18},c_2=\frac{
7}{81}, d_1=0, d_2=0$ for the coefficients of 
\[
\mathbf{h}\left(  x\right)  =:\left[
\begin{array}
[c]{c}%
h_{1}\left(u\right) \\
h_{2}\left(u\right) \\
h_{3}\left(u\right)
\\
h_{4}\left(u\right)
\end{array}
\right]  =\left[
\begin{array}
[c]{c}%
a_{1}u^{2}+a_{2}u^{3}+O\left(u^ {4}\right) \\
b_{1}u^{2}+b_{2}u^{3}+O\left(u^{4}\right) \\
c_{1}u^{2}+c_{2}u^{3}+O\left(u^{4}\right)
\\
d_{1}u^{2}+d_{2}u^{3}+O\left(u^{4}\right)
\end{array}
\right].
\] Substituting the values of the constants  $a_{1},b_{1},c_{1},...$ we
obtain for  $\gamma=1$ that the dynamics on the Center Manifold of the origin
is dictated by the same equation \eqref{rest1}. Hence, the origin  $u=0$ of
\eqref{rest1} is locally asymptotically unstable (saddle type). Then
$\mathbf{x}=\mathbf{0}$ is also a saddle point for the full 5D vector field.

\end{appendix}

\providecommand{\href}[2]{#2}\begingroup\raggedright
\bibstyle{apsrev4-1}
\endgroup
\end{document}